\documentclass[a4paper,11pt]{article}
\pdfoutput=1 
\usepackage[symbol]{footmisc} 
\usepackage{jheppub} 

\usepackage{color, colortbl}
\usepackage[T1]{fontenc} 
\usepackage[utf8]{inputenc}
\usepackage[T1]{fontenc}
\usepackage{epsfig,latexsym}
\usepackage{amsmath}
\usepackage{caption}
\usepackage{subcaption}
\usepackage[dvipsnames]{xcolor}
\usepackage{adjustbox}
\usepackage{tensor}
\usepackage{graphicx}
\usepackage{verbatim}
\graphicspath{ {Figs/} }
\usepackage{mathrsfs}
\usepackage{amssymb}
\usepackage{multirow}
\usepackage{epsfig}
\usepackage{color,colordvi}
\usepackage{appendix}
\usepackage{slashed}
\usepackage{array}
\usepackage{cancel}
\usepackage{float}
\usepackage[font=footnotesize, justification=centering]{caption}
\usepackage{epsf}
\usepackage{amsmath}
\usepackage{subdepth}
\usepackage{physics}
\usepackage{MnSymbol}
\usepackage{rotating}
\usepackage{multirow}
\usepackage{seqsplit}
\usepackage[normalem]{ulem}
\usepackage{tcolorbox}
\usepackage{bbold}
\usepackage{changepage}

\DeclareMathAlphabet{\mathpzc}{OT1}{pzc}{m}{it}
\allowdisplaybreaks

\newcommand{\beq}{\begin{equation}}
	\newcommand{\eeq}{\end{equation}}
\renewcommand{\[}{\left[}
\renewcommand{\]}{\right]}
\renewcommand{\(}{\left(}
\renewcommand{\)}{\right)}
\def\eq#1{{Eq.~(\ref{#1})}}

\newcolumntype{C}{>{\centering\arraybackslash}b{1.3cm}}
\newcommand{\be}{\begin{eqnarray}}
	\newcommand{\ee}{\end{eqnarray}}
\newcommand{\bea}{\begin{eqnarray}}
	\newcommand{\eea}{\end{eqnarray}}
\newcommand{\bi}{\begin{itemize}}
	\newcommand{\ei}{\end{itemize}}
\newcommand{\ben}{\begin{enumerate}}
	\newcommand{\een}{\end{enumerate}}
\def\bes{\begin{equation*}}
	\def\ees{\end{equation*}}
\def\bead{\begin{aligned}}
	\def\eead{\end{aligned}}
\def\bmat{\left(\begin{matrix}}
	\def\emat{\end{matrix}\right)}

\newcommand{\nocontentsline}[3]{}
\newcommand{\tocless}[2]{\vspace{10pt}\bgroup\let\addcontentsline=\nocontentsline#1{ #2}\egroup}
\newcommand{\toclesslab}[3]{\bgroup  \let\addcontentsline=\nocontentsline#1{#2\label{#3}}\egroup}

\def\Re{\text{Re}}
\def\Im{\text{Im}}
\def\diag{\text{diag}}

\def\cI{{\cal I}}

\def\cO{{\cal O}}
\def\cT{{\cal T}}
\def\CKM{\text{CKM}}
\def\Yu{{Y_{u\vphantom{d}}^{\vphantom{\dagger}}}}
\def\Yud{{Y_{u\vphantom{d}}^\dagger}}
\def\Yd{{Y_{d}^{\vphantom{\dagger}}}}
\def\Ydd{{Y_{d}^\dagger}}

\definecolor{Ecolor}{RGB}{106,157,235}
\definecolor{lightgrey}{RGB}{220,220,220}
\definecolor{newlightgrey}{RGB}{235,235,235}

\title{Opportunistic CP Violation}

\author[a,b,c]{Quentin Bonnefoy,}
\author[c,d,e,f]{Emanuele Gendy,}
\author[c,g]{Christophe Grojean,}
\author[h]{Joshua~T.~Ruderman}

\affiliation[a]{Berkeley Center for Theoretical Physics, University of California, Berkeley, CA 94720, USA}
\affiliation[b]{Theoretical Physics Group, Lawrence Berkeley National Laboratory, Berkeley, CA 94720, USA}
\affiliation[c]{Deutsches Elektronen-Synchrotron DESY, Notkestr. 85, 22607 Hamburg, Germany}
\affiliation[d]{Institute of Theoretical Physics, Universit\"at Hamburg, 22761 Hamburg, Germany}
\affiliation[e]{Kavli Institute for Theoretical Physics, UC Santa Barbara, Santa Barbara, CA 93106, USA}
\affiliation[f]{Technische Universit\"{a}t M\"{u}nchen, Physik-Department, 85748 Garching, Germany}
\affiliation[g]{Institut f\"ur Physik, Humboldt-Universit\"at zu Berlin, D-12489 Berlin, Germany}
\affiliation[h]{Center for Cosmology and Particle Physics, Department of Physics, New York University, New York, NY 10003, USA}

\emailAdd{q.bonnefoy@berkeley.edu}
\emailAdd{emanuele.gendy@tum.de}
\emailAdd{christophe.grojean@desy.de}
\emailAdd{ruderman@nyu.edu}

\abstract{In the electroweak sector of the Standard Model, CP violation arises through a very particular interplay between the three quark generations, as described by the Cabibbo--Kobayashi--Maskawa (CKM) mechanism and the single Jarlskog invariant $J_4$. Once  generalized to the Standard Model Effective Field Theory (SMEFT), this peculiar pattern gets modified by higher-dimensional operators, whose associated Wilson coefficients are usually split into CP-even and odd parts. However, CP violation at dimension four, i.e., at the lowest order in the EFT expansion, blurs this distinction: any Wilson coefficient can interfere with $J_4$ and mediate CP violation. In this paper, we study such interferences at first order in the SMEFT expansion, $\order{1/\Lambda^2}$, and we capture their associated parameter space via a set of 1551 linear CP-odd flavor invariants. This construction describes both new, genuinely CP-violating quantities as well as the interference between $J_4$ and CP-conserving ones. We call this latter possibility \textit{opportunistic CP violation}.
	Relying on an appropriate extension of the matrix rank to Taylor expansions, which we dub \emph{Taylor rank}, we define a procedure to organize the invariants in terms of their magnitude, so as to retain only the relevant ones at a given precision. We explore how this characterization changes when different assumptions are made on the flavor structure of the SMEFT coefficients. Interestingly, some of the CP-odd invariants turn out to be less suppressed than $J_4$, even when they capture opportunistic CPV, demonstrating that CP-violation in the SM, at dimension 4, is \textit{accidentally small.} 	}

\begin{document} 
	\begin{flushright}
		DESY 23-018\\
		HU-EP-22/39\\
		TUM-HEP-1453/23
	\end{flushright}
	\maketitle
	\flushbottom
	
	\section{Introduction}\label{section:intro}
	
	The  Lagrangian composed of $SU(3)_C \times SU(2)_L \times U(1)_Y$-invariant operators built with Standard Model fields and having dimension less than or equal to 4,
	which we denote by SM$_4$, has until now proven to be  in remarkable agreement with experiments. In particular, the observed CP violation (CPV) is consistent with the pattern expected from the SM$_4$, where it arises thanks to the simultaneous existence of up- and down-quarks, as well as the presence of (at least) three generations. This is the Cabibbo--Kobayashi--Maskawa (CKM) model, and besides describing how CP is broken, it also gives an explanation of why its breaking is so small, despite the phase associated to it being of order one. It is the consequence of both the phenomenological smallness of physical parameters in the Yukawa sector and of the fact that all of these parameters have to  come simultaneously into play for CP to be violated. 
	This fact can conveniently be captured in a flavor-invariant way, i.e.,  not affected by unitary rotations in flavor space, via a single quantity, the Jarlskog invariant~\cite{Jarlskog:1985ht, Jarlskog:1985cw, Bernabeu:1986fc}:
	\begin{align}
		J_4=\Im\Tr\left[\Yu \Yud,\Yd \Ydd \right]^3=3\,\Im\text{Det}\comm{\Yu\Yud}{\Yd\Ydd} \ .
	\end{align}
	This quantity can be interpreted as an order parameter for CPV in the SM$_4$, in the sense that CP is conserved if and only if $J_4=0$.
	
	If no new light degrees of freedom are assumed to lie below, or close to, the weak scale, then deviations from the SM$_4$ can be parametrized through its extension into an Effective Field Theory (EFT), whose Lagrangian supplements that of the SM$_4$ with an infinite tower of $SU(3)_C \times SU(2)_L \times U(1)_Y$-invariant operators of any dimension built with the SM$_4$ fields, i.e.
	\begin{align}
		\mathcal{L}=\mathcal{L}_{\text{SM}_4}+\sum_i\frac{C_i}{\Lambda^{d_i-4}}\mathcal{O}_i \ ,
	\end{align}
	where $\mathcal{O}_i$ is an operator of dimension $d_i>4$, $C_i$ a complex coefficient (sometimes called Wilson coefficient) generically of order one (modulo possible selection rules and after taking $\hbar$ dimensions into account) and $\Lambda$ is a dimensionful scale associated to heavy new physics.
	
	This procedure defines the so-called Standard Model Effective Field Theory (SMEFT).
	Clearly, the complex phases of the new coefficients $C_i$ can induce additional CPV beyond the SM$_4$ alone. 
	In particular, the $C_i$'s corresponding to operators containing fermionic fields have to be intended as matrices in flavor space, and as such contain a large number of a priori CP-odd phases. 
	In~\cite{Bonnefoy:2021tbt} we identified a set of 699 CP-odd flavor invariants linear in the coefficients\footnote{This number refers to operators containing fermions, only, and does not include the 6 CP-odd coefficients of fully bosonic operators. Since their flavor structure is trivial, we will not consider these operators in this work.} $C_i$ which, together with the SM$_4$ $J_4$ and the coefficients of the bosonic CP-odd operators at dimension six, represent the order parameters of CPV at order $\order{1/\Lambda^2}$ in observables, meaning that no CPV can be observed at this order if  and only if all the elements of this set vanish\footnote{Among the Wilson coefficients associated to dimension-six operators, only a subset contributes to observables at $\order{1/\Lambda^2}$ due to non-interference rules, assuming the limit of vanishing neutrino masses $m_\nu\to0$. In~\cite{Bonnefoy:2021tbt} and in this work, we neglect higher orders and the associated (possibly CP-odd) invariants non-linear in the $C_i$'s.}.
	
	Although this provides the correct characterization of the new sources of CPV at dimension six, it does not tell the full story. Indeed, since CP is broken in the SM$_4$ already, any real or complex dimension-six parameter can interfere with $J_4$ and produce additional CPV in observables. As it must be possible to express observables in terms of flavor-invariant objects, this means that we must be able to enlarge the set of CP-odd invariants linear in the $C_i$'s so that they capture the set of all appropriate Wilson coefficients. This procedure produces what we refer to as a \emph{maximal set}, containing 1551 invariants which can contribute to CPV observables at order $\order{1/\Lambda^2}$. 
	
	When $J_4\to 0$, this set would reduce again to the new sources of CPV found in~\cite{Bonnefoy:2021tbt}, which form the set of CPV order parameters, and which we dub here the \emph{minimal set}.
	This means that we can write, for any invariant $L$ in the maximal set,
	\beq
	L=\sum_{\tilde a\,\in \text{\,min.\,set}} \cI_{\tilde a}L_{\tilde a}+J_4R_4 \ ,
	\eeq
	where the $L_{\tilde a}$'s belong to the minimal set and are linear in the $C_i$, the $\cI_{\tilde a}$'s are CP-even flavor invariants which only depend on SM$_4$ coefficients, and $R_4$ is a CP-even flavor invariant linear in the $C_i$. That observables can be expressed in terms of invariants means that a similar decomposition exists for observables (except that $\cI_{\tilde a}$ and $R_4$ need not be invariants in this case). Since $J_4\ll 1$, one might expect that only the $L_{\tilde a}$ contribute significantly to CPV\@. However, this is incorrect, as we will show below, because often $R_4\gg 1$, and therefore invariants beyond the minimal set can contribute significantly to CPV observables.
	 This surprising result implies that many more invariants than those in the minimal set should be considered at a given level of precision, and a key result of this paper is the development of a formalism for doing so. We will refer to the sources of CP violation parametrized by the minimal set as \emph{direct CPV}, while, in a sense that will be made clear in the following, the difference between maximal and minimal sets will be what we will identify as \emph{opportunistic CPV}. 
	
	In principle, the large number of invariants in a maximal set, which follows from the large number of free parameters of SMEFT at dimension-six, makes it hard to study the set as a whole analytically. Fortunately, one can rely on a hierarchy within the invariants. This way, one has a handle to discriminate which invariants are the most important, and to restrict to appropriate subsets of invariants. This hierarchy follows from the flavor structure of the Yukawa sector of the SM$_4$ and from the phenomenological values of its parameters. In particular, one finds
	\begin{align}
		J_4\approx 8\times 10^{-24} \ .
	\end{align}
The same applies to our invariants which are built using Yukawa matrices and Wilson coefficients. For example, consider the modified Yukawa operator
	\begin{align}
		\mathcal{O}_{uH}=\frac{1}{\Lambda^2}C_{uH,\, ij}\left(H^\dagger H\right) \bar{Q}_iu_j\tilde{H}\ .
		\label{eq:modifiedYukawa}
	\end{align}
	Using its coefficient $C_{uH,\, ij}$ we can build e.g.\,the invariant
	\begin{align}
		\Im\Tr[C_{uH}Y_u^\dagger]\approx-3.7\,\text{Im}\,C_{uH,33} \ ,
		\label{eq:invexample1}
	\end{align}
	as well as
	\begin{align}
		\Im\Tr[C_{uH\vphantom{d}}^{\vphantom{\dagger}}Y_{\vphantom{d}u}^\dagger Y_d^{\vphantom{\dagger}}Y_d^\dagger]\approx 2.4\times10^{-4}\,\Im\Tr[C_{uH}Y_u^\dagger]
		-3.7\times10^{-5}\,\text{Im}\,C_{uH,23} \ .
		\label{eq:invexample2}
	\end{align}
	(The expressions above are evaluated in the so-called up-basis, to be defined below. Moreover, only the leading terms have been retained, under the flavor anarchic assumption that all Wilson coefficients are order one.) More analytically, we can track the size of any flavor invariant via an explicit power counting describing the flavor hierarchies, which is for instance achieved by the Wolfenstein parametrization and its expansion in terms of the small sine of the Cabibbo angle $\lambda\approx0.225$~\cite{Charles:2004jd}. We employ this expansion and ask how many independent invariants can contribute to an observable at a given level of precision, i.e.~at a given order in $\lambda$. To answer this question, one would naively truncate the expansion of each invariant at a certain order and check how many of them are linearly independent. However, this procedure should be carried out with care, and to this end we introduce in section~\ref{section:lambdarank} the new notion of {\it Taylor rank} of a matrix expanded in powers of a small parameter.
	As can be readily seen, the result is highly dependent on the flavor structure that is assumed to hold for the Wilson coefficients, in this case $C_{uH}$. Indeed, the size of the invariants in Eqs.~\eqref{eq:invexample1}-\eqref{eq:invexample2} is captured by the numerical prefactors for an anarchic flavor structure for the matrix $C_{uH,\, ij}$ whose entries are all taken to be $\order{1}$. 
	However, such an anarchic structure is greatly constrained by observations via measurements of mesons oscillations, electric dipole moments, and lepton flavor violation~\cite{Isidori:2010kg, ESPPPG:2019qin, Aebischer:2020dsw, Silvestrini:2018dos, Pruna:2014asa, Feruglio:2015yua} with lower bounds on the scale of New Physics (NP) on the order of $\order{10^3 \text{ TeV}}$. Were this the common suppression scale of the EFT, it would render its impact on observables irrelevant for collider physics, as well as imply a strong fine tuning of the scalar mass of the Higgs. On the other hand, we know that, already at dimension four, the flavorful parameters of the Standard Model are far from being anarchic, spanning a range that scans almost 6 orders of magnitude. Thus, different ansatzes have emerged addressing the flavor structure of SMEFT\@. These approaches have the twofold advantage of allowing us to bring the lower bounds on the NP scale down to the TeV region, and of reducing the number of relevant free parameters added by the EFT expansion. They usually proceed by either relating the SMEFT flavor structure to that of the SM$_4$ or deriving it from certain families of UV models. The archetype of the first kind of approach is represented by the Minimal Flavor Violation (MFV) ansatz~\cite{DAmbrosio:2002vsn, Isidori:2010kg}. 
	As another benchmark example, we will consider the $U(2)^5$ scenario~\cite{Barbieri:2011ci, Barbieri:2012uh, Blankenburg:2012nx, Faroughy:2020ina}, where the flavor symmetry is restricted to the first two generations of quarks and leptons. The last possibility we take into consideration is the one dictated by the so called Froggatt--Nielsen mechanism~\cite{Froggatt:1978nt}. 
	
	This paper is organized as follows: in section~\ref{section:opportunisctiCPV}, we motivate the need to define a formalism which systematically captures all independent physical quantities capable of contributing to CPV observables. In section~\ref{section:maximalset}, we then define maximal sets of CP-odd invariants and explain how we use them to describe the parameter space of CP-odd observables at $\order{1/\Lambda^2}$, providing examples for some dimension-six operators. In section~\ref{section:lambdarank}, we define the Taylor rank and explain how it helps in counting the number of independent invariants needed to span the parameter space of a CP-odd observable at fixed order in the $\lambda$ expansion. In section~\ref{section:flavorassumptions}, we explain the four different flavor scenarios that we take as benchmarks to understand how the results of section~\ref{section:lambdarank} change with them. 
	We then summarize our conclusions in section~\ref{section:conclusion}.
	Appendix \ref{appendix:singletrace} justifies the procedure we follow to build invariants, while in appendix~\ref{appendix:opportunisticRephasingInvariants} we illustrate features of opportunistic CPV at the level of rephasing invariants. Appendix~\ref{section:taylorrank} elaborates on the notion of Taylor rank introduced in section~\ref{section:lambdarank}. In appendix~\ref{section:taylorMFV} we provide more details on our use of the MFV expansion, and in appendix~\ref{section:U(2)5} we study the $U(2)^5$ ansatz using the tools of the Plethystic program. In Appendices~\ref{section:bilinearoperators} and~\ref{section:4fermioperators}, we list maximal sets for each operator of SMEFT at dimension-six. Finally, we collect in appendix~\ref{section:coeffvalues} the more cumbersome expressions for the explicit values of some quantities obtained in the main text.
	We always work in the limit of vanishing neutrino masses $m_\nu\to0$
	and use the so-called Warsaw basis~\cite{Grzadkowski:2010es}, listing all independent dimension-six operators in SMEFT, which we recap for convenience in Tables~\ref{Bilinearlist} and~\ref{4Fermilist}.
	
	\section{Opportunistic CP violation: an overview}\label{section:opportunisctiCPV}
		
	As mentioned in the introduction, in~\cite{Bonnefoy:2021tbt} we introduced the study of CP-violation in the Standard Model Effective Field Theory at order $1/\Lambda^2$ via CP-odd flavor invariants, each structured as single traces over the flavor indices containing only one power of a dimension-six operator coefficient. First, we briefly review our construction. We then explain why the minimal set of invariants that we introduced in~\cite{Bonnefoy:2021tbt} is not sufficient to describe the full parameter space of CPV at $\order{1/\Lambda^2}$.
	We begin by focusing on the simple example where the only operator turned on, beyond those of the SM$_4$, is
	\begin{align}
		\mathcal{O}_{HQ}^{(1)}=C_{HQ,ij}^{(1)}(H^\dagger i \overleftrightarrow{D}H)(\bar{Q_i}\gamma^\mu Q_j)\ .
	\end{align}

	\subsection{Parametrizing flavor}\label{conventionsSection}
	
	Before describing CPV flavor invariants in SMEFT, let us present the relevant flavor concepts and parametrizations that we use throughout this paper. First, notice that, in the unbroken phase, the kinetic part of the SM$_{4}$ Lagrangian is invariant under a $U(3)^5=U(3)_{Q_L}\times U(3)_{u_R}\times U(3)_{d_R}\times U(3)_{L_L}\times U(3)_{e_R}$ global flavor group, where each factor acts on the flavor indices of the associated fermion fields (we will drop the chirality indices in the following). Then, we can assign all coefficients in the Yukawa sector and of the SMEFT operators spurious transformation properties so that the whole SMEFT Lagrangian is formally invariant under the whole flavor group. 
	For the Yukawa couplings at dimension four, the transformation properties under the non-abelian part of the flavor group are as listed in Table~\ref{tab:ytrasmforma}. On top of that, each (anti-)fundamental representation has a charge $(-)1$ under the associated abelian group in the decomposition $U(3)_X=SU(3)_X\times U(1)_X$, with $X=Q,u,d,L,e$.
	\begin{table}[H]
		\centering
		\begin{tabular}{c|c|c|c|c|c}
			& $SU(3)_Q$ & $SU(3)_u$ & $SU(3)_d$ & $SU(3)_L$ & $SU(3)_e$\\\hline
			$Y_u$ &$\mathbf{3}$ & $ \mathbf{\bar{3}}$ & $\mathbf{1}$ &$\mathbf{1}$ &$\mathbf{1}$  \\[0.1cm]
			$Y_d$ &$\mathbf{3}$ & $ \mathbf{1}$ & $\mathbf{\bar{3}}$ &$\mathbf{1}$ &$\mathbf{1}$  \\[0.1cm]
			$Y_e$ &$\mathbf{1}$ & $ \mathbf{1}$ & $\mathbf{1}$ &$\mathbf{3}$ &$\mathbf{\bar{3}}$
		\end{tabular}
		\caption{Flavor transformation properties of the Yukawa matrices treated as spurions}
		\label{tab:ytrasmforma}
	\end{table}
	Using flavor transformations, one can reach flavor bases where the Yukawa matrices have a specific form, and which we will use in the following  to explicitly evaluate invariants. Mostly, we will use the {\it up basis}, defined so that 
	\beq
	Y_u=\diag(y_u,y_c,y_t) \ , \quad Y_d=V_\CKM \cdot \diag(y_d,y_s,y_b) \ , \quad Y_e=\diag(y_e,y_\mu,y_\tau) \ ,
	\label{eq:upBasis}
	\eeq
	and the  {\it down basis}
	\beq
	Y_{\smash{u}\vphantom{d}}^{\vphantom{\dagger}}=V_{\smash{\CKM}\vphantom{d}}^\dagger\cdot\diag(y_u,y_c,y_t) \ , \quad Y_d=\diag(y_d,y_s,y_b) \ , \quad Y_e=\diag(y_e,y_\mu,y_\tau) \ ,
	\label{eq:downBasis}
	\eeq
	where all $y$'s are real and positive and $V_\CKM$ is the CKM matrix. For later use, we define the combinations $X_{u,d,e}^{\vphantom{\dagger}}\equiv Y_{u,d,e}^{\vphantom{\dagger}}Y_{u,d,e}^\dagger$. When picking either of the above two bases, we exhaust the non-abelian part of the flavor group completely. Some of the remaining $U(1)$ factors can be used e.g.~to bring the CKM matrix into the following form~\cite{Chau:1984fp}:
	\beq
	V_\CKM=\bmat
	c_{12} c_{13} & c_{13} s_{12} & s_{13} e^{-i\delta_\CKM} \\
	-c_{23} s_{12}-c_{12} s_{13} s_{23} e^{i\delta_\CKM} & c_{12} c_{23}-s_{12} s_{13} s_{23} e^{i\delta_\CKM} & c_{13} s_{23} \\
	s_{12} s_{23}-c_{12} c_{23} s_{13} e^{i\delta_\CKM} & -c_{12} s_{23}-c_{23} s_{12} s_{13} e^{i\delta_\CKM} & c_{13} c_{23}
	\emat \ ,
	\label{paramCKM}
	\eeq
	where $c_{X},\, s_{X}=\cos(\theta_{X}),\, \sin(\theta_{X})$. In the quark sector, this choice leaves only the $U(1)_B$ factor corresponding to baryon number unbroken, while in the lepton sector the three $U(1)_{L_i}$, with $i=1,\,2,\,3$, are left unbroken. We also make use of the Wolfenstein parametrization, where the CKM matrix is expressed in terms of three $\order{1}$ parameters, $A\approx 0.81$, $\rho\approx 0.16$, and $\eta\approx 0.36$, as well a fourth parameter, $\lambda\approx0.225$, corresponding to the sine of the Cabibbo angle\footnote{These values are obtained in the $\overline{\text{MS}}$ scheme at a scale $\mu=M_t = 173.1 \text{ GeV}$, although, to a very good precision, only the parameter $A$ runs~\cite{Grossman:2022ehc}. (Only the contributions from the SM to the RG flow have been taken into account here, while the impact of dimension-6 operators has been neglected.) This choice, made here and for Eqs.~\eqref{eq:massesscalingintro}-\eqref{eq:massesscalingend}, is not an obligated one, and any scale $\mu\gtrsim M_Z$ is allowed. Different choices of $\mu$ would slightly affect the results presented in the following as they imply different $\lambda$ scaling. However, the presented framework and algorithms would stay unchanged.}~\cite{Charles:2004jd}, 
	\begin{multline}
		V_{\text{CKM}}=\\
		{
			\footnotesize\left(
			\begin{array}{ccc}
				\sqrt{1-\lambda ^2} \sqrt{1-A^2 \lambda ^6 \left(\eta ^2+\rho ^2\right)} & \lambda  \sqrt{1-A^2 \lambda ^6 \left(\eta ^2+\rho ^2\right)} & A \lambda ^3 (\rho -i \eta ) \\
				-\lambda  \sqrt{1-A^2 \lambda ^4}-A^2 \sqrt{1-\lambda ^2} \lambda ^5 (\rho +i \eta ) & \sqrt{1-\lambda ^2} \sqrt{1-A^2 \lambda ^4}-A^2 \lambda ^6 (\rho +i \eta ) & A \lambda ^2 \sqrt{1-A^2 \lambda ^6 \left(\eta ^2+\rho ^2\right)} \\
				A \lambda ^3-A \lambda ^3 \sqrt{1-\lambda ^2} \sqrt{1-A^2 \lambda ^4} (\rho +i \eta ) & -A \sqrt{1-\lambda ^2} \lambda ^2-A \lambda ^4 \sqrt{1-A^2 \lambda ^4} (\rho +i \eta ) & \sqrt{1-A^2 \lambda ^4} \sqrt{1-A^2 \lambda ^6 \left(\eta ^2+\rho ^2\right)} \\
			\end{array}
			\right)}\ ,
		\label{eq:ckmscaling}
	\end{multline} 
	reproducing the usual approximations when expanded in $\lambda$. With the same philosophy, the quark and lepton masses can also be assigned a $\lambda$-suppression. One possible choice is 
	\begin{align}
		\{y_u,y_c,y_t\}&=\{a_u\lambda^8, a_c\lambda^4,a_t\lambda^0\}\label{eq:massesscalingintro}	\\ 
		\{y_d,y_s,y_b\}&=\{a_d\lambda^7, a_s\lambda^5,a_b\lambda^3\}\\ 
		\{y_e,y_\mu,y_\tau\}&=\{a_e\lambda^9, a_\mu\lambda^5,a_\tau\lambda^3\}\ ,
	\end{align}
	with 
	\begin{align}
		\{a_u, a_c,a_t\}&\approx\{1.03, 1.33, 0.93\}\\ 
		\{a_d, a_s,a_b\}&\approx\{0.50, 0.51,1.36\}\\ 
		\{a_e, a_\mu,a_\tau\}&\approx\{1.88, 1.02,0.88\}\ ,
		\label{eq:massesscalingend}	
	\end{align}
	where the running masses in the $\overline{\text{MS}}$ scheme at the renormalization scale $\mu=M_t = 173.1 \text{ GeV}$ are taken~\cite{Huang:2020hdv}.
	Then, all the quantities sensitive to flavor can be expressed as an expansion in powers of $\lambda$ to consistently obtain an approximation for their magnitude. In particular, and relevantly for the case we are interested in here, the Jarlskog invariant $J_4$ can be expanded in terms of $\lambda$ to get
	\begin{align}
	\label{J4lambdaOrder}
		J_4=6 a_b^4 a_s^2 a_t^4a_c^2  A^2 \eta  \lambda ^{36}+\order{\lambda^{38}}\ .
	\end{align}

	\subsection{Minimal set: flavor-invariant order parameters for CPV at $\order{1/\Lambda^2}$}\label{minimalSection}
	
	Let us now review a consistent definition of order parameters for CP violation at $\order{1/\Lambda^2}$ in SMEFT\@. When performing a flavor transformation, the real and imaginary entries of a given coefficient are in general mixed, and CP is conserved if and only if one can find a flavor basis where all coefficients are real\footnote{Models with discrete symmetries yield caveats to this statement. See Ref.~\cite{Ivanov:2015mwl} for an example, or section 4.3 of Ref.~\cite{Trautner:2016ezn} for more details and references.}. Thus, rather than interpreting a single imaginary coefficient as a source of CPV, we characterize CP violation in SMEFT at leading order using flavor invariants. More specifically, mimicking the role of the Jarlskog invariant $J_4$ in the SM$_{4}$, we ask \emph{which flavor invariants vanish if and only if CP is conserved at leading order in SMEFT}\@. The set of these invariants, built as single traces in flavor space and with only one power of the dimension-six operator coefficient, is what we defined \emph{minimal set}. The choice of single trace invariants in this context may seem quite arbitrary. However, as we show in appendix~\ref{appendix:singletrace}, it actually represents the most general choice under the desired conditions.
	
	Let us give an example, using the $\mathcal{O}_{HQ}^{(1)}$ operator\footnote{Due to the linearity with respect to the $C_i$ of any observable at $\order{1/\Lambda^2}$, we can focus on a given SMEFT operator without assuming that the other ones are turned off.} defined above.
	After fixing a basis, the coefficient $C_{HQ}^{(1)}$ is represented, in flavor space, by a $3\times3$ hermitian matrix
	\begin{align}
		C_{HQ}^{(1)}=\begin{pmatrix}
			\rho_{11}&\rho_{12}&\rho_{13}\\
			\rho_{12}&\rho_{22}&\rho_{23}\\
			\rho_{13}&\rho_{23}&\rho_{33}\\
		\end{pmatrix}+i
		\begin{pmatrix}
			0&\eta_{12}&\eta_{13}\\
			-\eta_{12}&0&\eta_{23}\\
			-\eta_{13}&-\eta_{23}&0\\
		\end{pmatrix} \ .
		\label{eq:CHQparamaetrization}
	\end{align}
	One sees immediately that there are three independent complex phases in $C_{HQ}^{(1)}$. Consistently, in~\cite{Bonnefoy:2021tbt} we identified the minimal set for this operator as being composed of the following three invariants\footnote{The lower indices labeling the invariants are chosen to match the definitions of the subsequent Eq.~\eqref{eq:CHQmaxsetintro}.}
	\beq
	L_1=\Im\Tr(X_u^{\vphantom{\dagger}}X_{\smash{d}}^{\vphantom{\dagger}}C_{
		{HQ}}^{(1)}) ,\ \
	L_5=\Im\Tr(X_u^{2\vphantom{\dagger}}X_{\smash{d}}^{2\vphantom{\dagger}}C_{
		{HQ}}^{(1)}),\ \
	L_7=\Im\Tr(X_u^{\vphantom{\dagger}}X_{\smash{d}}^{\vphantom{\dagger}}X_u^{2\vphantom{\dagger}}X_{\smash{d}}^{2\vphantom{\dagger}}C_{
		{HQ}}^{(1)}) \ .
	\label{eq:minsetforCHQ}
	\eeq
	We recall the definitions $X_{u,d}^{\vphantom{\dagger}}\equiv Y_{u,d}^{\vphantom{\dagger}}Y_{u,d}^\dagger$.
	When $J_4=0$, the set in Eq.~\eqref{eq:minsetforCHQ} is enough to capture all CPV in the theory, so that CP is conserved if and only if the whole set vanishes. More precisely, this set is designed so that the statement is valid in whatever parametric limit we choose to reach $J_4=0$ (see~\cite{Bonnefoy:2021tbt} for more details).
	
	Extending this to all dimension-six SMEFT operators, we find 699 independent CPV order parameters coming from fermionic operators, which have to be set to zero together with $J_4$ and the 6 coefficients of CP-odd bosonic dimension-six operators for CP to be conserved at $\order{1/\Lambda^2}$ (see the next section for an example). This number is much smaller than the 1149 phases mentioned in~\cite{Alonso:2013hga}. To explain this, we notice that in either the up or the down basis, Eqs.~\eqref{eq:upBasis} and~\eqref{eq:downBasis}, the lepton sector of the flavor group still enjoys a $U(1)^3$ symmetry consisting of phase rotations acting on the three different lepton generations. Observables, too, have to respect this symmetry of the scattering states. The off-diagonal entries of the coefficients of dimension-six operators containing leptons are charged under this $U(1)^3$, and thus to enter observables they need to be multiplied with objects carrying opposite charge. Since no such object exists in the SM$_{4}$ Lagrangian, they will be multiplied by coefficients carrying at least an extra  ${1}/{\Lambda^2}$ suppression, and will thus contribute to observables starting from $\order{{1}/{\Lambda^4}}$. This consideration allowed us to distinguish between \emph{primary} sources of CPV in SMEFT, i.e.~those that can enter observables at order ${1}/{\Lambda^2}$ and are captured by our linear invariants, and the remaining \emph{secondary} CPV sources.
	
	\subsection{The need for a maximal set
	}\label{opportunisticDef}
	
	Using again the operator $\mathcal{O}_{HQ}^{(1)}$, we illustrate why we need to enlarge our definition of minimal set in order to capture the full parameter space of CP violation at $\order{1/\Lambda^2}$. We presented in Eq.~\eqref{eq:minsetforCHQ} the minimal set for $\mathcal{O}_{HQ}^{(1)}$, which describes the sources of CP violation which remain when $J_4=0$. We will refer to such sources of CPV as \emph{direct CPV}\@. We argued that, when $J_4=0$, they are enough to capture all CPV in the theory. However, in the real world $J_4$, although small, is apparently nonzero. This implies that, in order to predict the values of CPV observables, one needs strictly more information about the SMEFT Wilson coefficients than contained in \eqref{eq:minsetforCHQ}. Intuitively, this is because any physical parameter at $\order{1/\Lambda^2}$ can interfere with $J_4$ to produce additional CPV\@. Beyond the trivial cases where a CP-even flavor invariant multiplies $J_4$ to form a CPV quantity\footnote{We bar these ``factorized" invariants, which can be naturally factorized into (CP even)$\times$(CP odd), since they are such that the source of CPV from the CKM matrix does not communicate with the flavor structure of the Wilson coefficient. This explains why, as we will see below, all primary coefficients are captured by the maximal sets in the quark sector, but not in the lepton sector.}, one should study how the CKM phase affects flavor invariants which intertwine the flavor indices of the Wilson coefficients with those of Yukawa matrices. As far as CPV at $\cO(1/\Lambda^2)$ is concerned, and as we explain in Appendix~\ref{appendix:singletrace}, it suffices to consider CP-odd invariants of the form $\Im\Tr$ of a flavor-covariant monomial expression linear in the SMEFT Wilson coefficients. We label \emph{opportunistic CP violation} the sources of CPV described by such invariants modulo the minimal set. By definition of the minimal set, opportunistic CPV quantities vanish when either $J_4\to 0$ or $\Lambda\to \infty$. 
	
	As an example, let us consider the measure of direct CPV in kaon physics, $\epsilon'/\epsilon$, which receives BSM contributions at order $1/\Lambda^2$. In SMEFT at dimension six, several Wilson coefficients induce significant contributions, among which that of $C_{HQ}$ reads~\cite{Aebischer:2018quc,Aebischer:2018csl},
	\be
	\(\frac{\epsilon'}{\epsilon}\)_\text{BSM}\approx -191\frac{\text{Im}\left\{\(C_{HQ,12}^{(1)}\)\lambda_u\right\}}{\abs{\lambda_u}}\frac{(\text{TeV})^2}{\Lambda^2} \ ,
	\label{kaon1}
	\ee
	where the coefficients are written in the down basis and evaluated at the weak scale, $\lambda_u\equiv V_{\CKM,us}^*V_{\CKM,ud}$ (this parameter is real with the phase conventions of the Wolfenstein parametrization, which we always use below) and we assumed order one Wilson coefficients. When the minimal set of \eqref{eq:minsetforCHQ}, consisting of three independent CP-odd physical quantities, is sent to zero one does not find that $\text{Im} \(C_{HQ,ij}^{(1)}\)$ and $(\epsilon'/\epsilon)_\text{BSM}\to 0$ (as, by construction, would happen if $J_4$ were zero), but instead that
	\beq
	\text{Im} \(C_{HQ,12}^{(1)}\)=\eta_{12}\approx - \eta\(47\rho_{13} +9\rho_{23}\) \ ,
	\label{kaon2}
	\eeq
	 where $\eta$ is defined in Eq.~\eqref{eq:ckmscaling} and $\rho_{ij}$ in Eq.~\eqref{eq:CHQparamaetrization}. This relation is obtained by solving the set of equations $L_1=L_5=L_7=0$ perturbatively in $\lambda$, which yields expressions for all $\text{Im} \(C_{HQ,ij}^{(1)}\)$ coefficients, in particular for $\text{Im} \(C_{HQ,12}^{(1)}\)$. Inserting the numerical values for the Yukawa matrices (in the down basis) presented in Section~\ref{conventionsSection}, assuming that the $\text{Re} \(C_{HQ,ij}^{(1)}\)$ coefficients are all of order one and without accidental cancellations between them, and restricting to the two largest contributions, one finds the above expression. Cancelling the minimal set is therefore not sufficient to cancel $(\epsilon'/\epsilon)_\text{BSM}$: there are strictly more than three independent CP-odd physical quantities at dimension six when $J_4\propto \eta \neq 0$. As we will show in the following, although this contribution originates from an interference with the SM$_4$ $J_4$, it is not as suppressed as $J_4$ itself.
	
	Instead of looking for several observables to probe the space of CP-odd physical quantities, we can directly work with flavor invariants. Consider therefore a fourth invariant $L_2$,
	\begin{align}
		L_2&\equiv\Im\Tr(X_u^{2\vphantom{\dagger}}X_{\smash{d}}^{\vphantom{\dagger}}C_{{HQ}}^{(1)}) \ .
	\end{align}
	By evaluating it e.g.~in the standard parametrization in the down-basis, we can check that, as long as $J_4\neq0$, $L_2$ is an independent object with respect to the remaining three. When $J_4\to0$, however, this cannot be the case anymore, as we proved that $L_{1,5,7}$ span the whole parameter space of CPV observables in this limit. 
	This line of reasoning leads us to write
	\begin{align}
		\cI_2 L_2=\cI_1L_1+\cI_5L_5+\cI_7L_7+J_4 R_4\ ,
		\label{eq:maxvsminsketchedexample}
	\end{align}
	where $R_4$ is a CP-even invariant, still linear in the coefficient $C_{HQ}^{(1)}$, although not necessarily expressed as a single trace of a polynomial expression. The $\cI_i$ coefficients, on the other hand, are combinations of the 10 independent CP-even invariants that can be built with $X_{u,d}$~\cite{Jenkins:2009dy}, namely
	\begin{align}
		\begin{array}{ll}
			I_{1,0}=\text{Tr}\left(X_u^{\vphantom{2}}\right), &\quad I_{0,1}=\text{Tr} \left(X_d^{\vphantom{2}}\right), \\
			I_{2,0}=\text{Tr}\left(X_u^2\right), &\quad I_{1,1}=\text{Tr} (X_u^{\vphantom{2}} X_d^{\vphantom{2}}), 
			\\
			I_{0,2}=\text{Tr}\left(X_d^2\right), &\quad I_{3,0}=\text{Tr}\left(X_u^3\right), \\
			I_{2,1}=\text{Tr}\left(X_u^2 X_d^{\vphantom{2}}\right), &\quad I_{1,2}= \text{Tr}\left(X_u^{\vphantom{2}} X_d^2\right), \\
			I_{0,3}=\text{Tr}\left(X_d^3\right), &\quad I_{2,2}=\text{Tr}\left(X_u^2 X_d^2\right).
		\end{array}
		\label{eq:SMinvariants}
	\end{align}
	As a matter of fact, a  relation in the form of Eq.~\ref{eq:maxvsminsketchedexample} can be found explicitly. Since the coefficients are quite involved, however, and their explicit form does not add any insight to the discussion, we stick to the generic expression in Eq.~\eqref{eq:maxvsminsketchedexample} and display explicit expressions for the coefficients in Appendix~\ref{section:coeffvalues}\@.
	 In any case, the existence of the independent invariant $L_2$ confirms the lesson learnt by dealing with $\epsilon'/\epsilon$: our minimal set is not enough to capture all CP violation at $\order{1/\Lambda^2}$ when $J_4\neq 0$, due to opportunistic CP violation. It only captures direct CPV, the set of CP-odd quantities that remain nonzero when $J_4\to0$ and only vanish in the limit $\Lambda\to\infty$. With the notation of Eq.~\eqref{eq:maxvsminsketchedexample}, an example of opportunistic CPV would be the quantity
	\begin{align}
		\cI_2 L_2-(\cI_1L_1+\cI_4L_4+\cI_6L_6)=J_4 R_4\ .
	\end{align}
	Equivalently, since the generated span is the same, we may as well just consider $L_2$ as capturing this additional source of CPV\@. 
	Carrying on along this line of reasoning, we can add as many invariants as possible on top of the minimal set until any other additional invariant is not independent from the other ones. For $C_{HQ}^{(1)}$, we can for example pick the set
	\begin{align}
		L_1&=\Im\Tr(X_u^{\vphantom{\dagger}}X_{\smash{d}}^{\vphantom{\dagger}}C_{
			{HQ}}^{(1)}),\nonumber&
		L_2&=\Im\Tr(X_u^{2\vphantom{\dagger}}X_{\smash{d}}^{\vphantom{\dagger}}C_{
			{HQ}}^{(1)}),\nonumber\\
		L_3&=\Im\Tr(X_u^{\vphantom{\dagger}}X_{\smash{d}}^{2\vphantom{\dagger}}C_{
			{HQ}}^{(1)}),\nonumber&
		L_4&=\Im\Tr(X_u^{\vphantom{\dagger}}X_{\smash{d}}^{\vphantom{\dagger}}X_u^{2\vphantom{\dagger}}C_{
			{HQ}}^{(1)}),\nonumber\\
		L_5&=\Im\Tr(X_u^{2\vphantom{\dagger}}X_{\smash{d}}^{2\vphantom{\dagger}}C_{
			{HQ}}^{(1)}),\nonumber&
		L_6&=\Im\Tr(X_{\smash{d}}^{\vphantom{\dagger}}X_u^{\vphantom{\dagger}}X_{\smash{d}}^{2\vphantom{\dagger}}C_{
			{HQ}}^{(1)}),\nonumber\\	
		L_7&=\Im\Tr(X_u^{\vphantom{\dagger}}X_{\smash{d}}^{\vphantom{\dagger}}X_u^{2\vphantom{\dagger}}X_{\smash{d}}^{2\vphantom{\dagger}}C_{
			{HQ}}^{(1)}),\nonumber&	
		L_8&=\Im\Tr(X_u^{\vphantom{\dagger}}X_{\smash{d}}^{2\vphantom{\dagger}}X_u^{2\vphantom{\dagger}}X_{\smash{d}}^{\vphantom{\dagger}}C_{
			{HQ}}^{(1)}),\nonumber\\
		L_9&=\Im\Tr(X_u^{2\vphantom{\dagger}}X_{\smash{d}}^{\vphantom{\dagger}}X_u^{\vphantom{\dagger}}X_{\smash{d}}^{2\vphantom{\dagger}}C_{
			{HQ}}^{(1)}).\label{eq:CHQmaxsetintro}
	\end{align}
	We refer to this as a \emph{maximal set} for $C_{HQ}^{(1)}$. Its cardinal matches the number of free coefficients in $C_{HQ}^{(1)}$, which shows, as announced, that they can all participate in CPV\@. This also trivially implies that fixing all the invariants of the maximal set to zero suffices to make $\(\epsilon'/\epsilon\)_\text{BSM}$ vanish, and that it can be decomposed along the maximal set. Below, we exhibit the combination of maximal set invariants which captures the leading contribution in \eqref{kaon1}. In section~\ref{section:maximalset}, we generalize this construction to all dimension-six operators of SMEFT\@.
	
	Interestingly, opportunistic CPV, as captured by the invariants of the maximal set, does not need to be as suppressed as the $J_4$ with which it interferes. As we mentioned in the introduction, we can expand the invariants in powers of the small parameter $\lambda\approx0.225$, in order to compare their magnitudes. If we perform our expansion on the invariants $L_1-L_9$ defined in Eq.~\eqref{eq:CHQmaxsetintro}, we get, in the up or down basis,
\beq
\bead
L_1&=-4 A a_b^2 a_t^2 \eta_{23} \lambda^8+...\\
L_2&=...+4 a_b^2 a_c^2 a_t^2 A\(\rho\eta_{13}+\eta\rho_{13}\) \lambda^{17}+...\\
L_3&=...-4 a_s^2a_b^2 a_t^2 A\eta_{13}\lambda^{19}+...\\
L_4&=...+4 a_c^2 a_t^4\(\(a_s^2 + a_b^2A^2\rho\)\eta_{12} + 
   a_b^2 A^2 \eta\rho_{12}\)\lambda^{19}+...\\
L_5&=...-4 a_s^2a_b^2a_c^2 a_t^2\eta_{12}\lambda^{25}+...\\
L_6&=...+4 a_s^2a_b^4 a_t^2A^2\eta\(\rho_{11}-\rho_{22}\)\lambda^{28}+...\\
L_7&=...-4 a_s^2a_b^4a_c^2 a_t^4A\eta\rho_{23}\lambda^{34}+...\\
L_8&=...+4 a_s^2a_b^4a_c^2 a_t^4A^2\eta\rho_{22}\lambda^{36}+...\\
L_9&=...+4 a_s^2a_b^4a_c^2 a_t^4A^2\eta\rho_{33}\lambda^{36}+...
\eead
\label{eq:CHQmaxsetexpanded1}
\eeq
where, in each line, we only display the leading new independent contribution. For instance, the second line corresponds to the leading term in $L_2-\Tr(X_u)L_1$, which projects out the content of $L_2$ aligned with $L_1$. This projection can be done step by step, as shown in Appendix \ref{section:coeffvalues}; we see in particular that the leading contribution to $\(\epsilon'/\epsilon\)_\text{BSM}$ given in Eq.~\eqref{kaon1} is captured by the following combination of invariants (up to a SMEFT-independent large numerical factor),
\beq
\(\frac{\epsilon'}{\epsilon}\)_\text{BSM}\propto L_5 + \Tr\(X_u\)\Tr\(X_d\) L_1 - \Tr\(X_d\)L_2-   \Tr\(X_u\)L_3\propto \eta_{12} \ .
\eeq
The last two contributions are not part of the minimal set of Eq.~\eqref{eq:minsetforCHQ}, which explains why the overall expression does not need to vanish when the minimal set does. It vanishes as it should when $J_4\to 0$ in addition, since in this situation $\eta_{12}$ becomes proportional to $\eta$, as shown in Eq.~\eqref{kaon2}.

It also becomes clear from Eq.~\eqref{eq:CHQmaxsetexpanded1} that seven contributions dominate $J_4$ (which is of order $\cO(\lambda^{36})$, as shown in Eq.~\eqref{J4lambdaOrder}), although only three do not require interference with it: setting $\eta=0$ in Eq.~\eqref{eq:CHQmaxsetexpanded1} reduces the number of independent quantities strictly larger than $\cO(\lambda^{36})$ from seven to three. (Eq.~\eqref{eq:CHQmaxsetexpanded1} only displays the leading order contributions to the different invariants, but we have checked that the claim holds to all orders.) 
In order to assess such behavior more systematically, we can ask: supposing that we have enough precision to resolve the invariants up to some fixed order $n$ in the $\lambda$-expansion, then how many independent sources of CP violation are we able to distinguish? We will come back to this question in section~\ref{section:lambdarank}. Notice also that the $\lambda$-scaling in \eqref{eq:CHQmaxsetexpanded1} was obtained under the assumption that all the entries of the matrix $C_{HQ}^{(1)}$ are of $\order{1}$, meaning that they do not carry additional $\lambda$ suppression. This does not need to be the case, and in section~\ref{section:flavorassumptions} we will see how this result changes when this hypothesis is modified and different flavor scenarios are adopted.
	
	At this stage, we ought to make two remarks. First, the  expressions above only encapsulate the suppression coming from the flavor structure. Clearly, we still have to consider the $1/\Lambda^2$ suppression coming from the EFT scale, where the power of the cutoff $\Lambda$ is chosen to match that accompanying the dimension-six SMEFT operators considered in this paper. Rather, we can play with the two different contributions and ask, for example, what the EFT scale must be for each invariant to be comparable to $J_4$. Assuming $\eta_{23}=\order{1}$, we get for example that $L_1$ is comparable with $J_4$, i.e. $\frac{v^2}{\Lambda^2}\frac{L_1}{J_4}\approx 1$, with $v=246$ GeV the electroweak scale, when $\Lambda\approx5\times 10^{11}\text{ GeV}$\@. Second, one should be aware of the following caveat: although, as we argued above, the invariants in Eq.~\eqref{eq:CHQmaxsetintro} are enough to parametrize the whole parameter space of CP violation generated by the operator $\mathcal{O}_{HQ}^{(1)}$, we have no say in how big the coefficients relating them to observables actually are. It may well be that, in particular observables of interest, loop factors, logs or mass factors come to modify the relative importance of invariants between themselves or the relative importance of the $\order{1/\Lambda^2}$ invariants and $J_4$. Nevertheless, the invariant analysis signals that there exist simple CP-odd physical quantities larger than the naive expectation, $J_4\times (\text{polynomial CP-even quantity})$, although they only exist if $J_4\neq 0$.
		
	\section{Maximal sets: capturing all CPV contributions}
	\label{section:maximalset}
	
	\subsection{Maximal set}
	
	As we explained in the example of the previous section, the construction of~\cite{Bonnefoy:2021tbt} exhausts the characterization of new \emph{direct} sources of CP violation at the leading SMEFT order, but it does not suffice to describe the full parameter space of CP-odd observables: CP is broken already at dimension four, and the source of its breaking can interfere with real entries of dimension-six parameters of SMEFT to produce additional CPV, in the sense that it vanishes in the limit where $\Lambda\to\infty$. As explained, however, it must be possible to parametrize observables through flavor-invariant quantities, as the physics should not depend on the flavor basis. Thus, we have to be able to describe the parameter space of leading order CP-odd observables through a larger set, where each CP-odd invariant captures a quantity that is responsible for new CP violation either on its own or by interference with the SM$_{4}$ one. This is what we define as a \emph{maximal set}. 
	Expanding on~\cite{Bonnefoy:2021tbt}, we build such invariants by taking the imaginary part of a single trace\footnote{Again, a justification for why this is the most general choice is shown in appendix~\ref{appendix:singletrace}.} of products of Yukawa matrices with one power of one dimension-six operator coefficient, generically denoted as $C^{(6)}$. Because of the imposed linearity, we can define a set of invariants for each dimension-six operator independently. 
	
	The quantities we consider take the following form
	\begin{align}
		L_a(C^{(6)})=\text{Im}\Tr\left(M_a C^{(6)}\right),
	\end{align}
	where, if $C^{(6)}$ belongs to a representation $\mathbf{r}$ of the flavor group, $M_a$ is a matrix built out of products of Yukawa that belongs to the conjugate representation $\bar{\mathbf{r}}$. A generalization to 4-Fermi operators follows similar lines and will be presented later.
	If we fix a basis in flavor space, we can define a vector containing all the entries of $C^{(6)}$ 
	\begin{align}
		\vec{C}^{(6)}_i\equiv \left(\left(\Re\, C^{(6)}\right)_1,\,\left(\Re\,C^{(6)}\right)_2,\, \dots\left(\Im\, C^{(6)}\right)_1,\,\ldots\right).
	\end{align}
	Then, again thanks to linearity, we can always find a matrix $\mathcal{T}_{ai}$, that we call a {\it transfer matrix}, such that
	\begin{align}
		L_a(C^{(6)})=\mathcal{T}_{ai}\vec{C}^{(6)}_i=T^R_{ai}\left(\Re\, C^{(6)}\right)_i+T^I_{ai}\left(\Im\, C^{(6)}\right)_i\ ,
		\label{eq:transfermatrixdef}
	\end{align}
	where we explicitly separated its action on the real and imaginary entries in a given basis, so that it takes the block form
	\begin{align}
		\mathcal{T}=\bigg(T^R \hspace{0.3cm}T^I\bigg).
	\end{align}
	
	In~\cite{Bonnefoy:2021tbt}, we used the transfer matrix to define minimal sets in the following way: when $J_4=0$, a minimal set is a set of flavor invariants such that if we add to it any other invariant the rank of $\mathcal{T}$ is unchanged, while if we remove any the rank decreases. With this formulation, it is easy to define maximal sets as: 
	\begin{tcolorbox}
		A set of flavor invariants is a maximal set iff, when $J_4\neq0$, the rank of the transfer matrix $\mathcal{T}$ does not increase by adding any flavor invariant to the set and decreases if any flavor invariant is removed from the set.
	\end{tcolorbox}
	Here, too, we can distinguish between \emph{primary} and \emph{secondary} entries of a dimension-six operator coefficient as those that can or cannot enter observables at $\order{1/\Lambda^2}$. Since secondary quantities cannot be arranged in linear invariants by definition, the maximal set only parametrizes primary ones. In the same way as for minimal sets, the number of invariants in a maximal set must be larger than or equal to the number of real and imaginary primary entries of a dimension-six operator coefficient in a given basis (up to a subtlety for leptonic operators, which we discuss below). For the sets presented in the next section and in appendices~\ref{section:bilinearoperators} and~\ref{section:4fermioperators}, we find that the equality holds for all operators of dimension 6 in SMEFT, as expected since the invariants are linear in the dimension-six coefficients. 
	
	\subsection{Examples}
	Here we present some examples of maximal sets of invariants in SMEFT at dimension-six. As explained, because of linearity, we can treat each operator independently. In addition, because of the Cayley--Hamilton theorem, we are sure that the set of all possible (polynomial single trace) CP-odd linear invariants we can build using $3\times3$ Yukawa matrices is finite. 
	
	\subsubsection{Fermionic bilinear operators}
	We can start by looking at SMEFT operators which are bilinear in fermion fields. For such operators, the relevant single trace invariants, linear in their coefficient $C$, take the form
	\begin{align}
		L_{abcd}(\tilde C)\equiv \Im\Tr(X_u^aX_d^bX_u^cX_d^d\tilde C) \ , \text{ with } a,b,c,d=0,1,2 \text{ and } a\neq c, b\neq d,
		\label{bilinearFormulaquark}
	\end{align}
	for quark operators, and 
	\begin{align}
		L_{a}(\tilde C)\equiv \Im\Tr(X_e^a\tilde C) \ , \text{ with } a=0,1,2,
		\label{bilinearFormulaleptons}
	\end{align}
	for the lepton ones, where $\tilde C=C,CY_{f=u,d,e}^\dagger$ or $Y^{\mathstrut}_fCY_{f}^\dagger$, depending on the chiral structure of the operator under study (see below for explicit formulae).
	
	As a first example, we consider $C_{uH}$. It is the coefficient of a non-hermitian operator bilinear in fermion fields, thus it contains, in a fixed basis, 9 real and 9 imaginary coefficients. Since none of them is charged under the leptonic $U(1)^3$ mentioned above, all of them are physical at $\order{1/\Lambda^2}$, and its corresponding maximal set can be chosen to be
	\begin{align}
		\mathcal{I}_a\left[C_{uH}\right]=\left\{ \ 
		\begin{array}{cccccc}
			\colorbox{newlightgrey}{$L_{0000}\left(C_{uH}Y_u^{\dagger }\right)$} & \colorbox{newlightgrey}{$L_{1000}\left(C_{uH}Y_u^{\dagger }\right) $}& L_{2000}\left(C_{uH}Y_u^{\dagger }\right) \\\colorbox{newlightgrey}{$ L_{0100}\left(C_{uH}Y_u^{\dagger }\right)$} & \colorbox{newlightgrey}{$L_{1100}\left(C_{uH}Y_u^{\dagger }\right) $}& \colorbox{newlightgrey}{$L_{0110}\left(C_{uH}Y_u^{\dagger }\right) $}\\
			L_{2100}\left(C_{uH}Y_u^{\dagger }\right) & L_{0120}\left(C_{uH}Y_u^{\dagger }\right) & L_{1120}\left(C_{uH}Y_u^{\dagger }\right) \\ L_{0200}\left(C_{uH}Y_u^{\dagger }\right) & L_{1200}\left(C_{uH}Y_u^{\dagger }\right) & L_{0210}\left(C_{uH}Y_u^{\dagger }\right) \\
			\colorbox{newlightgrey}{$L_{2200}\left(C_{uH}Y_u^{\dagger }\right) $}& \colorbox{newlightgrey}{$L_{0220}\left(C_{uH}Y_u^{\dagger }\right) $}& \colorbox{newlightgrey}{$L_{1220}\left(C_{uH}Y_u^{\dagger }\right) $} \\ L_{0112}\left(C_{uH}Y_u^{\dagger }\right) & \colorbox{newlightgrey}{$L_{0122}\left(C_{uH}Y_u^{\dagger }\right) $}& L_{1122}\left(C_{uH}Y_u^{\dagger }\right) \\
		\end{array}
		\ \right\}	\ ,
	\end{align}
	where the invariants highlighted in gray are those already included in the minimal set. 
	Let us now look at an operator, $\mathcal{O}_{eH}$, containing two leptonic fields, namely 
	\begin{align}
		\mathcal{O}_{eH}=\frac{1}{\Lambda^2}C_{eH,\, ij}\left(H^\dagger H\right) \bar{L}_ie_jH.
		\label{eq:modifiedYukawalepton}
	\end{align}
	Now, the off-diagonal entries of $C_{eH}$, in a given basis, are all charged under the leptonic $U(1)^3$. Thus, we expect only the 3 diagonal real and 3 diagonal imaginary entries to be captured by our maximal set. However, using the definition in Eq.~\eqref{bilinearFormulaleptons}, we see that we can only build invariants of the form
	\begin{align}
		L_a(C_{eH}Y_e^\dagger)\ , \text{ with } a=0,1,2.
	\end{align}
	This is because, for $3\times 3$ matrices, the Cayley–Hamilton theorem ensures that any $L_a(C_{eH}Y_e^\dagger)$ with $a>2$ is redundant.
	Thus a possible maximal set is simply
	\begin{align}
		\mathcal{I}_a[C_{eH}]=\left\{ \ \begin{matrix}
			L_0\(C_eY_e^\dagger\)&
			L_1\(C_eY_e^\dagger\)&
			L_2\(C_eY_e^\dagger\)
		\end{matrix} \ \right\}.
	\end{align}

	More straightforwardly, as there is no $J_4$ in the lepton sector, there is nothing a real entry can interfere with to produce CP violation, so only the 3 imaginary entries are captured. 
	One can also directly verify that, by adding to this set any $L_a(C_{eH}Y_e^\dagger)$ with $a>2$, the rank of the transfer matrix defined as in Eq.~\eqref{eq:transfermatrixdef} does not increase. Despite this, the 3 real diagonal entries are still considered as \emph{primary} coefficients, as they do enter observables at $\order{1/\Lambda^2}$, but they can only affect \emph{CP-odd} observables through the multiplication of $J_4$ and objects of the form $\Re\Tr(X_e^aC_{eH})$, $a=0,1,2$. As mentioned at the beginning of section~\ref{opportunisticDef}, we choose to ignore this trivial possibility. 
	
\subsubsection{Four-Fermi operators}
	Let us now look at 4-Fermi operators, i.e.~those quartic in fermionic fields. We start by considering $\mathcal{O}_{uu}=(\bar u_i \gamma_\mu u_j)(\bar u_k \gamma^\mu u_l)$. Because of its symmetry properties, i.e.
	\begin{align}
		C_{uu,\,ijkl}=C_{uu,\,klij} \quad \text{and} \quad C_{uu,\,ijkl}=C_{uu,\,jilk}^*\ ,
	\end{align}
	its coefficient contains, in a given basis, 27 real entries and 18 imaginary ones. None of them carries lepton quantum numbers, so all of them are physical at $\order{1/\Lambda^2}$ and are expected to interfere non-trivially with the CKM phase. Hence, all of them need to be captured by an invariant in the maximal set.
	As in~\cite{Bonnefoy:2021tbt}, we use the following definitions
	\beq
	\Tr_A\(M^{(1)},M^{(2)},C\)\equiv M^{(1)}_{ji}M^{(2)}_{lk}C_{ijkl}^{\vphantom{(2)}}, \quad \Tr_B\(M^{(1)},M^{(2)},C\)\equiv M^{(1)}_{li}M^{(2)}_{jk}C_{ijkl}^{\vphantom{(2)}},
	\label{eq:trAandtrB}
	\eeq
	and
	\beq
	\bead
	A^{abcd}_{efgh}(C)=&\Im\Tr_A\(X_u^aX_d^bX_u^cX_d^d,X_u^eX_d^{\smash{f}}X_u^gX_d^h,C\), \\
	B^{abcd}_{efgh}(C)=&\Im\Tr_B\(X_u^aX_d^bX_u^cX_d^d,X_u^eX_d^{\smash{f}}X_u^gX_d^h,C\).
	\label{eq:L4Fermi}
	\eead\eeq
	Moreover, we define the notation
	\beq 
	C_{\tilde u\tilde uuu,ijkl}^{\vphantom{\dagger}}\equiv \sum_{m,n}Y_{u,im}^{\vphantom{\dagger}}Y_{u,nj}^\dagger C_{uu,mnkl}^{\vphantom{\dagger}},
	\label{eq:tildeCoeffs}
	\eeq
	and similarly for $C_{\tilde uuu\tilde u},C_{u\tilde u\tilde uu},C_{uu\tilde u\tilde u}$, $C_{\tilde u\tilde u\tilde u\tilde u}$, and for the down quark versions.
	We find, then, that the maximal set for $C_{uu}$ can be expressed as
	\begin{align}
		\mathcal{I}_a\left[C_{uu}\right]=\left \{
		\begin{matrix}
			\colorbox{newlightgrey}{$A_{1100}^{0000}\(C_{\smash{dd\tilde{d}\tilde{d}}}\) $}&\colorbox{newlightgrey}{$B_{1100}^{0000}\(C_{\smash{dd\tilde{d}\tilde{d}}}\) $}&A_{2100}^{0000}\(C_{\smash{dd\tilde{d}\tilde{d}}}\) \\
			B_{2100}^{0000}\(C_{\smash{dd\tilde{d}\tilde{d}}}\) &A_{1200}^{0000}\(C_{\smash{dd\tilde{d}\tilde{d}}}\) &B_{1200}^{0000}\(C_{\smash{dd\tilde{d}\tilde{d}}}\) \\
			\colorbox{newlightgrey}{$A_{1100}^{1000}\(C_{\smash{dd\tilde{d}\tilde{d}}}\) $}&\colorbox{newlightgrey}{$B_{1100}^{1000}\(C_{\smash{dd\tilde{d}\tilde{d}}}\) $}&\colorbox{newlightgrey}{$A_{2200}^{0000}\(C_{\smash{dd\tilde{d}\tilde{d}}}\) $}\\
			A_{2100}^{1000}\(C_{\smash{dd\tilde{d}\tilde{d}}}\) &B_{2100}^{1000}\(C_{\smash{dd\tilde{d}\tilde{d}}}\) &\colorbox{newlightgrey}{$A_{2000}^{1100}\(C_{\smash{dd\tilde{d}\tilde{d}}}\) $}\\
			A_{2100}^{2000}\(C_{\smash{dd\tilde{d}\tilde{d}}}\) &A_{1120}^{0000}\(C_{\smash{dd\tilde{d}\tilde{d}}}\) &B_{1120}^{0000}\(C_{\smash{dd\tilde{d}\tilde{d}}}\) \\
			\colorbox{newlightgrey}{$A_{1100}^{0100}\(C_{\smash{dd\tilde{d}\tilde{d}}}\) $}&B_{1100}^{0100}\(C_{\smash{dd\tilde{d}\tilde{d}}}\) &A_{2100}^{0100}\(C_{\smash{dd\tilde{d}\tilde{d}}}\) \\
			\colorbox{newlightgrey}{$B_{2100}^{0100}\(C_{\smash{dd\tilde{d}\tilde{d}}}\) $}&A_{1200}^{1000}\(C_{\smash{dd\tilde{d}\tilde{d}}}\) &\colorbox{newlightgrey}{$A_{1100}^{1100}\(C_{\smash{dd\tilde{d}\tilde{d}}}\) $}\\
			\colorbox{newlightgrey}{$A_{2200}^{1000}\(C_{\smash{dd\tilde{d}\tilde{d}}}\) $}&B_{2200}^{1000}\(C_{\smash{dd\tilde{d}\tilde{d}}}\) &A_{2100}^{1100}\(C_{\smash{dd\tilde{d}\tilde{d}}}\) \\
			\colorbox{newlightgrey}{$B_{2000}^{1200}\(C_{\smash{dd\tilde{d}\tilde{d}}}\) $}&\colorbox{newlightgrey}{$A_{1122}^{0000}\(C_{\smash{dd\tilde{d}\tilde{d}}}\) $}&A_{1221}^{0000}\(C_{\smash{dd\tilde{d}\tilde{d}}}\) \\
			A_{2112}^{0000}\(C_{\smash{dd\tilde{d}\tilde{d}}}\) &A_{1120}^{1100}\(C_{\smash{dd\tilde{d}\tilde{d}}}\) &A_{2110}^{1100}\(C_{\smash{dd\tilde{d}\tilde{d}}}\) \\
			A_{2200}^{2000}\(C_{\smash{dd\tilde{d}\tilde{d}}}\) &A_{2100}^{2100}\(C_{\smash{dd\tilde{d}\tilde{d}}}\) &A_{1120}^{2100}\(C_{\smash{dd\tilde{d}\tilde{d}}}\) \\
			A_{2110}^{2100}\(C_{\smash{dd\tilde{d}\tilde{d}}}\) &A_{1120}^{1000}\(C_{\smash{dd\tilde{d}\tilde{d}}}\) &A_{1120}^{2000}\(C_{\smash{dd\tilde{d}\tilde{d}}}\) \\
			A_{0112}^{0000}\(C_{\smash{dd\tilde{d}\tilde{d}}}\) &A_{1200}^{0100}\(C_{\smash{dd\tilde{d}\tilde{d}}}\) &A_{1100}^{0200}\(C_{\smash{dd\tilde{d}\tilde{d}}}\) \\
			\colorbox{newlightgrey}{$A_{2200}^{1100}\(C_{\smash{dd\tilde{d}\tilde{d}}}\) $}&\colorbox{newlightgrey}{$A_{1122}^{1000}\(C_{\smash{dd\tilde{d}\tilde{d}}}\) $}&\colorbox{newlightgrey}{$A_{1220}^{1100}\(C_{\smash{dd\tilde{d}\tilde{d}}}\) $}\\
			\colorbox{newlightgrey}{$A_{2110}^{1200}\(C_{\smash{dd\tilde{d}\tilde{d}}}\) $}&\colorbox{newlightgrey}{$A_{0122}^{2100}\(C_{\smash{dd\tilde{d}\tilde{d}}}\) $}&\colorbox{newlightgrey}{$A_{1220}^{2200}\(C_{\smash{dd\tilde{d}\tilde{d}}}\) $}
		\end{matrix}
		\ \right \}.
	\end{align}

	The list of all maximal sets for each fermionic operator at dimension six can be found in Appendices~\ref{section:bilinearoperators} and~\ref{section:4fermioperators}\@. 
	We wish to stress that the lists of maximal sets are not unique. In our case, we picked them with the constraint that they contain the minimal number of Yukawa matrices and that they are maximized when the observed values of fermion masses and CKM entries are plugged in.

	\subsection{Maximal vs. minimal set}\label{section:vs}

	In this section, we wish to show in full generality a property of maximal sets that already appeared in the form of Eq.~\eqref{eq:maxvsminsketchedexample}.
	Specifically, we want to show that it must be possible to express every invariant $L_a$ of a maximal set in the following form:
	\begin{align}
		L_a=\sum_{\tilde a}\cI_{a\tilde a}L_{\tilde a}+J_4R_{4a},
		\label{eq:maxvsmin}
	\end{align}
	where $\{L_{\tilde a}\}$ are taken from the minimal set, the $\cI_{a\tilde a}$ are Yukawa-dependent flavor-invariant coefficients and $R_{4a}$ is a CP-even invariant. In particular, $R_{4a}$ does not need to be a single trace. However, by the same arguments we present in appendix \ref{appendix:singletrace}, since it needs to be linear in the Wilson coefficient $C^{(6)}$, such a coefficient will only appear in single traces, possibly multiplied by a factor constituted by some invariant combination of the Yukawa matrices. 
	The proof of this statement goes as follows: consider the limit $J_4\to 0$. In this limit, the set of independent CP-odd invariants is reduced to the minimal set, so that every invariant from the maximal set can be expressed as a linear combination of the minimal ones:
	\begin{align}
		L_a=\sum_{\tilde a}\cI_{a\tilde a}L_{\tilde a},
		\label{eq:maxvsminj4=0}
	\end{align}
	where the $\cI_{a\tilde a}$'s are invariants built with $X_{u,d}$ only. Now, the combination \begin{align}
		L'_a\equiv L_a-\sum_{\tilde a}\cI_{a\tilde a}L_{\tilde a}
		\label{eq:vanishingcomb}
	\end{align}
	has to vanish when $J_4\to0$. This means that, picking a flavor basis, Eq.~\eqref{eq:vanishingcomb} has to vanish when any of the factors appearing in $J_4$ (expressed in this basis) vanishes. Thus, we have to be able to extract from $L'_a$ a full $J_4$ factor. The rest has to be CP-even, flavor-invariant and linear in the Wilson coefficient (and, by construction, it remains finite for most ways of taking $J_4\to 0$). 
	
	Using the current notation, the remark made around \eqref{eq:CHQmaxsetexpanded1} means that\footnote{Analytically computing $R_{4a}$ is possible, as exemplified in Appendix~\ref{section:coeffvalues}.} $R_{4a}\gg 1$ is to be expected, but the expressions are lengthy and one needs to evaluate them numerically. For the case of Eq.~\eqref{eq:maxvsminsketchedexample}, assuming all entries of $C_{HQ}^{(1)}$ are of the same order of magnitude, we find for example $R_4/\cI_2\sim 10^{14}$ for flavor anarchic Wilson coefficients. More enlightening formulae can be found by restricting to CP-odd rephasing invariants (namely, singlets of the $U(1)^9$ group associated to the vectorlike phase shifts of each fermion mass eigenstate), which makes sense when assuming that all quark masses are non-degenerate and choosing a field basis which diagonalizes the masses. We expand on this in Appendix~\ref{appendix:opportunisticRephasingInvariants}.
	
	It is worth stressing that the validity of Eq.~\eqref{eq:maxvsmin} is strictly dependent on having picked the correct minimal set. Indeed, such an expression has to be valid for all values of the matrices involved, and in particular for the special points where masses are degenerate and/or the mixing angles assume special values, which we extensively studied in~\cite{Bonnefoy:2021tbt}. On any of those points, $J_4=0$, so Eq.~\eqref{eq:maxvsmin} reduces to Eq.~\eqref{eq:maxvsminj4=0}. Had we picked wrongly the minimal set, e.g.~had it not captured all sources of CPV in all degenerate cases, we could not argue that a full $J_4$ can be pulled out of \eqref{eq:vanishingcomb}, and Eq.~\eqref{eq:maxvsminj4=0} is not guaranteed anymore. This subtlety forces us to make seemingly unnatural choices for the minimal set, which is often not optimized to describe a given observable. Consider for illustration the leading two-loop Barr--Zee contribution of $C_{uH}$ to the electron EDM~\cite{Barr:1990vd,Brod:2013cka,Panico:2018hal},
	\be
	\frac{d_e}{e}\bigg\vert_\text{Barr--Zee}=-\frac{vm_e}{48\pi^2m_h^2}\sum_{i
	}\frac{m_{u,i}\Im(C_{uH,ii})}{\Lambda^2}F_1\(\frac{m_{u,i}^2}{m_h^2},0\),
	\label{barrZeeElectronOneGenRealMass}
	\ee
	where we chose a flavor basis which diagonalizes $Y_u$, $m_h$ and $v$ are the Higgs mass and vacuum expectation value (vev) respectively, and
	\be
	F_1\(a,0\)=\int_0^1dx\frac{\ln(\frac{a}{x(x-1)})}{a-x(x-1)} \, \, . 
	\ee
	This expression is most naturally expressed in terms of the three invariants $L_{n000}\(C_{uH}Y_u^\dagger\)=\sum_{i}m_{u,i}^{2n+1}\Im\(C_{uH,ii}\)$, however only two out of those belong to our minimal set, while the third is decomposed as in \eqref{eq:maxvsmin}. This is due to the fact that they are not independent when two up-type quark masses are degenerate, whereas all three $\Im\(C_{uH,ii}\)$ remain three direct sources of CPV~\cite{Bonnefoy:2021tbt}. Therefore, the three $L_{i000}\(C_{uH}Y_u^\dagger\)$ do not capture all sources of CPV whenever $J_4=0$, which can be achieved in particular by making two masses degenerate. Nevertheless, they belong to our maximal set and permit a natural expression of ${d_e}/{e}$ in terms of that set. Similarly to the discussion of CPV in kaon mixings in \eqref{kaon1}-\eqref{kaon2}, we find that the above contribution to the electron EDM does not vanish when the minimal set of $C_{uH}$ does. Instead, it reads
	\beq
	\frac{d_e}{e}\bigg\vert_\text{Barr-Zee}=-1.3\times 10^{-28}\(\frac{\text{TeV}}{\Lambda}\)^2\Re\(C_{uH,13}\) \text{ cm} \ ,
	\eeq
	where we used the observed values of quark masses and mixings.
	
	\section{The parameter space at fixed precision: the Taylor rank}
	\label{section:lambdarank}
	
	As reviewed above, the (linear) CP-odd flavor invariants that we build are in one-to-one correspondance with primary combinations of dimension-six SMEFT coefficients, namely those which can contribute to (CPV) observables truncated at $\cO(1/\Lambda^2)$, i.e.~at the first non-trivial order in the expansion in inverse powers of the cutoff. On the other hand, as seen from Eq.~\eqref{eq:CHQmaxsetexpanded1}, we can rely on an additional power counting, associated to other small parameters in SMEFT, which are the ratios of fermion masses to that of the top quark, and the mixing angles in the CKM matrix. As reviewed in section~\ref{conventionsSection}, their smallness can be expressed in terms of a single small parameter $\lambda$. Therefore, we can only probe the full parameter space of CPV when the precision we have is enough to resolve all of the invariants, i.e.~when we expand to high enough powers in $\lambda$ such that all invariants are independent. One could then ask: how many invariants are required to predict the value of any observable? 
	
	First of all, the answer depends on the precise relation between a given observable $O$ and the invariants $L_a$. At $\cO(1/\Lambda^2)$, the invariants parametrize the set of physical CPV quantities, so such a relation must exist and be linear,
	\beq
	O=\sum_a o_aL_a \ ,
	\eeq
	where the $o_a$ depend on parameters of the dimension-four Lagrangian, but not on dimension-six Wilson coefficients. However, that fact does not give us any quantitative information on the $o_a$: they could be large or small, they need not be analytic, etc\footnote{Let us consider an explicit example, the leading contribution of the dimension-six Yukawa operators to the electron EDM, displayed in Eq.~\eqref{barrZeeElectronOneGenRealMass}. In the up basis, one can easily work out the relationship between the three $\Im\, C_{uH,ii}$ and the $\cI_{j=1,2,3}=\Im\Tr\(X_u^{j-1}C_{uH}Y_u^\dagger\)$:
	\bes
	m_{u_i}\Im\, C_{uH,ii}=\frac{\cI_3-\cI_2\sum_{j\neq i}m_{u_j}^2+\cI_1\prod_{j\neq i}m_{u_j}^2}{\prod_{j\neq i}\(m_{u_i}^2-m_{u_j}^2\)} \ .
	\ees
	 Plugging these relations in Eq.~\eqref{barrZeeElectronOneGenRealMass}, one easily identifies the three $o_i$, respectively associated to $\cI_i$. They are non-analytic with respect to the Yukawas and large, $o_{1,2,3}\sim\lambda^{3,-5,-5}$. In the equation above, one also sees that the natural size of Wilson coefficients, $C_{uH,ii} \leq \cO(1)$, and the smallness of quark masses impose that there must be relations between invariants at low orders in $\lambda$. These relations are exactly captured by the Taylor rank which we now discuss.}. The ideal way to tackle this would be to identify a set of CPV observables in one-to-one correspondence with our invariants, to work out the associated $o_a$, and to derive from them the sensitivity of CPV observables to the SMEFT parameter space. 
	This is beyond the scope of this paper.
	Instead, we focus on the complete maximal sets of flavor invariants which we identified above. Those are physical quantities, hence we use them as a proxy for the general set of observables, and as a way to develop our formalism. Nonetheless, we stress again that relating them to the kind of CPV observables experimentalists usually deal with would be a very insightful endeavor, which we leave for future work. 
	
	Now, the answer to our initial question still depends on the degree of precision which is aimed for, which motivates us to introduce the {\it Taylor rank}. It addresses the following question:	how many invariants should one be given to know them all at $\cO(\lambda^n)$? This number, which we denote $r_n$ and call the Taylor rank, is not necessarily equal to the number of invariants in the maximal set. Indeed, when $n$ is not too large, one is able to generate the whole maximal set as a linear combination of one of its subsets $\{L_1,...,L_{r_n}\}$,
	\beq
	\forall a \, , \ L_a=\sum_{b=1}^{r_n}c^{(n)}_{ab}L_b +\cO(\lambda^{n+1}) \ , \quad c^{(n)}_{ab}\leq\cO(1) \ ,
	\label{expansionTaylorRank}
	\eeq
	where $r_n$ is the smallest cardinality of a subset for which this decomposition can be achieved. Such a subset captures the physical degrees of freedom at the chosen degree of precision $\cO(\lambda^n)$.
	
	Since the invariants are linear in $C_{HQ}^{(1)}$, one can phrase the condition \eqref{expansionTaylorRank} in terms of the transfer matrix $\cT$ and deal with a linear algebra problem. More precisely, \eqref{expansionTaylorRank} captures the fact that one can find several linear relations between the columns of $\cT$, provided one is allowed to drop or add small $\cO(\lambda^{n+1})$ coefficients. Therefore, the Taylor rank is the smallest rank encountered in the equivalence class  of $\cT$, defined by the equivalence relation
	\beq
	\cT \sim \cT' \text{ if } \cT-\cT'=\cO(\lambda^{n+1}) \ ,
	\eeq
	where it is understood that all entries of $\cT,\cT'$ are Taylor-expanded up to a common order $\cO(\lambda^n)$. The Taylor rank generalizes the notion of rank,  which would tell us how many independent invariants there are overall, to situations where one neglects high orders in $\lambda$. Importantly, the Taylor rank does not coincide with the rank of $\cT$ truncated at $\cO(\lambda^n)$, which is usually larger than $r_n$. In Appendix~\ref{section:taylorrank}, we expand slightly on the definition and properties of the Taylor rank, and we describe our algorithm to compute it. For the invariants of the maximal set of $C_{HQ}^{(1)}$ displayed in Eq.~\eqref{eq:CHQmaxsetintro}, we obtain the result in Table~\ref{tab:TaylorrankCHQ}.
	
	\begin{table}[h]
		\centering
		Taylor rank for $C_{HQ}$\\[5pt]
		\begin{tabular}{c|c|c|c|c|c|c|c}
			$\lambda$ order& 8 & 17  & 19 & 25 & 28 & 34 & 36 \\\hline
			Taylor rank&1&2&4&5&6&7&9  \\
		\end{tabular}
		\caption{Result of the Taylor rank procedure for the invariants associated to $C_{HQ}^{(1)}$. Each entry in the Taylor rank row must be interpreted as the number of different sources of CPV coming from $C_{HQ}^{(1)}$ that we are able to identify as independent at the corresponding order in $\lambda$.}
		\label{tab:TaylorrankCHQ}
	\end{table}
	
	We see that those numbers are consistent with \eqref{eq:CHQmaxsetexpanded1}, and with the observation made below it: at $\order{\lambda^{35}}$, i.e.~before we are even able to resolve $J_4$, we can identify 7 independent sources of CP violation, four of which come from opportunistic CPV\@.

\begin{figure}[H]
		\centering
		\captionsetup{singlelinecheck=off}
		\textbf{Taylor rank at each order in $\lambda$ for $\mathcal{O}_{HQ}^{(1)}$}\\\medskip
		\includegraphics[width=1\textwidth]{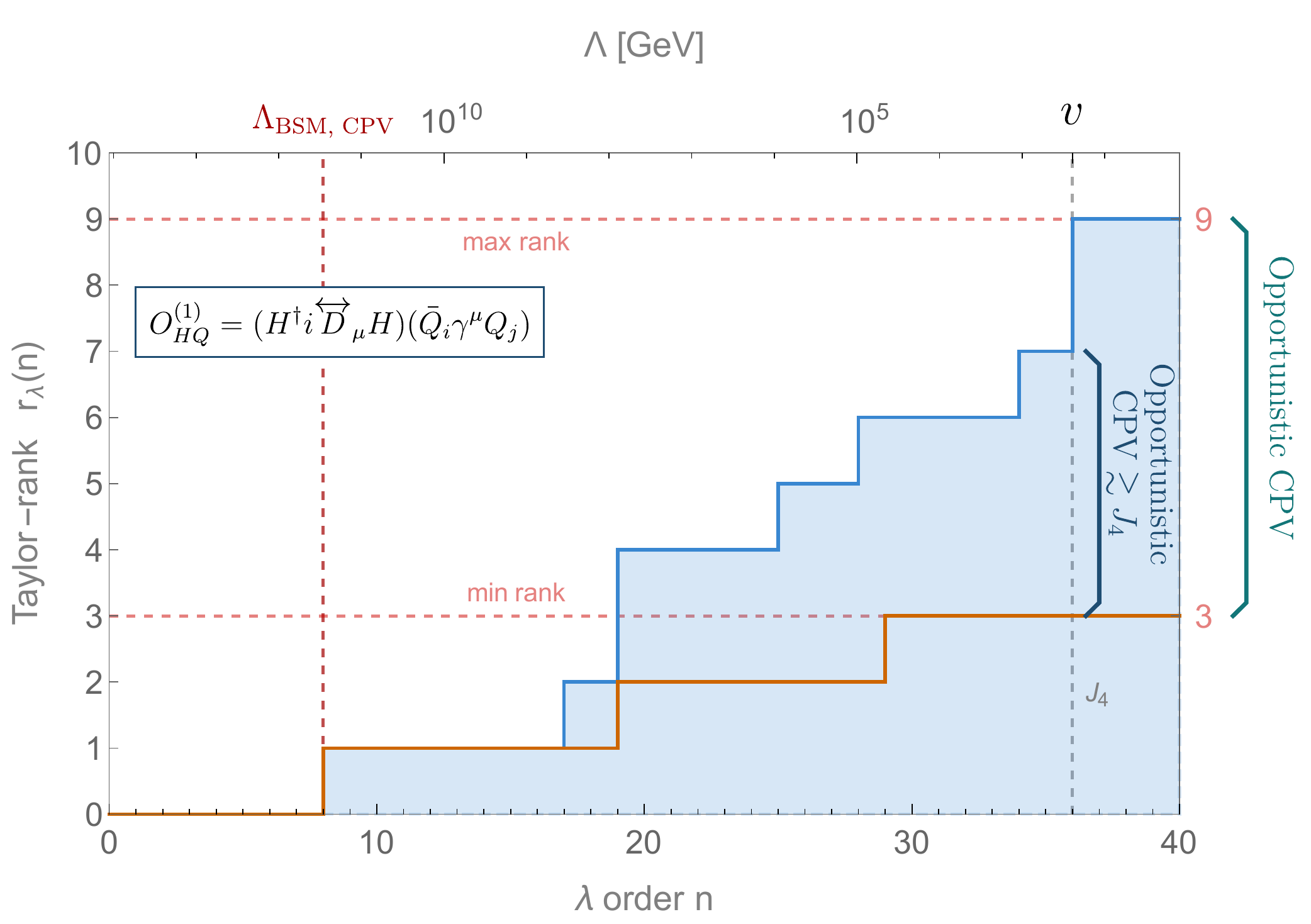}
		\caption[thiscaption]{Number of independent invariants from the maximal (blue step-wise line) and the minimal (dark yellow step-wise line) sets, denoted as Taylor rank $r_n$ in the text, at each order $n$ in the $\lambda$ expansion for the $\mathcal{O}_{HQ}^{(1)}$ operator. At a fixed order in $\lambda$, the top $x$-axis shows the value of $\Lambda$ for which an invariant appearing at such an order would be comparable to $J_4$, assuming it scales as $v^2/\Lambda^2$, with $v\sim246\text{ GeV} $ the vev of the Higgs field. This plot serves as a guide on how to extract the most relevant information from the similar ones that follow, and as such all the main point of interest have been highlighted:
			\begin{enumerate}
				\item the vertical grey dashed line marks the order $\lambda^{36}$ (corresponding to $\Lambda=v$, as indicated on the top $x$-axis) where the SM$_4$ $J_4$ shows up,
				\item the horizontal red lines mark the values of maximal and minimal ranks, also labeled on the right $y$-axis for readability,
				\item the scale $\Lambda_{\text{BSM, CPV}}$ indicates the highest value of $\Lambda$ where at least one invariant is comparable to $J_4$
				, here $\Lambda= 5\times 10^{11} \text{ GeV}\leftrightarrow \lambda^8$ (this is the value that we already showed at the end of Section~\ref{opportunisticDef}),
				\item we indicate on the right the number of sources of opportunistic CPV as the difference between the maximal and minimal sets,
				\item we also highlight the number of sources of opportunistic CPV $\gtrsim J_4$ as the difference between the maximal and minimal set at $\lambda^{36}$.
			\end{enumerate}
		}
		\label{fig:lambdaranksdescription}
	\end{figure}

\begin{figure}[H]
		\centering
		\textbf{Taylor rank at each order in $\lambda$ for all bilinear operators}\\\medskip
		\includegraphics[width=1\textwidth]{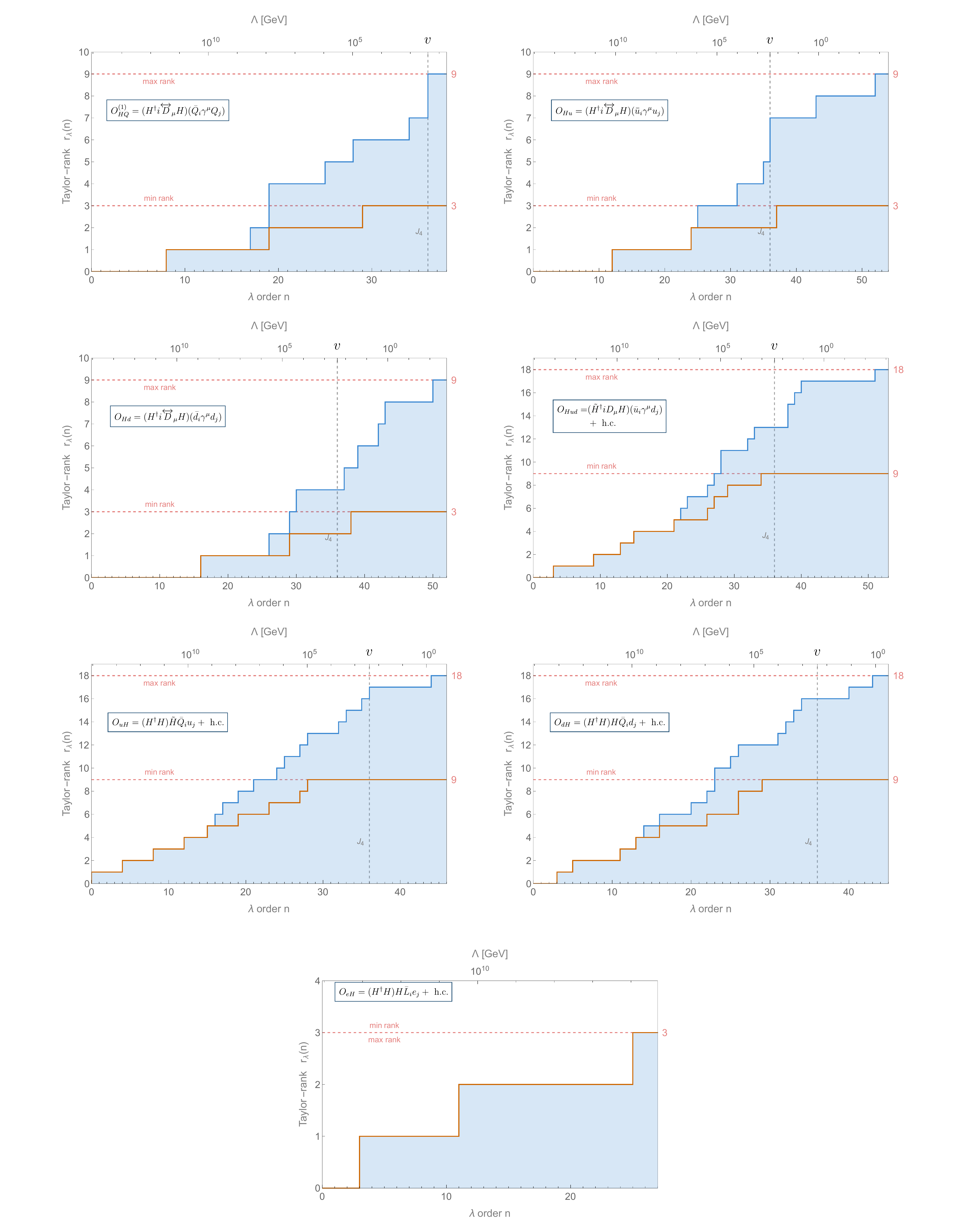}
		\caption{Number of independent invariants from the maximal (blue step-wise line) and the minimal (dark yellow step-wise line) sets, denoted as Taylor rank $r_n$ in the text, at each order $n$ in the $\lambda$ expansion for all bilinear operators. At a fixed order in $\lambda$, the top $x$-axis shows the value of $\Lambda$ for which an invariant appearing at such order would be comparable to $J_4$, assuming it scales as $v^2/\Lambda^2$, with $v\sim246\text{ GeV} $ the vev of the Higgs field.
			The vertical dashed line marks the order $\lambda^{36}$ (corresponding to $\Lambda=v$, as indicated) where the SM$_4$ $J_4$ shows up, while the horizontal lines mark the values for the maximal and minimal rank, also labeled on the right $y$-axis. Each plot corresponds to a group of operators in Table~\ref{Bilinearlist} (excluding those with 0 maximal and minimal sets), of which only one is chosen as a representative.}
		\label{fig:lambdaranksbilinears}
	\end{figure}

\begin{figure}[H]
			\vspace*{-1cm}
		\centering
		\textbf{Taylor rank at each order in $\lambda$ for all 4-Fermi operators}\\\medskip
		\hspace*{-1.5cm}
		\includegraphics[width=1.15\textwidth]{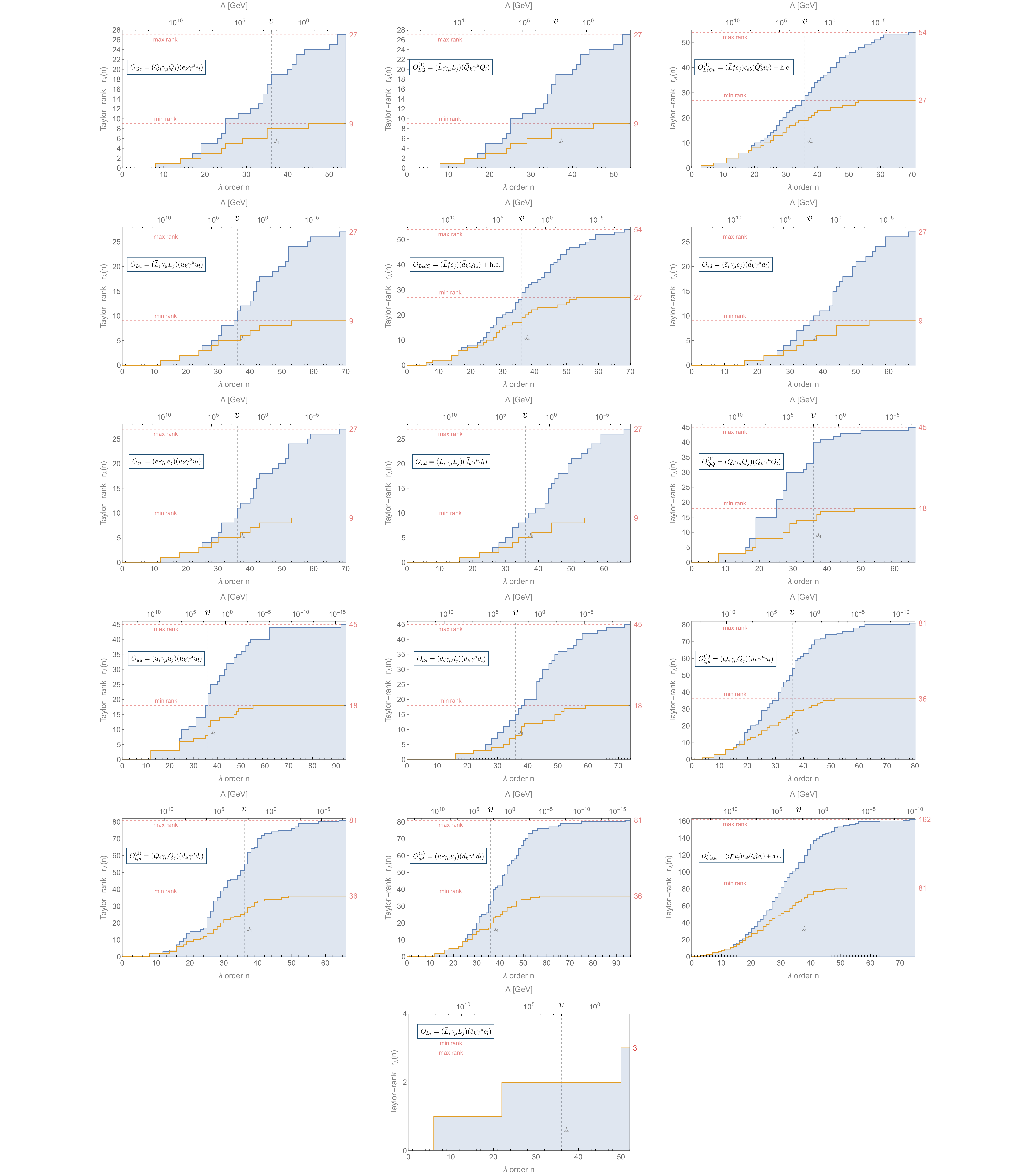}
		\caption{
			Number of independent invariants from the maximal (blue step-wise line) and the minimal (dark yellow step-wise line) sets, denoted as Taylor rank $r_n$ in the text, at each order $n$ in the $\lambda$ expansion for all bilinear operators. At a fixed order in $\lambda$, the top $x$-axis shows the value of $\Lambda$ for which an invariant appearing at such order would be comparable to $J_4$, assuming it scales as $v^2/\Lambda^2$, with $v\sim246\text{ GeV} $ the vev of the Higgs field.
			The vertical dashed line marks the order $\lambda^{36}$ (corresponding to $\Lambda=v$, as indicated) where the SM$_4$ $J_4$ shows up, while the horizontal lines mark the values for the maximal and minimal rank, also labeled on the right $y$-axis. Each plot corresponds to a group of operators in Table~\ref{fig:lambdaranks4Fermi} (excluding those with 0 maximal and minimal sets), of which only one is chosen as a representative.}
		\label{fig:lambdaranks4Fermi}
	\end{figure}

	We can apply the above to every operator in the dimension-six SMEFT\@. The results are plotted in Figures~\ref{fig:lambdaranksbilinears} and~\ref{fig:lambdaranks4Fermi}.  In Figure~\ref{fig:lambdaranksdescription} we present  an example how to extract all relevant information from this kind of plot. As advertised previously,
	the choice of maximal set is not unique, and different choices would only lead to slightly different results for these plots. In particular, our choice was to arbitrarily prioritize the invariants with smallest leading $\lambda$ power. However, another possibility that we did not pursue would have been to optimize the set w.r.t. the produced plots, e.g.~by choosing the one for which the maximal rank is reached earlier in the $\lambda$-rank procedure. In any case, this choice does not affect the nature of our conclusions. 
	\newpage

	\begin{table}[H]
		\centering
		\renewcommand{\arraystretch}{1.35}
		\resizebox{1\columnwidth}{!}{%
			\begin{tabular}{c|c|c|c|c|c}
				\multicolumn{6}{c}{\textbf{Bilinear operators}}\\\hline
				Label & Operator&\begin{minipage}{0.15\columnwidth}\centering
					Minimal rank\vspace*{0.1cm}
				\end{minipage}  & \begin{minipage}{0.15\columnwidth}\centering
					Maximal rank\vspace*{0.1cm}
				\end{minipage}  &\begin{minipage}{0.23\columnwidth}\centering
					\# imaginary entries (of which primary)\vspace*{0.1cm}
				\end{minipage} &\begin{minipage}{0.23\columnwidth}\centering
					\# real entries \\  (of which primary)\vspace*{0.1cm}
				\end{minipage} \\\hline
				$\mathcal{O}_{H Q}^{(1)}$&$(H^\dag i\overleftrightarrow{D}_\mu H)(\bar Q_i \gamma^\mu Q_j)$&\multirow{2}{0.1\textwidth}{\centering 3}&\multirow{2}{0.1\textwidth}{\centering 9}&\multirow{2}{0.1\textwidth}{\centering 3 (3)}&\multirow{2}{0.1\textwidth}{\centering 6 (6)}\\
				$\mathcal{O}_{H Q}^{(3)}$&$(H^\dag i\overleftrightarrow{D}^I_\mu H)(\bar Q_i \tau^I \gamma^\mu Q_j)$&&&&\\\hline
				$\mathcal{O}_{H u}$&$(H^\dag i\overleftrightarrow{D}_\mu H)(\bar u_i \gamma^\mu u_j)$&3&9&3 (3)&6 (6)\\\hline
				$\mathcal{O}_{H d}$&$(H^\dag i\overleftrightarrow{D}_\mu H)(\bar d_i \gamma^\mu d_j)$&3&9&3 (3)&6 (6)\\\hline
				$\mathcal{O}_{H u d}$&$i(\widetilde H ^\dag D_\mu H)(\bar u_i \gamma^\mu d_j)+\text{h.c.}$&9&18&9 (9)&9 (9)\\\hline
				$\mathcal{O}_{uH}$&$(H^\dag H)(\bar Q_i u_j \widetilde H )+\text{h.c.}$&\multirow{4}{0.1\textwidth}{\centering 9}&\multirow{4}{0.1\textwidth}{\centering 18}&\multirow{4}{0.1\textwidth}{\centering 9 (9)}&\multirow{4}{0.1\textwidth}{\centering 9 (9)}\\
				$\mathcal{O}_{uG}$&$(\bar Q_i \sigma^{\mu\nu} T^A u_j) \widetilde H \, G_{\mu\nu}^A+\text{h.c.}$&&&&\\
				$\mathcal{O}_{uW}$&$(\bar Q_i \sigma^{\mu\nu} u_j) \tau^I \widetilde H \, W_{\mu\nu}^I+\text{h.c.}$&&&&\\
				$\mathcal{O}_{uB}$&$(\bar Q_i \sigma^{\mu\nu} u_j) \widetilde H \, B_{\mu\nu}+\text{h.c.}$&&&&\\\hline
				$\mathcal{O}_{dH}$&$(H^\dag H)(\bar Q_i d_j H)+\text{h.c.}$&\multirow{4}{0.1\textwidth}{\centering 9}&\multirow{4}{0.1\textwidth}{\centering 18}&\multirow{4}{0.1\textwidth}{\centering 9 (9)}&\multirow{4}{0.1\textwidth}{\centering 9 (9)}\\
				$\mathcal{O}_{dG}$&$(\bar Q_i \sigma^{\mu\nu} T^A d_j) H\, G_{\mu\nu}^A+\text{h.c.}$&&&&\\
				$\mathcal{O}_{dW}$&$(\bar Q_i \sigma^{\mu\nu} d_j) \tau^I H\, W_{\mu\nu}^I+\text{h.c.}$&&&&\\
				$\mathcal{O}_{dB}$&$(\bar Q_i \sigma^{\mu\nu} d_j) H\, B_{\mu\nu}+\text{h.c.}$&&&&\\\hline
				$\mathcal{O}_{eH}{}^*$&$(H^\dag H)(\bar L_i e_j H)+\text{h.c.}$&\multirow{3}{0.1\textwidth}{\centering 3}&\multirow{3}{0.1\textwidth}{\centering 3}&\multirow{3}{0.1\textwidth}{\centering 9 (3)}&\multirow{3}{0.1\textwidth}{\centering 9 (3)}\\
				$\mathcal{O}_{eW}{}^*$&$(\bar L_i \sigma^{\mu\nu} e_j) \tau^I H W_{\mu\nu}^I+\text{h.c.}$&&&&\\
				$\mathcal{O}_{eB}{}^*$&$(\bar L_i \sigma^{\mu\nu} e_j) H B_{\mu\nu}+\text{h.c.}$&&&&\\\hline
				$\mathcal{O}_{H L}^{(1)}{}^*$&$(H^\dag i\overleftrightarrow{D}_\mu H)(\bar L_i \gamma^\mu L_j)$&\multirow{3}{0.1\textwidth}{\centering 0}&\multirow{3}{0.1\textwidth}{\centering 0}&\multirow{3}{0.1\textwidth}{\centering 3 (0)}&\multirow{3}{0.1\textwidth}{\centering 6 (3)}\\
				$\mathcal{O}_{H L}^{(3)}{}^*$&$(H^\dag i\overleftrightarrow{D}^I_\mu H)(\bar L_i \tau^I \gamma^\mu L_j)$&&&&\\
				$\mathcal{O}_{H e}{}^*$&$(H^\dag i\overleftrightarrow{D}_\mu H)(\bar e_i \gamma^\mu e_j)$&&&&\\\hline
				\multicolumn{2}{c|}{Total}&102&207&129 (102)&150 (123) \\\hline
			\end{tabular}%
		}
		\caption{\label{Bilinearlist}The list of dimension-6 fermionic bilinear operators of SMEFT, as given in
			Ref.\,\cite{Grzadkowski:2010es}, together with their corresponding minimal and maximal rank, as explained in the text. In parentheses we indicate how many of the entries in each cell are primary. When $+\hbox{ h.c.}$ is specified, the hermitian conjugate of the operator must be included in the Lagrangian too. We indicate with $i,j,k,l$ the flavor indices and with $a,b$ indices in the fundamental of $SU(2)_{L}$. $T^A$, $A=1,\ldots,8$ are the generators of the gauge $SU(3)_c$, while $\tau^I=\frac{\sigma^I}{2}$, $I=1,2,3$ are the generators of $SU(2)_L$ in the fundamental representation, with $\sigma^I$ the Pauli matrices. The operators are grouped so that each block corresponds to one of the plots of Fig.~\ref{fig:lambdaranksbilinears} (except for the operators with empty maximal and minimal sets). 
			Notice that for the last 6 operators, the sum of primary entries in the third and fourth columns do not coincide with the maximal rank (second column). Indeed, since there is no CP violation in the lepton sector, their real primary entries cannot interfere with anything to produce additional CPV\@. We indicate with a $*$ those operators for which this happens.}
	\end{table}

	\begin{table}[H]
		\centering
		\renewcommand{\arraystretch}{1.15}
		\resizebox{1\columnwidth}{!}{%
			\begin{tabular}{c|c|c|c|c|c}
				\multicolumn{6}{c}{\textbf{4-Fermi}}\\\hline
				Label & Operator&\begin{minipage}{0.15\columnwidth}\centering
					Minimal rank\vspace*{0.1cm}
				\end{minipage}  & \begin{minipage}{0.15\columnwidth}\centering
					Maximal rank\vspace*{0.1cm}
				\end{minipage}  &\begin{minipage}{0.23\columnwidth}\centering
				\vspace*{0.1cm}
					\# imaginary entries (of which primary)\vspace*{0.1cm}
				\end{minipage} &\begin{minipage}{0.23\columnwidth}\centering
				\vspace*{0.1cm}
					\# real entries \\  (of which primary)\vspace*{0.1cm}
				\end{minipage} \\\hline
				$\mathcal{O}_{Qe}$            & $(\bar Q_i \gamma_\mu Q_j)(\bar e_k \gamma^\mu e_l)$ &9&27&36 (9)&45 (18)\\\hline
				$\mathcal{O}_{LQ}^{(1)}$   & $(\bar L_i \gamma_\mu L_j)(\bar Q_k \gamma^\mu Q_l)$ &\multirow{2}{0.1\textwidth}{\centering 9}&\multirow{2}{0.1\textwidth}{\centering 27}&\multirow{2}{0.1\textwidth}{\centering 36 (9)}&\multirow{2}{0.1\textwidth}{\centering 45 (18)}\\
				$\mathcal{O}_{LQ}^{(3)}$              & $(\bar L_i \gamma_\mu \tau^I L_j)(\bar Q_k \gamma^\mu \tau^I Q_l)$&&&&\\\hline
				$\mathcal{O}_{LeQu}^{(1)}$& $(\bar L_i^a e_j) \epsilon_{ab} (\bar Q_k^b u_l)+\text{h.c.}$&\multirow{2}{0.1\textwidth}{\centering 27}&\multirow{2}{0.1\textwidth}{\centering 54}&\multirow{2}{0.1\textwidth}{\centering 81 (27)}&\multirow{2}{0.1\textwidth}{\centering 81 (27)}\\
				$\mathcal{O}_{LeQu}^{(3)}$& $(\bar L_i^a \sigma_{\mu\nu} e_j) \epsilon_{ab} (\bar Q_s^k \sigma^{\mu\nu} u_t)+\text{h.c.}$&&&&\\\hline
				$\mathcal{O}_{Lu}$   & $(\bar L_i \gamma_\mu L_j)(\bar u_k \gamma^\mu u_l)$&9&27&36 (9)&45 (18)\\\hline
				$\mathcal{O}_{LedQ}$ & $(\bar L_i^a e_j)(\bar d_k Q_{la})+\text{h.c.}$&27&54&81 (27)&81 (27)\\\hline
				$\mathcal{O}_{ed}$ & $(\bar e_i \gamma_\mu e_j)(\bar d_k\gamma^\mu d_l)$&9&27&36 (9)&45 (18)\\\hline
				$\mathcal{O}_{eu}$ & $(\bar e_i \gamma_\mu e_j)(\bar u_k \gamma^\mu u_l)$&9&27&36 (9)&45 (18)\\\hline
				$\mathcal{O}_{Ld}$  & $(\bar L_i \gamma_\mu L_j)(\bar d_k \gamma^\mu d_l)$&9&27&36 (9)&45 (18)\\\hline
				$\mathcal{O}_{QQ}^{(1)}$  & $(\bar Q_i \gamma_\mu Q_j)(\bar Q_k \gamma^\mu Q_l)$&\multirow{2}{0.1\textwidth}{\centering 18}&\multirow{2}{0.1\textwidth}{\centering 45}&\multirow{2}{0.1\textwidth}{\centering 18 (18)}&\multirow{2}{0.1\textwidth}{\centering 27 (27)}\\
				$\mathcal{O}_{QQ}^{(3)}$  & $(\bar Q_i \gamma_\mu \tau^I Q_j)(\bar Q_k \gamma^\mu \tau^I Q_l)$ &&&&\\\hline
				$\mathcal{O}_{uu}$   & $(\bar u_i \gamma_\mu u_j)(\bar u_k \gamma^\mu u_l)$&18&45&18 (18)&27 (27)\\\hline
				$\mathcal{O}_{dd}$   & $(\bar d_i \gamma_\mu d_j)(\bar d_k \gamma^\mu d_l)$&18&45&18 (18)&27 (27)\\\hline
				$\mathcal{O}_{Qu}^{(1)}$ & $(\bar Q_i \gamma_\mu Q_j)(\bar u_k \gamma^\mu u_l)$&\multirow{2}{0.1\textwidth}{\centering 36}&\multirow{2}{0.1\textwidth}{\centering 81}&\multirow{2}{0.1\textwidth}{\centering 36 (36)}&\multirow{2}{0.1\textwidth}{\centering 45 (45)}\\
				$\mathcal{O}_{Qu}^{(8)}$& $(\bar Q_i \gamma_\mu T^A Q_j)(\bar u_k \gamma^\mu T^A u_l)$ &&&&\\\hline
				$\mathcal{O}_{Qd}^{(1)}$& $(\bar Q_i \gamma_\mu Q_j)(\bar d_k \gamma^\mu d_l)$ &\multirow{2}{0.1\textwidth}{\centering 36}&\multirow{2}{0.1\textwidth}{\centering 81}&\multirow{2}{0.1\textwidth}{\centering 36 (36)}&\multirow{2}{0.1\textwidth}{\centering 45 (45)}\\
				$\mathcal{O}_{Qd}^{(8)}$& $(\bar Q_i \gamma_\mu T^A Q_j)(\bar d_k \gamma^\mu T^A d_l)$ &&&&\\\hline
				$\mathcal{O}_{ud}^{(1)}$ & $(\bar u_i \gamma_\mu u_j)(\bar d_k \gamma^\mu d_l)$&\multirow{2}{0.1\textwidth}{\centering 36}&\multirow{2}{0.1\textwidth}{\centering 81}&\multirow{2}{0.1\textwidth}{\centering 36 (36)}&\multirow{2}{0.1\textwidth}{\centering 45 (45)}\\
				$\mathcal{O}_{ud}^{(8)}$               & $(\bar u_i \gamma_\mu T^A u_j)(\bar d_k \gamma^\mu T^A d_l)$&&&&\\\hline
				$\mathcal{O}_{QuQd}^{(1)}$& $(\bar Q_i^a u_j) \epsilon_{ab} (\bar Q_k^b d_l)+\text{h.c.}$&\multirow{2}{0.1\textwidth}{\centering 81}&\multirow{2}{0.1\textwidth}{\centering 162}&\multirow{2}{0.1\textwidth}{\centering 81 (81)}&\multirow{2}{0.1\textwidth}{\centering 81 (81)}\\
				$\mathcal{O}_{QuQd}^{(8)}$& $(\bar Q_i^a T^A u_j) \epsilon_{ab} (\bar Q_k^b T^A d_l)+\text{h.c.}$ &&&&\\\hline
				$\mathcal{O}_{LL}{}^*$&$(\bar L_i \gamma_\mu L_j)(\bar L_k \gamma^\mu L_l)$ &0&0&18 (0)&27 (9)\\\hline
				$\mathcal{O}_{ee}{}^*$ & $(\bar e_i \gamma_\mu e_j)(\bar e_k \gamma^\mu e_l)$&0&0&15 (0)&21 (6)\\\hline
				$\mathcal{O}_{Le}{}^*$ & $(\bar L_i \gamma_\mu L_j)(\bar e_k \gamma^\mu e_l)$&3&3&36 (3)&45 (12)\\\hline
				\multicolumn{2}{c|}{Total}& 597&1344&1014 (597)&1191 (774)\\\hline
				\multicolumn{2}{c|}{Bilinears + 4-Fermi}&699&1551&1143 (699)&1341 (897)\\\hline
			\end{tabular}%
		}
		\caption{\label{4Fermilist}
		The list of dimension-6 4-Fermi operators of SMEFT, as given in
		Ref.\,\cite{Grzadkowski:2010es}, together with their corresponding minimal and maximal rank, as explained in the text. In parentheses we indicate how many of the entries in each cell are primary. When $+\hbox{ h.c.}$ is specified, the hermitian conjugate of the operator must be included in the Lagrangian too. We indicate with $i,j,k,l$ the flavor indices and with $a,b$ indices in the fundamental of $SU(2)_{L}$. $T^A$, $A=1,\ldots,8$ are the generators of the gauge $SU(3)_c$, while $\tau^I=\frac{\sigma^I}{2}$, $I=1,2,3$ are the generators of $SU(2)_L$ in the fundamental representation, with $\sigma^I$ the Pauli matrices. The operators are grouped so that each block corresponds to one of the plots of Fig.~\ref{fig:lambdaranks4Fermi} (except for the operators with empty maximal and minimal sets). 
		Notice that for the last 3 operators, the sum of primary entries in the third and fourth columns do not coincide with the maximal rank (second column). Indeed, since there is no CP violation in the lepton sector, their real primary entries cannot interfere with anything to produce additional CPV\@. We indicate with a $*$ those operators for which this happens.
}
	\end{table}

	\begin{figure}[h]
		\vspace*{-1cm}
		\centering
		\textbf{Taylor rank at each order in $\lambda$ for all operators}\\\medskip
		\begin{subfigure}[b]{0.45\textwidth}
			\includegraphics[width=\textwidth]{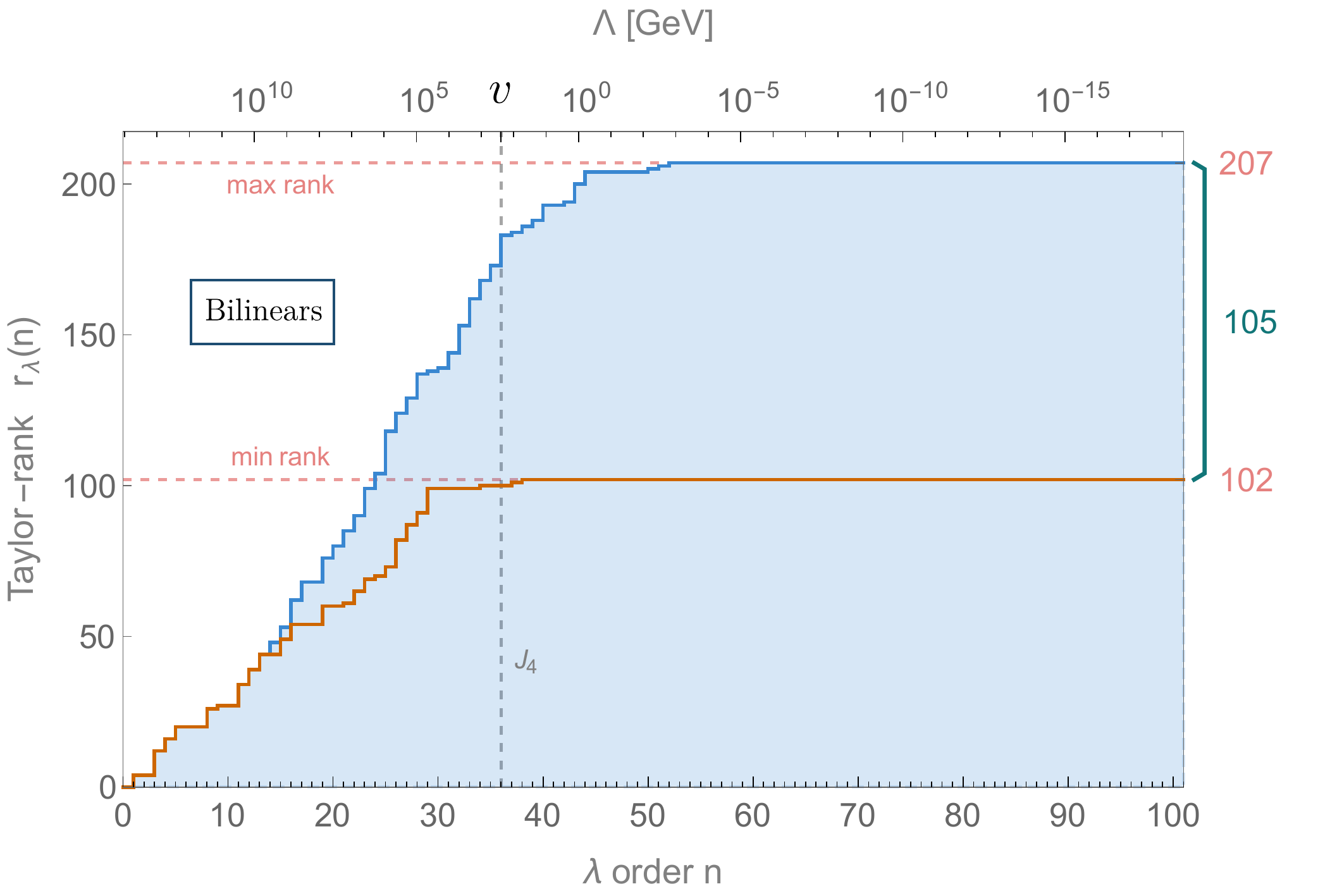}
			\caption{Bilinears}
			\label{fig:totbil}
		\end{subfigure}
		~ 	
		\begin{subfigure}[b]{0.45\textwidth}
			\includegraphics[width=\textwidth]{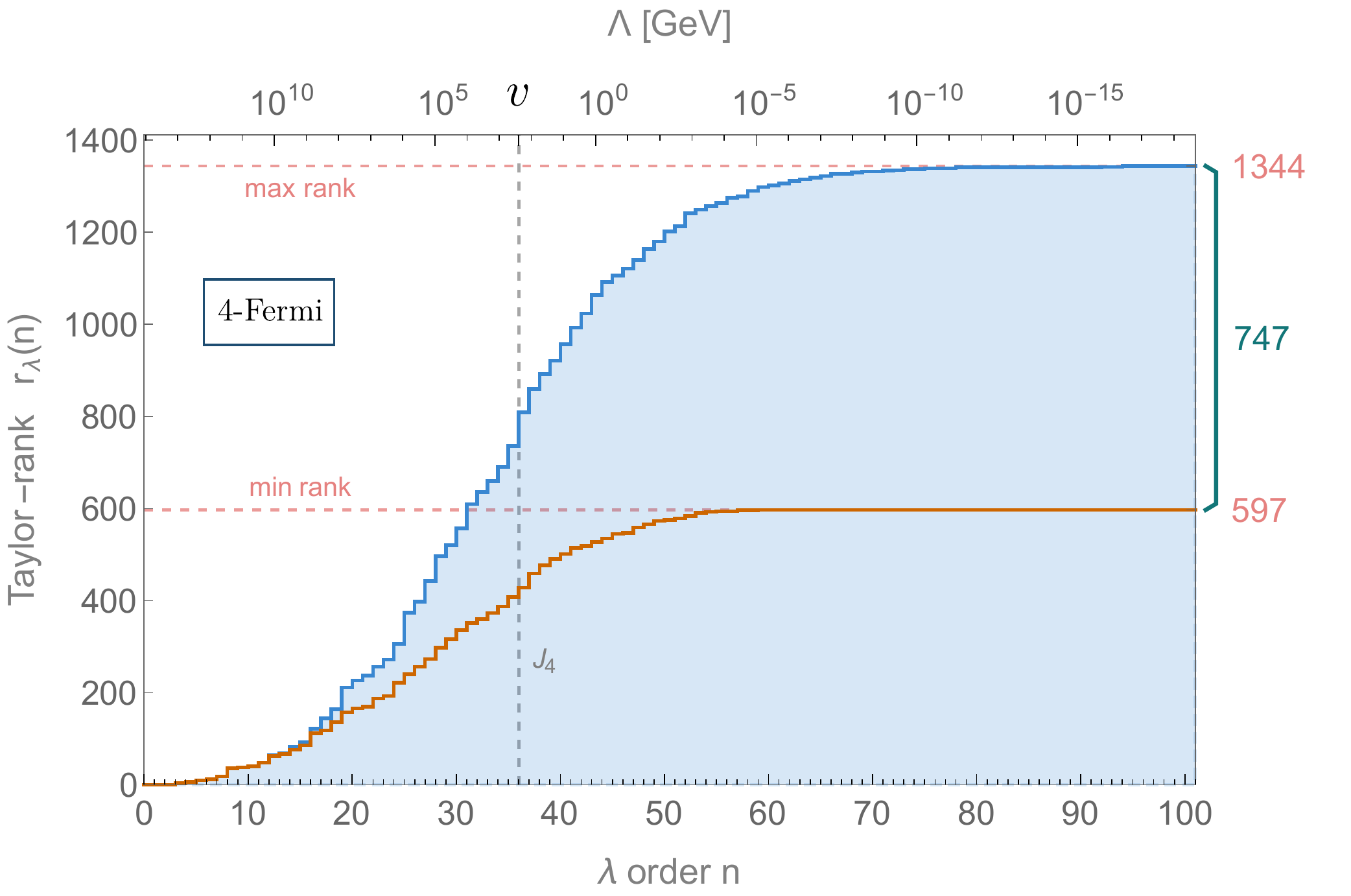}
			\caption{4-Fermi}
			\label{fig:tot4fermi}
		\end{subfigure}
		~ 
		\begin{subfigure}[b]{0.5\textwidth}
			\includegraphics[width=\textwidth]{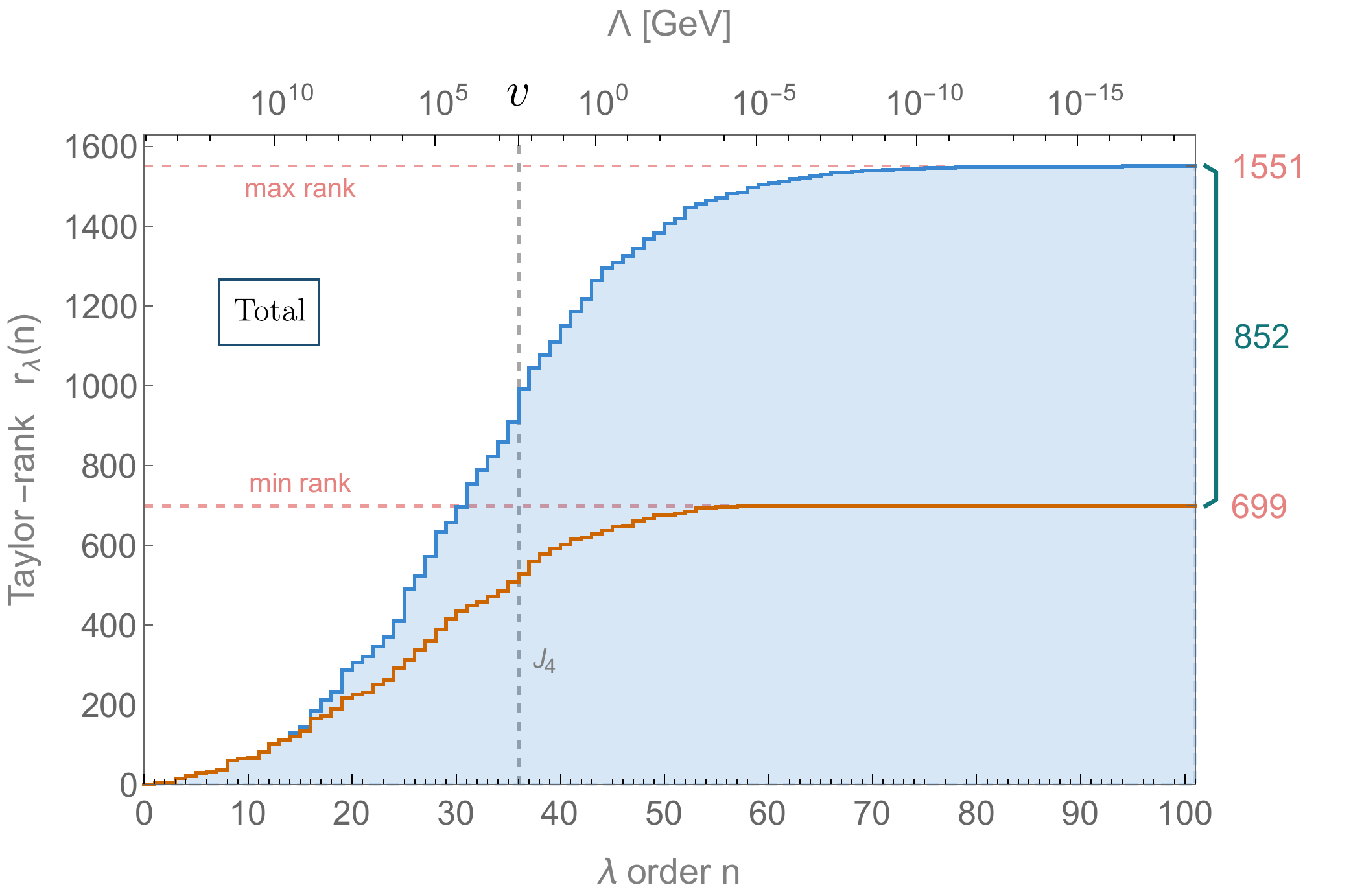}
			\caption{Total}
			\label{fig:tottot}
		\end{subfigure}
		\caption{Number of independent invariants from the maximal (blue step-wise line) and the minimal (dark yellow step-wise line) sets, denoted as Taylor rank $r_n$ in the text, at each order $n$ in the $\lambda$ expansion for the sum of all bilinear operators (\subref{fig:totbil}), 4-Fermi operators (\subref{fig:tot4fermi}), and all operators (\subref{fig:tottot}). At a fixed order in $\lambda$, the top $x$-axis shows the value of $\Lambda$ for which an invariant appearing at such order would be comparable to $J_4$, assuming it scales as $v^2/\Lambda^2$. The vertical dashed line marks the order $\lambda^{36}$ (corresponding to $\Lambda=v$, as indicated) where the SM$_4$ $J_4$ shows up, while the horizontal lines mark the values for the maximal and minimal rank, also labeled on the right $y$-axis. Finally, in each plot we highlighted the sources of opportunistic CPV as the difference between maximal and minimal ranks.}\label{fig:lambdarankstotal}
	\end{figure}
	
	\label{section:lambdaforall}
	\section{Flavor scenarios}
	\label{section:flavorassumptions}
	
	In the previous section, we explained how to compute the Taylor ranks of a set of CP-odd invariants associated to a dimension-six operator. 
	 In order to do this, an understanding of the $\lambda$-scaling of the building blocks of each invariant was needed. For the Yukawa matrices, this is done by means of the parametrization in Eqs.~\eqref{eq:ckmscaling}-\eqref{eq:massesscalingend}. On the other hand, the flavor structure of the Wilson coefficients is obviously unknown, as it can only be specified when measured or when a specific UV model is selected. To get the results displayed in Figures~\ref{fig:lambdaranksbilinears},~\ref{fig:lambdaranks4Fermi}, and~\ref{fig:lambdarankstotal}, we adopted an anarchic assumption, where all coefficient entries are assumed to be $\order{1}$.
	However, different ansatzes, appropriately justified, can be made on such coefficients. 
	In the next sections, we consider four of these scenarios, starting from the anarchic one used in the results above. We first summarize their characteristics, and, in order to compare them, we apply our Taylor rank algorithm on the maximal set of all bilinear operators. Our goal is to understand how the results of section~\ref{section:lambdaforall}, on the number of independent invariants at a given order in the $\lambda$-expansion, actually depend on the flavor assumptions.
	
	\subsection{Flavor anarchic scenario}
	
	The simplest assumption one can make on the flavor structure of the SMEFT coefficients is the anarchic or generic one, consisting in just taking all entries of a flavorful coefficient to be $\order{1}$. For $C_{uH}$, this means simply
	\begin{align}
		C_{uH}&=
		\left(
		\begin{array}{ccc}
			\rho _{11}+i \eta _{11} &  \rho _{12}+i \eta _{12} &  \rho _{13}+i \eta _{13} \\
			  \rho _{21}+i \eta _{21} &  \rho _{22}+i \eta _{22} &   \rho _{23}+i \eta _{23} \\
			  \rho _{31}+i \eta _{31} &   \rho _{32}+i \eta _{32} & \rho _{33}+i \eta _{33} \\
		\end{array}
		\right) &\rho_{ij}&\sim\eta_{ij}\sim\order{1}\ . 
		\label{eq:anarchicCuH}
	\end{align}

	If this is true, since in the SM$_{4}$ off-diagonal flavor entries are suppressed with respect to diagonal ones, the off-diagonal entries of a flavorful higher-dimensional coefficient will have a relatively larger impact on flavor-violating observables, such as Flavor Changing Neutral Currents. However, these observables are among the best studied and constrained. Indeed bounds arising from quantities related to flavor violation either in the quark or lepton sector~\cite{Isidori:2010kg, ESPPPG:2019qin, Aebischer:2020dsw, Silvestrini:2018dos, Pruna:2014asa, Feruglio:2015yua} place the scale of New Physics to be at least of $\order{10^6\text{ TeV}}$\@. Bounds coming from the electron dipole moment~\cite{Kley:2021yhn} also push the NP scale to these values if the anarchic assumption is taken. By looking at Figure~\ref{fig:lambdarankstotal}, then, we can infer that for $\Lambda\gtrsim 10^6\text{ TeV}$ we still have $\sim{70}$ invariants for the bilinear and $\order{100}$ invariants for the 4-Fermi operators that could have the same flavor suppression as the SM$_4$ $J_4$.
	
	\subsection{Minimal Flavor Violation}\label{sec:MFV}
	
	As a first non-anarchic assumption on the flavor structure of SMEFT we consider the scenario of Minimal Flavor Violation (MFV)~\cite{DAmbrosio:2002vsn, Isidori:2010kg}. If we turn off the Yukawa matrices, the kinetic sector of the SM$_4$ Lagrangian is invariant under the global $U(3)^5$ group of flavor transformations acting on quarks and leptons. If we assign spurious transformation properties to the Yukawa matrices as in Table~\ref{tab:ytrasmforma}, the Yukawa sector of  SM$_4$ is invariant too. Minimal Flavor Violation is then the requirement that any higher-dimensional operator has to be built out of $Y_{u,d,e}$ matrices and SM fields, so as to be formally invariant under the flavor group, taking into account the transformation properties in Table~\ref{tab:ytrasmforma}. For example, for $C_{uH}$, which lies in the same representation as $Y_u$:
	\begin{align}
		C_{uH}=\left(\rho_{uH}+i\eta_{uH}\right)\Yu + \order{Y_{u\vphantom{d}}^{\vphantom{\dagger}2}, Y_d^{\vphantom{\dagger}2}} \ ,
		\label{eq:MFVCuH}
	\end{align}
	where $\rho_{uH},\eta_{uH}$ are unknown numbers (without flavor indices). Looking at such an expression, it is clear how this scenario has a strong practical appeal, since it drastically reduces the number of free parameters entering the Lagrangian at each mass-dimension and at low orders in $\lambda$~\cite{Faroughy:2020ina, Greljo:2022cah}. 
	In addition, by tying the amount of flavor violation to that already present in the SM$_4$, it is more easily compatible with observables, bringing the lower bounds on the NP scale down to the TeV region~\cite{Kley:2021yhn,Isidori:2010kg,Workman:2022ynf}. However, a clear shortcoming of this method resides in the largest eigenvalue, $y_t$, of the up-Yukawa matrix, $Y_u$, being of order~1, making a controlled expansion in powers of $Y_u$ poorly defined. For example, we could multiply from the left the leading order term showed in Eq.~\eqref{eq:MFVCuH} with any power of $X_u$, and obtain a quantity with the same $\lambda$ suppression. 
	A number of solutions have been proposed to bypass this issue. First of all, one can notice that, because of the Cayley--Hamilton theorem, not all powers of $X_u$ are actually independent, so that all coefficients obtained with $X_u^n$ with $n>3$ can be reabsorbed by some appropriate redefinition~\cite{Greljo:2022cah}. Alternatively, one can postulate a flavor symmetry breaking in two steps: the top-Yukawa coupling is turned on first and breaks the $U(3)_Q\times U(3)_u $ sector of the flavor symmetry down to $U(2)_Q\times U(2)_u\times U(1)_t$. Then, one can build an EFT expansion around this vev, and the flavor symmetry is realized only non-linearly~\cite{Kagan:2009bn, Feldmann:2008ja}. This construction leaves an unbroken $U(2)^2\times U(1)\times U(3)^3$ which is linearly realized; it can be shown~\cite{Kagan:2009bn} to be a restriction of the $U(2)^5$ symmetry which we treat below.
	
\begin{figure}[t]
		\vspace*{-1cm}
		\centering
		\textbf{Taylor rank for $C_{uH}$ at each step in the MFV expansion}\\\medskip
		\includegraphics[width=1\textwidth]{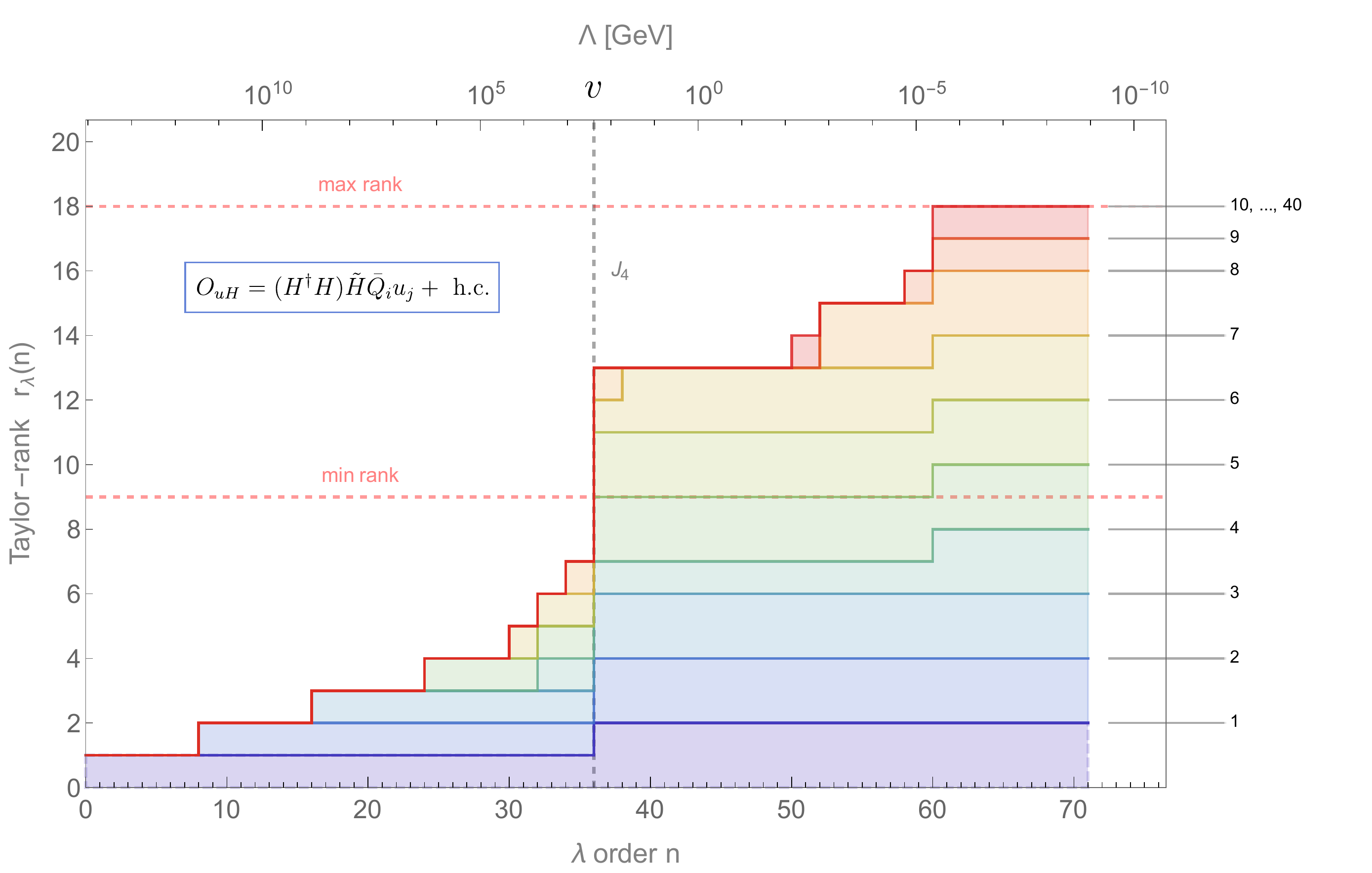}
		\caption{Illustration of the algorithm described in appendix~\ref{section:taylorMFV} applied to the bilinear coefficients $C_{uH}$: at each step, labeled on the right, we plot the result of the Taylor rank. We see that the algorithm saturates quite fast to the final result, and we stopped the iterations after the resulting plot was unchanged for 30 steps. 
			The vertical dashed line marks the order $\lambda^{36}$ where the SM$_4$ $J_4$ shows up, while the horizontal ones show the minimal and maximal ranks of $C_{uH}$. The final line, constituting the envelope of the plot, reproduces the result relative to the MFV scenario for $C_{uH}$ indicated in red in Figure~\ref{fig:CuHlambdaranks}.} 
		\label{fig:MFValgorithm}
	\end{figure}

	Our invariant based treatment, however, can also help to shed a new light on this problem. Focusing on $C_{uH}$, we sketch here a procedure to obtain a consistent expansion, leaving the details for appendix~\ref{section:taylorMFV}. We start from the first possible term we can write that is compatible with the MFV assumption:
	\begin{align}
		C_{uH}=(\rho_0+i\eta_0)Y_u
	\end{align}
	and compute the Taylor rank of this expression. Clearly, the rank is going to be 1, corresponding to $\eta_0$, as long as we expand to order $\lambda^n$, $n<36$, and becomes 2 at some point for $n\geq 36$, i.e.~after we start to be sensitive to the SM CP violation, $J_4\sim\lambda^{36}$. Then, we add another term, for example
	\begin{align}
		C_{uH}=(\rho_0+i\eta_0)Y_u+(\rho_1+i\eta_1)X_u Y_u
	\end{align}
	and compute the Taylor rank again\footnote{Clearly, as there is no obvious way to decide a hierarchy between the terms, an arbitrary ordering has to be chosen. A possibility is shown in appendix~\ref{section:taylorMFV}.}. At each step, we produce a plot like those in Figures~\ref{fig:lambdaranksbilinears} and~\ref{fig:lambdaranks4Fermi}. After a number of iterations, if we find that the plot does not change regardless of what other terms we add to the expansion, then  we stop. In this way, we have obtained an expansion for $C_{uH}$, specified by a set of parameters ($\rho_i$, $\eta_i$), such that any further term we could add would provide no additional information. 
	The result of this procedure for $C_{uH}$ is presented in Figure~\ref{fig:MFValgorithm}.
	
	\subsection{$U(2)^5$}\label{section:U(2)5maxset}
	As already mentioned, one of the proposed workarounds to avoid the power counting ambiguity carried by MFV consists in picking only a subgroup of the full $U(3)^5$. Here, we will focus on the $U(2)^5=U(2)_Q\times U(2)_u\times U(2)_d\times U(2)_L\times U(2)_e$ case~\cite{Barbieri:2011ci}, as a benchmark. This is not the only choice that can be made, and a larger set of possibilities has been studied in the literature~\cite{Faroughy:2020ina, Greljo:2022cah}. This approach has a number of advantages. First of all, similarly to the full MFV,
	flavor and CP bounds allow $\Lambda$, the scale where the EFT expansion breaks down, to be as low as the TeV region~~\cite{Kley:2021yhn, Barbieri:2012uh, Barbieri:2011fc}.
	To realize $U(2)^5$, we assume that the first two generations of the SM quark and lepton fields transform under the different factors of the flavor group as in Table~\ref{tab:doubletU2^5transform}, with $a=1,\,2$, while $Q_3$,  $u_3$, $d_3$, $L_3$, and $e_3$ are $U(2)^5$ singlets.
	\begin{table}[H]
		\centering
		\begin{tabular}{C|C|C|C|C|C}
			& $U(2)_Q$ & $U(2)_u$ & $U(2)_d$ & $U(2)_L$ & $U(2)_e$\\\hline\vspace{0.1cm}
			$Q_a$ &$\mathbf{2}$ & $ \mathbf{1}$ & $\mathbf{1}$ &$\mathbf{1}$ &$\mathbf{1}$  \\
			$u_a$ &$\mathbf{1}$ & $ \mathbf{2}$ & $\mathbf{1}$ &$\mathbf{1}$ &$\mathbf{1}$  \\
			$d_a$ &$\mathbf{1}$ & $ \mathbf{1}$ & $\mathbf{2}$ &$\mathbf{1}$ &$\mathbf{1}$  \\
			$L_a$ &$\mathbf{1}$ & $ \mathbf{1}$ & $\mathbf{1}$ &$\mathbf{2}$ &$\mathbf{1}$  \\
			$e_a$ &$\mathbf{1}$ & $ \mathbf{1}$ & $\mathbf{1}$ &$\mathbf{1}$ &$\mathbf{2}$  \\
		\end{tabular}
		\captionsetup{width=.75\textwidth}
		\caption{Transformation properties of the first two generations of each fermion SM field under the $U(2)^5$ flavor symmetry ansatz.}
		\label{tab:doubletU2^5transform}
	\end{table}
	
	Then, to reproduce the observed Yukawa couplings, one needs to break the flavor symmetry via the spurions $\{V_l,\, V_q,\,\Delta_u,\,\Delta_d,\,\Delta_e\}$, transforming as in Table~\ref{tab:spurionstransform}.
	\begin{table}[H]
		\centering
		\begin{tabular}{C|C|C|C|C|C}
			& $U(2)_Q$ & $U(2)_U$ & $U(2)_D$ & $U(2)_L$ & $U(2)_E$\\\hline\vspace{0.1cm}
			$V_l$ &$\mathbf{1}$ & $ \mathbf{1}$ & $\mathbf{1}$ &$\mathbf{2}$ &$\mathbf{1}$  \\
			$V_q$ &$\mathbf{2}$ & $ \mathbf{1}$ & $\mathbf{1}$ &$\mathbf{1}$ &$\mathbf{1}$  \\
			$\Delta_e$ &$\mathbf{1}$ & $ \mathbf{1}$ & $\mathbf{1}$ &$\mathbf{2}$ &$\mathbf{\bar{2}}$  \\
			$\Delta_u$ &$\mathbf{2}$ & $ \mathbf{\bar{2}}$ & $\mathbf{1}$ &$\mathbf{1}$ &$\mathbf{1}$  \\
			$\Delta_d$ &$\mathbf{2}$ & $ \mathbf{1}$ & $\mathbf{\bar{2}}$ &$\mathbf{1}$ &$\mathbf{1}$  \\
		\end{tabular}
		\captionsetup{width=.75\textwidth}
		\caption{Transformation properties of the independent spurions under the $U(2)^5$ flavor symmetry ansatz.}
		\label{tab:spurionstransform}
	\end{table}
	Using the available $U(2)^5$ flavor transformations, we can bring the quark Yukawas in the form~\cite{Greljo:2022cah}:
	\begin{align}
		Y_u=\begin{pmatrix}
			\Delta_u & V_q\\
			0 & y_t
		\end{pmatrix}
		\qquad
		Y_d=\begin{pmatrix}
			\Delta_d & 0\\
			0 & y_b
		\end{pmatrix}
		\label{eq:yukawasinU2^5}
	\end{align}
	with $y_{t,b}$ singlets of $U(2)^3$. With this parametrization, the Yukawa sector is clearly formally invariant under the $U(2)^5$ flavor group. Then, we extend the assumption that the spurions in Table~\ref{tab:doubletU2^5transform} are the only source of flavor symmetry breaking to the whole SMEFT, so that every flavorful coefficient has to be appropriately built with them. For example, the $C_{uH}$ coefficient can be expanded as
	\begin{align}
		C_{uH} = \begin{bmatrix}
			m(a_1^u,a_2^u,a_3^u,a_4^u)\Delta_u + \dots &
			m(b_1^u,b_2^u,b_3^u,b_4^u) V_q + \ldots \\
			V_q^\dagger m(c_1^u,c_2^u,c_3^u,c_4^u)  \Delta_{u} + \ldots &
			\Tr\left(m(b_1^u,b_2^u,b_3^u,b_4^u)\right)+ \ldots
		\end{bmatrix}\ ,
	\label{eq:U(2)5CuH}
	\end{align}
	where we defined
	\begin{align}
		x_{u,d}\equiv \Delta_{u,d}^{\vphantom{\dagger}}\Delta_{u,d}^\dagger\qquad
		m(a,b,c,d)\equiv a \mathbb{1}_{2\times 2}+bx_u+c x_d+d (V_q^{\vphantom{\dagger}}\otimes V_q^\dagger)\ ,
	\end{align}
	and the ellipsis indicate higher orders.
	In contrast with MFV, the performed expansion is now meaningful, as the involved spurions have eigenvalues $\order{10^{-2}}\ll1$. Nonetheless, we could still proceed as for MFV and remain agnostic about the expansion, consider the series in its entirety, and then retain only the coefficients that maximize the rank of the transfer matrix at each fixed $\lambda$ order. Then, again, we can directly use the invariants to characterize the parameter space. 
	
	Here, to assign the appropriate $\lambda$ scaling to the spurions, one needs to relate the expression in Eq.~\eqref{eq:yukawasinU2^5} to a basis where the actual Yukawas take that form. Alternatively, and in a sense more straightforwardly, one can compute the 10 invariants in Eq.~\eqref{eq:SMinvariants} both in the Wolfenstein parametrization Eqs.~\eqref{eq:ckmscaling}-\eqref{eq:massesscalingend} and in the one of Eq.~\eqref{eq:yukawasinU2^5} and relate the two. This is an unambiguous procedure, as it is flavor-basis independent. To prove that it is possible, we also need to study the invariants that can be built using the spurions in Table~\ref{tab:spurionstransform}. A detailed explanation of how this can be done is given in Appendix~\ref{section:U(2)5}\@.
	In particular, we show that 8 independent invariants can be built out of the spurions $\Delta_{u,d}$ and $V_q$. In addition, one has to consider the two singlets (one real and one complex) appearing in the $(3,3)$ position of $Y_u$. Since the maximal set will be ultimately built out of invariants of these objects, one may find that it is quite coincidental that their number is just enough as to make the transfer matrix have rank 18. More specifically, it may seem particularly remarkable that the 18 invariants we picked for the maximal set, which were chosen for the rather unrelated reasons explained above, end up picking all of the 18 independent invariant objects we can build out of  $\Delta_{u,d}$, $V_q$ and a singlet. However, one should note that, although there are 9 independent invariants, the relations that link redundant invariants to them are non-linear, and a linear object, such as the transfer matrix, cannot be sensitive to them. Thus, effectively, all of the linearly independent invariants are allowed. For more details, see footnote~\ref{foot:LinVsAlgIndependent} in Appendix~\ref{section:coeffvalues}.
	
\subsection{Froggatt--Nielsen}
	The last case we consider is that of a horizontal symmetry, i.e.~the Froggatt--Nielsen mechanism. In their seminal work~\cite{Froggatt:1978nt}, Froggatt and Nielsen noticed that the large hierarchies spanned by the quark masses and CKM matrix entries could be explained as coming from different powers of a symmetry breaking order parameter. Explicitly, this can be obtained by adding a complex scalar field $\phi$ to the Standard Model, taken to be a singlet of the SM $SU(3)_C \times SU(2)_L \times U(1)_Y$ gauge group. An abelian $U(1)_{FN}$ global symmetry is postulated, under which the scalar field has charge -1:
	\begin{align}
		FN(\phi)=-1 \ .
	\end{align}
	On the other hand, quarks can be assigned non-negative\footnote{This assumption can actually be relaxed, as long as the Yukawa terms in Eq.~\eqref{eq:fnyukawas} are invariant under $U(1)_{FN}$.} $U(1)_{FN}$ charges
	\begin{align}
		FN(Q_i)=f_{Q_i}\geq0 \quad FN(u_i^c)=f_{u_i}\geq0 \quad FN(d_i^c)=f_{d_i}\geq0 
	\end{align}
	where we used the charge conjugated fields\footnote{in the Dirac representation for the $\gamma$ matrices~\cite{Peskin:1995ev}.} $q_i^c=-i\gamma_0\gamma_2 \bar{q}^T$ so as to deal with left-handed fermions, only. With the aid of a global hypercharge transformation, the Higgs field can be taken to be neutral.
	This symmetry forbids renormalizable Yukawa terms. Instead, only higher-dimensional terms of the form
	\begin{align}
		-\mathcal{L}_{FN}\supset \sum_{ij}C_{ij}^{(u)}\left(\frac{\phi^*}{\Lambda_{FN}}\right)^{f_{u_i}+f_{Q_j}}\bar{Q}_ju_i\tilde{H}+\sum_{ij}C_{ij}^{(d)}\left(\frac{\phi^*}{\Lambda_{FN}}\right)^{f_{d_i}+f_{Q_j}}\bar{Q}_jd_iH+\text{ h.c.}
		\label{eq:fnyukawas}
	\end{align} 
	are allowed, where $C_{ij}^{(u,d)}$ are matrices with $\order{1}$ entries and $\Lambda_{FN}\gg v$ is the scale at which this theory needs to be UV completed.  
	Through an appropriate scalar potential, the scalar field acquires a vev
	\begin{align}
		\frac{\expval{\phi}}{\Lambda_{FN}}=\lambda
		\label{eq:lambdaspurion}
	\end{align}
	which spontaneously breaks the symmetry. Then, the entries of the quark Yukawa matrices are determined by the operators in Eq.~\eqref{eq:fnyukawas} as
	\begin{align}
		Y_{u, \, ij}=C_{ij}^{(u)}\lambda^{f_{u_i}+f_{Q_j}} \quad Y_{d, \, ij}=C_{ij}^{(d)}\lambda^{f_{d_i}+f_{Q_j}}\ .
	\end{align}
	Appropriate choices for the values of $\{f_{Q_i}, \,f_{u_i}, \,f_{d_i}\}$ allow us to reproduce the observed hierarchies in the quark sector. A suitable choice is
	\begin{align}
		f_{Q_i}=\{3,\,2,\,0\} \quad  f_{u_i^c}=\{5,\,2,\,0\} \quad  f_{d_i^c}=\{1,\,0,\,0\}\ .
	\end{align}
	This construction can be extended to the whole SMEFT~\cite{Bordone:2019uzc} in the same way: flavorful operators are multiplied by appropriate powers of $\phi/\Lambda_{FN}$ so as to respect the $U(1)_{FN}$ symmetry, and their entries are suppressed correspondingly when $\phi$ freezes to its vev.
	For example, the now familiar coefficient of the modified Yukawa operator, $C_{uH}$, becomes
	\begin{align}
		C_{uH}=
		\left(
		\begin{array}{ccc}
			\lambda ^8 \left(\rho _{11}+i \eta _{11}\right) & \lambda ^5 \left(\rho _{12}+i \eta _{12}\right) & \lambda ^3 \left(\rho _{13}+i \eta _{13}\right) \\
			\lambda ^7 \left(\rho _{21}+i \eta _{21}\right) & \lambda ^4 \left(\rho _{22}+i \eta _{22}\right) & \lambda ^2 \left(\rho _{23}+i \eta _{23}\right) \\
			\lambda ^5 \left(\rho _{31}+i \eta _{31}\right) & \lambda ^2 \left(\rho _{32}+i \eta _{32}\right) & \lambda^0(\rho _{33}+i \eta _{33}) \\
		\end{array}
		\right) \ ,
		\label{eq:FNCuH}
	\end{align}
where the 18 numbers $\rho_{i,j}, \eta_{ij}$ are all supposed to be of order 1.	
	Although we introduced the Froggatt--Nielsen model using the language of a UV completion,
	one alternatively could postulate a horizontal symmetry~\cite{Leurer:1992wg} in the IR, i.e.~that the flavorful coefficients have to be built out of the spurion in Eq.~\eqref{eq:lambdaspurion} so as to respect the $U(1)_{FN}$ symmetry.

	\subsection{Selected results and discussion}
\begin{figure}[H]
		\vspace*{-1cm}
		\centering
		\textbf{Taylor rank at each order in $\lambda$ for all bilinear operators: flavor scenarios}\\\medskip
		\includegraphics[width=0.95\textwidth]{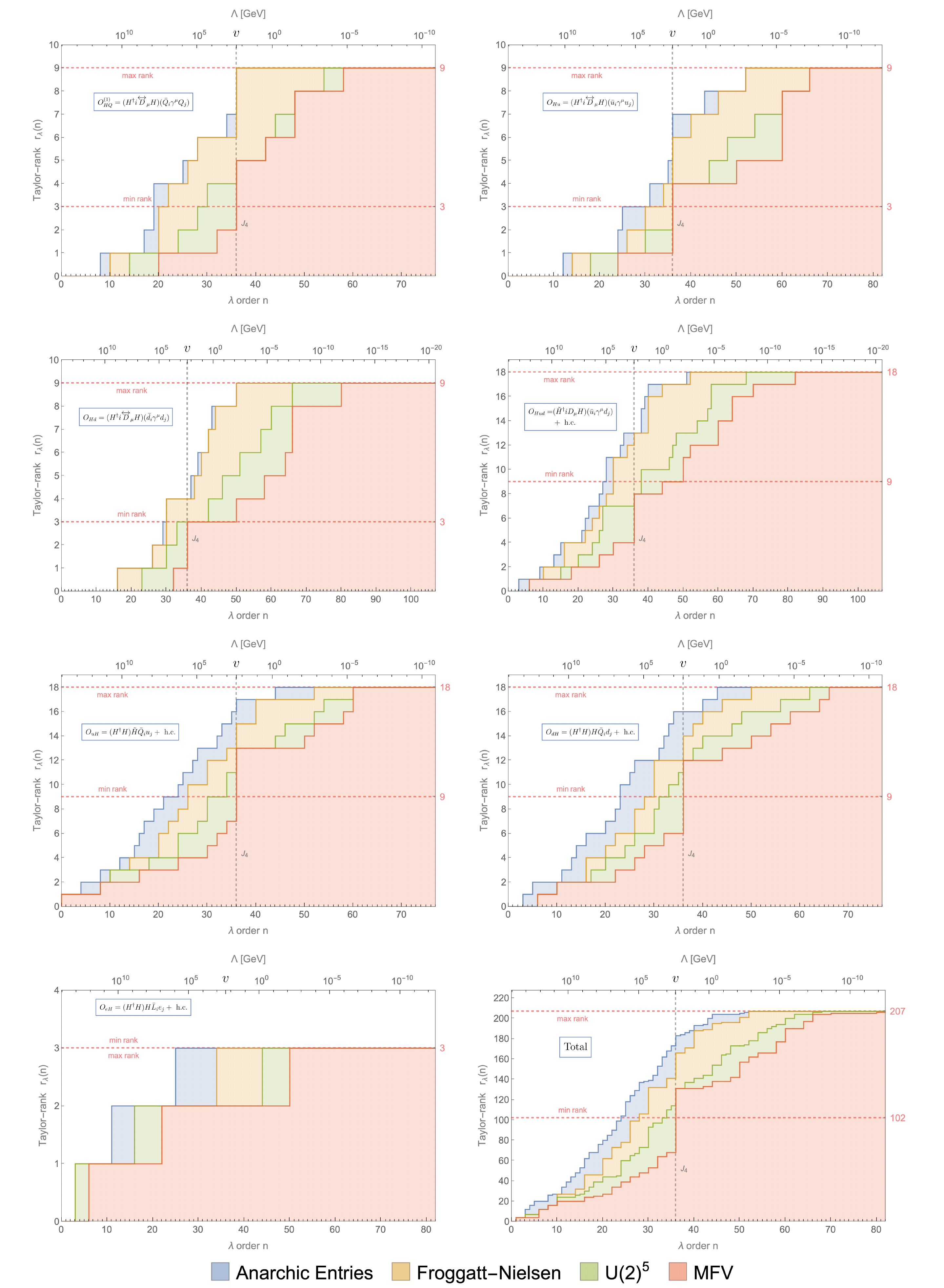}
		\caption{Plot comparing the Taylor rank obtained for bilinear operators at each order in $\lambda$ for the four different flavor scenarios described in the text: anarchic entries, Minimal Flavor Violation, $U(2)^5$ and Froggatt--Nielsen. The dashed vertical line marks the order $\lambda^{36}$ that characterizes the Jarlskog invariant $J_4$, while horizontal lines have been placed to indicate the minimal and maximal rank for each operator. Each of the first 7 plots corresponds to a group of operators in Table~\ref{Bilinearlist} (excluding those with 0 maximal and minimal sets), only one of which is chosen as a representative, while the last one is the total Taylor rank for all bilinear operators. 
			Notice that the line corresponding to MFV only becomes larger than the minimal rank after reaching $\order{\lambda^{36}}$. Indeed, the only way to start resolving real entries is via interference with a CP-odd quantity. As the MFV flavorful building blocks are just $Y_{u,d}$, the first possible object is the familiar $J_4$, appearing at $\lambda^{36}$.}
		\label{fig:CuHlambdaranks}
	\end{figure}

	As we have shown, each flavor scenario dictates a specific structure of the Wilson coefficients, as exemplified by Eqs.~\eqref{eq:anarchicCuH}, \eqref{eq:MFVCuH}, \eqref{eq:U(2)5CuH} and~\eqref{eq:FNCuH} for $C_{uH}$. We generalized this procedure to all bilinear operators, parametrizing their coefficients according to each of the four scenarios above, and computed their Taylor ranks. The result is visualized in Figure~\ref{fig:CuHlambdaranks}. The first feature to catch the eye is the fact that there appears to be a clear hierarchy between the different possibilities. Stated loosely, the more restrictive the symmetry, the harder it gets to increase the rank, as illustrated by the curves never crossing each other, although they coincide in some points. This result gives a more strong footing to the idea that assuming a symmetry reduces the number of parameters needed to describe some observables at a given precision, as it shows that it is a flavor-invariant statement.
	As mentioned in the discussion about MFV and $U(2)^5$, this approach offers a way out of power counting ambiguities: at a fixed $\lambda$ order, we can clearly discriminate which parameters contribute and which do not.
	
	The plots in Figure~\ref{fig:CuHlambdaranks}, however, tell us even more than this. Indeed, as we have stated, the different flavor scenarios imply different bounds on the scale of new physics $\Lambda$. If we assume that the invariants scale as $\sim v^2/\Lambda^2$, then the top $x$ axis indicates how many independent $\order{{1}/{\Lambda^2}}$ invariants are comparable with $J_4$ at each value of $\Lambda$. By looking at constraints from FCNC \cite{Isidori:2010kg}, for example, the bound on $\Lambda$ for the flavor anarchic scenario is $\Lambda \gtrsim 10^9\text{ GeV}$, which means, by intersecting the corresponding blue line, that there are $\lesssim 70$ invariants of the maximal set that could be comparable with $J_4$. On the other hand, the bound on MFV is milder, $\Lambda \gtrsim 10^4\text{ GeV}$. However, the much slower growth for the MFV curve is not compensated by this difference of the NP scale, as this constraint implies here that only $\lesssim 50$ invariants can be approximately as large as $J_4$. We performed this counting for all four flavor scenarios analyzed here, and we summarize the result in Table \ref{tab:flavorscenariosbounds}.
	\begin{table}[H]
		\centering
		\begin{tabular}{c|c|c}
			Flavor Scenario & NP scale bounds& $\#$ of invariants $\sim J_4$ \\[0.1cm]\hline
			Anarchic &$\Lambda\gtrsim 10^9$ GeV& $\lesssim 120$ \\[0.1cm]
			FN&$\Lambda\gtrsim 10^7$ GeV& $\lesssim 90$    \\[0.1cm]
			$U(2)^5$ &$\Lambda\gtrsim 10^5$ GeV& $\lesssim 70$    \\[0.1cm]
			MFV &$\Lambda\gtrsim 10^4$ GeV& $\lesssim 50$    \\[0.1cm]
		\end{tabular}
		\captionsetup{width=\textwidth}
		\caption{Number of invariants that could be comparable with $J_4$ for the different flavor scenarios analyzed in Sec.~\ref{section:flavorassumptions}. The third column is obtained by counting the number of invariants $\mathcal{I}$ that respect $\left(v^2/\Lambda^2\right)\mathcal{I}\gtrsim J_4$, with $\Lambda$ the lowest scale of new physics compatible with the flavor constraints in the different scenarios. We only take bilinear operators into account and we assume the invariants scale as $v^2/\Lambda^2$. The bounds are derived from FCNC and EDM data \cite{Isidori:2010kg, Leurer:1992wg, Leurer:1993gy, Bordone:2019uzc, Altmannshofer:2022aml}.  For the FN case, the bounds depend strongly on the specific model \cite{Tsumura:2009yf, Pascoli:2016wlt, Berger:2014gga, Bauer:2016rxs, Altmannshofer:2022aml}: we quote here the most stringent one.}
		\label{tab:flavorscenariosbounds}
	\end{table}

	\section{Conclusion}\label{section:conclusion}
	
	In this paper, we explored the parameter space of CP-violating observables at order $\order{1/\Lambda^2}$ in the Standard Model Effective Field Theory, employing and expanding on the formalism of linear CP-odd flavor invariants developed in~\cite{Bonnefoy:2021tbt}. Since CP is already violated by the Standard Model in the electroweak sector at dimension four, characterizing this parameter space in its entirety requires that we account for interference between the Standard Model CP-odd invariant $J_4$ and CP-even $\Lambda^{-2}$ suppressed objects, which we dubbed {\it opportunistic CPV}\@. To do this, we defined a larger set of invariants with respect to~\cite{Bonnefoy:2021tbt}, which we called a \emph{maximal set}, and presented such a set for all operators of SMEFT at dimension-six in the Warsaw basis. We then studied the relation that links this set to the minimal one defined when $J_4=0$. Using the Wolfenstein parametrization, one can provide a hierarchy within the invariants, as well as a counting at each order in the $\lambda$ expansion.
	We found that opportunistic CPV, although dependent on the fact that $J_4\neq 0$, manifests itself at a $\lambda$-order much lower than $J_4$. We also discussed how the $\lambda$-counting is affected by the underlying flavor assumption on the flavorful coefficients, focusing on four specific ansatzes as benchmarks. As our construction allows us to relate invariants and Lagrangian parameters in a linear way, it incidentally provides a way to clearly identify which of the latter can be resolved, while bypassing at the same time power counting ambiguities, usually connected to e.g.~the MFV ansatz.

	This work still leaves room for exploration in a number of directions. Our formalism allows one to treat the interference of CP-even, $\Lambda^{-2}$ suppressed quantities with the SM$_4$ $J_4$ to form CP-violating objects. Moreover, we managed to infer a hierarchy between the invariants, so that, at a given precision, we can reduce the number of parameters needed to span the parameter space. As such, it would be interesting to see how to employ our construction for phenomenological applications, for example to check whether one could obtain competitive bounds on CP-even quantities using CP-odd observables such as, e.g., the electron EDM\@. As we stressed, directly running our Taylor rank analysis at the level of a set of observables, instead of using invariants as a proxy, would be very useful in order to get the most accurate picture. Such an endeavor would result in a solid assessment of the sensitivity of CPV observables to the SMEFT parameter space.
	
	Another aspect worth exploring would be the extension of the Standard Model to include right-handed neutrinos to account for neutrino masses. In that case, additional sources of CP-violation with respect to $J_4$ are already present in the renormalizable SM$_4$, appearing as phases in the PMNS matrix~\cite{Maki:1962mu, Pontecorvo:1957qd}. These three phases, then, can provide additional sources of interference, and the PMNS matrix can be used as an additional building block to build our invariants \cite{Branco:1986gr,Branco:1998bw,Dreiner:2007yz,Yu:2019ihs,Wang:2021wdq}. In particular, as the $U(1)^3$ residual symmetry in the lepton sector is now broken, most of the secondary phases become physical again at $\order{1/\Lambda^2}$. The case of Majorana masses induced by the Weinberg operator is also interesting (see Refs.~\cite{Yu:2021cco,Yu:2022nxj,Yu:2022ttm} for recent progress in that direction).
	
	Moreover, in this work we limited ourselves to the case of SMEFT, which is built on the assumption that the Higgs field belongs to a $SU(2)_L$ doublet and the electroweak symmetry is linearly realized. Relaxing these hypotheses leads to the so-called Higgs Effective Field Theory (HEFT) \cite{Feruglio:1992wf,Alonso:2012px, Buchalla:2013rka, Brivio:2013pma, Brivio:2016fzo,Sun:2022ssa,Sun:2022snw,Sun:2022aag}. It would be valuable to understand how the work performed here generalizes to that context. 
	Finally, our work could be extended to study the parameter space of CP violating observables in those approaches that rely on an on-shell description \cite{Durieux:2019eor, Durieux:2019siw, Durieux:2020gip, Dong:2022jru,Chang:2022crb}, bypassing and complementing the lagrangian picture that we took as our starting point.

	\acknowledgments
	We thank G.~Branco, V.~Cort\'es, J.~Kley, A.~Trautner, N. Weiner and C.~Yao for inspiring discussions. This work is supported by the Deutsche Forschungsgemeinschaft under Germany's Excellence Strategy  EXC 2121 ``Quantum Universe'' - 390833306.  JTR is supported by National Science Foundation grants PHY-19154099 and PHY-2210498, and by an award from the Alexander von Humboldt Foundation.  EG is supported in part by the National Science Foundation under Grant No. NSF PHY-1748958 and by the Collaborative Research Center SFB1258 and the Excellence Cluster ORIGINS, which is funded by the Deutsche Forschungsgemeinschaft (DFG, German Research Foundation) under Germany’s Excellence Strategy – EXC-2094-390783311. QB is supported by the Office of High Energy Physics of the U.S. Department of Energy under Contract No. DE-AC02-05CH11231.  CG and JTR performed part of this work at the Aspen Center for Physics, which is supported by National Science Foundation grant PHY-2210452.
	
	\appendix
	
	\section{Single trace invariants}
	\label{appendix:singletrace}
	In both this work and~\cite{Bonnefoy:2021tbt}\footnote{We thank David Hogg and Ben Blum-Smith for some useful insights on the subject of this section.} we built the flavor invariants relevant to our analysis as the imaginary part of a single trace built out of one instance of a dimension-6 coefficient, generically labeled $C^{(6)}$, and an arbitrary number of Yukawa matrices. In our previous work, this procedure, which was also introduced earlier in the literature \cite{Botella:1994cs}, was presented as a choice. However, we show here that this choice turns out to be the most generic one, under some suitable assumptions. Specifically, we require our invariants to be:
	\begin{enumerate}
		\item singlets of the full $U(3)^5$ flavor group, \label{prop1}
		\item polynomials in $C^{(6)}$, $Y_{u,d,e}$ and their hermitian conjugates, \label{prop2}
		\item suppressed, at least naively, as $1/\Lambda^2$, meaning $C^{(6)}$ can only appear once,\label{prop3}
		\item CP-odd. \label{prop4}
	\end{enumerate} 
	Notice that, once an invariant $I$ realizing~\ref{prop1}-\ref{prop3} is built,~\ref{prop4} is fulfilled by taking $\Im{(I)}$. Thus, it is enough to focus on the properties of those $I$ that realize~\ref{prop1}-\ref{prop3}.
	We show here that these assumptions are sufficient to restrict to the type of invariants used in the main text. 
	
	A theorem due to Weyl (\cite{Weyl1939} Section II.A.9) \footnote{The formulation used here is taken from~\cite{Villar:2021wnx}.} states that if $f$ is a $U(n)$-invariant scalar function of a set $v_1,\dots, v_r$ of vectors of $U(n)$, then it can be written as a function of only the scalar products of the $v_i$'s. This means that there is a function $g$ such that
	\begin{align}
		f(v_1,\dots, v_r)=g(V^\dagger \cdot V)=g(v^\dagger_i\cdot v_j)
	\end{align}
	where $V$ is the matrix with the $v_i$'s as columns. In other words, an invariant function is just a function of invariants. 
	We can adapt this theorem for our scopes in the following way. Consider a $U(3)_Q\times U(3)_u$-invariant function $f$ of $Y_u$. We can consider the matrix $Y_u$ as formed by three row belonging to the anti-fundamental of $U(3)_u$, and the previous theorem implies the existence of a function $g$ such that
		\begin{align}
		f(Y_u)=g(Y_uY_u^\dagger )=g(X_u)\ ,
	\end{align}
	where now we only need to impose that $g(X_u)$ is invariant under $U(3)_Q$. 
	However, the theorem exposed above for vector representations can be generalized to arbitrary tensors (Theorem 9.3 in Ref.~\cite{Popov1994}), so that there exists a function $h$ of the invariants built with $X_u$ such that 
	\begin{align}
	f(Y_u)=g(X_u)=h(\Tr[X_u],\Tr[X_u^2],\Tr[X_u^3])\ .
	\end{align}
	The argument can be generalized to functions of $C^{(6)}, Y_u,Y_d,Y_e$, which are our building blocks here, to show that an invariant function under the flavor group is a function of invariants\footnote{It is worth noticing that the mentioned theorem by Weyl also shows that if $f$ is polynomial, then $g$ is polynomial, too. This also holds for its generalization to tensors \cite{Popov1994}, which guarantees that property~\ref{prop2} is fulfilled.}.
	
	With this in mind, let us focus for the moment on quark bilinear operators, whose associated $C^{(6)}$ transform under some representation of $U(3)_Q\times U(3)_u \times U(3)_d$. 
	In this case, the only quantity transforming under the leptonic factor of the flavor group $U(3)_L \times U(3)_e$ is $Y_e$. Our invariant $I$ then forcibly factorizes as
	\begin{align}
		I(C^{(6)}, Y_u,Y_d,Y_e)=f(C^{(6)}, Y_u,Y_d)g(Y_e)\ ,
	\end{align}
	where $f(C^{(6)}, Y_u,Y_d)$ is a $U(3)_Q\times U(3)_u \times U(3)_d$ singlet and $g(Y_e)$ a $U(3)_L\times U(3)_e$ singlet. Since $g(Y_e)$ represents an overall factor, independent on $C^{(6)}$, we can safely drop it and focus on $f(C^{(6)}, Y_u,Y_d)$. From what we showed above, the invariant function $f$ is equal to a function of the invariants built with $C^{(6)}$, $Y_u$ and $Y_d$. We wish to understand how it can look like.
	
	We can contract the indices of the matrices only using the two invariant tensors of each $SU(3)$ factor, $\delta_{a}^b$ and $\epsilon_{abc}$. Since we wish, additionally, invariance under the full $U(3)^5$, we can do without the latter, as an odd number of $\epsilon_{abc}$ is forbidden, while even combinations of them can be reduced to combinations of $\delta_{a}^b$'s.
	
	Given property~\ref{prop3}, we can start building a generic invariant from the single bilinear coefficient ${C^{(6)}}^a_b$ we have at our disposal: if it has either two $U(3)_{u}$ or $U(3)_{d}$ indices (one upper and one lower), we can either contract them with each other, in which case we're done, or not. 
	If not, these indices must be saturated by contraction with $Y_{u,d}$, so we can always consider the resulting combination belonging to $\mathbf{3}_Q\otimes\bar{\mathbf{3}}_Q$ (e.g.~$C_{uH}^{\vphantom{\dagger}}\to C_{uH}^{\vphantom{\dagger}}Y_{\vphantom{d}u}^\dagger$ as in Eq.\eqref{eq:invexample1}). Now, its remaining free $SU(3)_Q$ indices must be contracted with some product of Yukawas also belonging to the same representation (or within themselves). As the $SU(3)_{u,d}$ indices of the Yukawas must, too, be contracted with instances of $\delta_a^b$, we are forced to use products of $X_{u,d}$s to build a singlet. This leaves us with a single linear trace. Obviously, we could again multiply such single trace with some other invariant quantity $g_q(Y_u,Y_d)$. However, this quantity cannot depend on $C^{(6)}$, since the result would not be linear, and is then just a combinations of Yukawa matrices. As such, they can be discarded: $g_q(Y_u,Y_d)$ can either yield irrelevant prefactors in front of some $\Im \, I$, or generate $J_4$ when the imaginary part acts on it. As alluded to in Section~\ref{opportunisticDef}, this second possibility generates a trivial kind of opportunistic CPV, which we do not include in our analysis. In particular, such invariants are always more suppressed than $J_4$ itself.
	
Clearly, there is nothing special about the $U(3)_Q$ indices, and we could have as well started by contracting them and then proceed to contract the $U(3)_{u,d}$ ones. A generalization to the lepton case is straightforward. Finally, a similar analysis can be performed for 4-Fermi operators, leading to the fact that the indices of their Wilson coefficients should be paired and inserted in a matrix trace together with an appropriate sequence of $X_{e,u,d}$. This results in two generalized trace structures, shown in Eq.~\eqref{eq:trAandtrB}.

	\section{Opportunistic CPV with rephasing invariants}
	\label{appendix:opportunisticRephasingInvariants}
	
	In this appendix, we expand on the discussion of section~\ref{section:vs} and illustrate some aspects of opportunistic CPV, using rephasing invariants (that is, singlets under the $U(1)^9$ group associated to the vectorlike phase shifts of each fermion mass eigenstate), for which formulae are simpler than for complete flavor invariants.
	
	Consider the CP-odd rephasing invariants 
	\beq
	\label{eq:rephasingInvariantsCuH}
	\text{Im}\(V_{\text{CKM},ki}^*V_{\text{CKM},ji}C_{uH,kj}\) \ ,
	\eeq
	where $i,j$ are fixed and $k$ is summed over, and $C_{uH}$ is the Wilson coefficient introduced in \eq{eq:modifiedYukawa}. The one with $i=j=3$ for instance shows up as the leading contribution to CPV observables in single top+Higgs processes at the LHC~\cite{Faroughy:2019ird}, when one modifies the top Yukawa coupling. To see opportunistic CPV at play, let us assume that $C_{uH,ij}=\delta_{i1}\delta_{j3}C_{uH,13}$. Accounting for CKM unitarity, we find that the invariants of \eq{eq:rephasingInvariantsCuH} capture only two quantities,
		\beq
		Q_1\equiv\text{Im}\(V_{\text{CKM},11}^*V_{\text{CKM},31}C_{uH,13}\) \ , \quad Q_2\equiv\text{Im}\(V_{\text{CKM},12}^*V_{\text{CKM},32}C_{uH,13}\) \ .
		\eeq
		If $J=0$, where $J\equiv\text{Im}\(V_{\text{CKM},12}V_{\text{CKM},32}^*V_{\text{CKM},11}^*V_{\text{CKM},31}\)$ is the CKM part of the Jarlskog invariant $J_4$, those two quantities degenerate to one ($\text{Im}\(C_{uH,13}\)$, in a basis where the CKM matrix is real). This is consistent with the rephasing-invariant version of \eq{eq:maxvsmin}: one can write
		\beq
		\label{eq:largebeta}
		Q_2=\alpha Q_1+J \beta \ , \text{ with } \beta=\frac{\text{Re}\(V_{\text{CKM},11}^*V_{\text{CKM},31}C_{uH,13}\)}{\abs{V_{\text{CKM},11}V_{\text{CKM},31}}^2} \ ,
		\eeq
		where $\beta$ does not diverge for most ways of taking $J\to 0$. Nevertheless, when $J\neq 0$, $Q_{1,2}$ are independent and allow one to reconstruct the full $C_{uH,13}=\text{Re}\(C_{uH,13}\)+i\,\text{Im}\(C_{uH,13}\)$. This is an avatar of opportunistic CPV: the two CP-odd invariants $Q_{1,2}$ capture two coefficients which are often classified as CP-even and CP-odd, via the interference with the CKM phase. Furthermore, it can be checked that those coefficients are captured at order $\cO(\lambda^3)$, i.e.~the Taylor rank of the set $\{Q_1,Q_2\}$ reaches $2$ at order $\cO(\lambda^3)$, despite $J\sim \lambda^6$. This is consistent with \eq{eq:largebeta}, since $\beta\sim \lambda^{-3} \gg 1$. This is an equivalent statement to the fact that $R_{4a}$ in \eq{eq:maxvsmin} can be much larger than one.
		 
	\section{More on the Taylor rank}
	\label{section:taylorrank}
	
	In this appendix, we elaborate on the notion of Taylor rank that we introduced in the main text. 
	
	Although we focussed on the transfer matrix previously, the Taylor rank can be defined for any matrix $M$, whose entries are expanded in a Taylor series in a small parameter $x$ up to a common fixed order $n$. As anticipated in Section~\ref{section:lambdarank}, its Taylor rank $r_n(M)$ corresponds to the smallest rank encountered in the equivalence class  of $M$, defined by the equivalence relation
	\beq
	M \sim N \text{ if } M-N=\cO(x^{n+1}) \ .
	\eeq
	It is smaller or equal to the rank that $M$ has when truncated to $\cO(x^n)$. For instance,
	\begin{align}
		r_9(M)=1 \text{ for } 
		M=\left(
		\begin{array}{ccc}
			x& x^2 & x^4 \\
			x ^3 & x ^4 & x ^6 \\
			x ^7 & x ^8 &0\\
		\end{array}
		\right)+\order{x^{10}}\ .
		\label{rankExample}
	\end{align}
	This matrix, when taken at face value, has rank 2: the first two rows are proportional to each other, while the third one is not. However, when computing $r_9(M)$, we should consider adding to any entry of $M$ any number which is 0 at $\order{x^9}$, in particular we can add $x^{10}$ in the $(3,3)$ position. Doing so, we reduce the rank to 1, which is therefore the Taylor-rank we are after. 
	
	Scanning the full equivalence class becomes computationally expensive rather quickly as the dimension of the corresponding matrix increases, which is why we rely on an algorithmic Gaussian elimination approach. The latter proceeds via a refined version of the usual Gaussian elimination, as follows: pick the largest entry in the matrix, say $M_{ij}$, i.e.~the one starting at the lowest power in $x$, say $x^m$:
	\beq
	M_{ij}\big\vert_{\cO(x^m)}\neq 0 \ , \quad \forall k,l,\forall p<m \ , \ M_{kl}\big\vert_{\cO(x^p)}= 0 \ .
	\eeq
	If there is more than one of such entries, it suffices to randomly pick one of them. If there is no non-zero entry, we stop here, and the rank is clearly $r_n(M)=0$. If there is, then $r_n(M)\geq1$. Now switch the first row with the $i$-th and the first column with the $j$-th, so that the largest entry is now in the $(1,1)$ position. Next, subtract to each row $i\neq 1$
	\beq
	M_{ik}\to M_{ik}-\frac{M_{i1}}{M_{11}}M_{1,k}
	\eeq
	and expand all entries to order $n$ in $x$ to stay consistent with the expansion. This means clearly that $M_{i1}\to 0$, so that now the first column is composed of 0s everywhere but in the first spot. In addition, this manipulation does not introduce large numbers, since by definition $M_{11}= \cO(x^m)$ and $M_{i1}\leq\cO(x^m)$, therefore $\frac{M_{i1}}{M_{11}}\leq \cO(1)$. Next, focus on the sub-matrix $M_{ik}$ with $i>1$. If this sub-matrix is identically 0, then we are done and $r_n(M)=1$. If not, pick the largest entry of this sub-matrix, again choosing a random one if there are repetitions and repeat the same steps. At the end, we will have a matrix in row echelon form, and we define $r_n(M)$ as the number of non-zero rows.
	
	We can explicitly evaluate this algorithm in the specific example of \eqref{rankExample}. The largest entry is $x$ and it is already in the $(1,1)$ position. Now we subtract the first row from the second and third ones, multiplied by the first entry of each and divided by $M_{11}=x$:
	\begin{align}
		M\to\left(
		\begin{array}{ccc}
			x& x^2 & x^4 \\
			0 & 0 & 0 \\
			0 & 0 & 0\\
		\end{array}
		\right)+\order{x^{10}}\ .
	\end{align}
	Since we reached the desired row echelon form, we can stop here and we obtained $r_{10}(M)=1$. Notice that, crucially, we neglected a $x^{10}$ term which would have appeared in the $(3,3)$ position after the last step, as it was negligible in the expansion considered here. 
	
	In section~\ref{section:lambdarank}, we applied this algorithm to $x=\lambda$ and $M=\cT$, the transfer matrix connecting the invariants $L_a$ and the SMEFT parameters $C_i$ as defined in Eq.~\eqref{eq:transfermatrixdef}. At each step, we therefore perform\footnote{It may seem that \eqref{redefinitionInvariantsLambdaRank} does not define new flavor invariants, since $\frac{\cT_{i1}}{\cT_{11}}$ is not an invariant notion. However, neither is $\lambda$, so in this expression, one should really understand "$\frac{\cT_{i1}}{\cT_{11}}$ computed in a basis where $\lambda$ is explicitly defined" (such as the up- or down-basis), which turns $\frac{\cT_{i1}}{\cT_{11}}$ into an invariant concept, since it is computed in a well-defined basis to which all flavor bases are implicitly related. In particular, any mass or mixing angle appearing in $\frac{\cT_{i1}}{\cT_{11}}$ can be expressed in terms of the invariants of \eqref{eq:SMinvariants}, making the whole procedure invariant.}
	\beq
	L_{a>1}\to L_a-\frac{\cT_{a1}}{\cT_{11}}L_1 \ ,
	\label{redefinitionInvariantsLambdaRank}
	\eeq
	so that after the last, $r_n(\cT)$-th step, we obtain
	\beq
	L_{a>r_n(\cT)}=\cO(\lambda^{n+1}) \ ,
	\eeq
	i.e.~all invariants lie in the span of $\{L_1,...,L_{r_n(\cT)}\}$, up to corrections of $\cO(\lambda^{n+1})$, as announced in \eqref{expansionTaylorRank}.
	
	Let us close this appendix by mentioning other intuitive but incorrect approaches to the Taylor rank computation. One could first think of truncating the matrix $M$ to its $\cO(x^n)$ and compute its rank. This does not work, as already mentioned above and exemplified by \eqref{rankExample}. Another tentative characterization consists in evaluating all minors of $M$ truncated at $\cO(x^n)$, and identifying the maximal size of a non-vanishing such truncation of a minor. However, this also fails, as can be seen from
	\bes
	M=\left(\begin{array}{ccc}
		x^6  & 0 & 0 \\
		0 & x^9 & 0 \\
		0 & 0 &0\\
	\end{array}\right)+\order{x^{10}}
	\ees
	for which $r_{10}(M)=2$ but all 2x2 minors vanish at $\order{x^{10}}$.
	
	\section{Refining the MFV expansion}\label{section:taylorMFV}
	In this section, we provide more details for the simple algorithm introduced in Section~\ref{sec:MFV}, which we use to obtain a consistent MFV expansion to plug into our maximal set and perform the Taylor rank reduction described in appendix~\ref{section:taylorrank}. We use again the coefficient $C_{uH}$ as a benchmark and define a shorthand for an MFV-compatible expansion of it:
	\begin{align}
		C_{uH}[a_1b_1a_2b_2\dots a_n b_n]\equiv
		(\rho_{a_1b_1a_2b_2\dots a_n b_n}+i\eta_{a_1b_1a_2b_2\dots a_n b_n})X_{u\vphantom{d}}^{a_{1}}X_d^{b_{1}}\dots X_{u\vphantom{d}}^{a_{n}}X_d^{b_{n}}Y_{u\vphantom{d}}^{\vphantom{b_{i,\,1}}}
		\label{eq:CuHMFVasvector}
	\end{align}
	Since we are working with $3\times 3$ matrices, we can restrict ourselves to $a_i,b_i<3$, as higher powers could be reabsorbed in the traces we will take in the maximal set by a (Yukawa dependent) redefinition of the $\rho$ and $\eta$ coefficients\footnote{Modulo cancellations, since all the entries of $X_{u,d}$ are $\lesssim1$, this will not spoil the assumption that all the $\rho$'s and $\eta$'s are of $\order{1}$.}.
	
	First of all, let us define a procedure to obtain from an integer $n$ a list of vectors we can use to plug into Eq.~\eqref{eq:CuHMFVasvector}. Given such $n$, we compute its partitions and only keep those composed exclusively of integers smaller than 3. Then, we first permute them in all possible ways, and to the obtained set we add the one we get by shifting all values to the right by one place and putting 0 in the first position. For clarity, let us look at $n=4$. Its partitions are
	\begin{align}
		4\to 4,\,31,\,22,\,211,\,1111\ ,
	\end{align}
	of which we only keep 
	\begin{align}
		4\to 22,\,211,\,1111\ .
	\end{align}
	By performing all possible permutations we obtain
	\begin{align}
		4\to 22,\,211,\,121,\,112,\,1111\ ,
	\end{align}
	and with the final step we add
	\begin{align}
		4\to 22,\,211,\,121,\,112,\,1111,\,022,\,0211,\,0121,\,0112,\,01111\ .
		\label{eq:orderingpermutations}
	\end{align}
	For each $n$, the result will be kept in the order shown here, that can be in general obtained by interpreting the lists of numbers as the decimal digits of a rational number with 0 integer value, ordered in decreasing order, e.g.~in Eq.~\eqref{eq:orderingpermutations} $0.22>0.211>0.121>\dots>0.1111$. If we start from $n=0$, this procedure defines an ordering of all the possible acceptable values for the list of integers in Eq.~\eqref{eq:CuHMFVasvector}\footnote{Such choice is arbitrary, but the result does not depend on it anyway.}. We call this ordered list $I$. Thus, we define an MFV-compatible $C_{uH}^i$ as the sum of all $C_{uH}[a_1b_1a_2b_2\dots a_n b_n]$ defined with the first $i$ lists in $I$. For example for $i=6$
	\begin{align}
		C_{uH}^6= C_{uH}[0]+C_{uH}[1]+C_{uH}[01]+C_{uH}[2]+C_{uH}[11]+C_{uH}[02]\ .
	\end{align}
	Now to the final step: for each $i$, we compute the Taylor rank for $C_{uH}^{i}$, obtaining each time a plot like those in Figure~\ref{fig:lambdaranksbilinears}. When the plot does not change for a large enough amount of steps, say 20, we stop. At this point, we should start removing $\rho$'s and $\eta$'s from the first $C_{uH}^i$ for which the plot we stopped at was obtained, and do so in all possible ways that do not alter the plot. At the end, we should obtain the 9 $\rho$'s and 9 $\eta$'s which we should retain, and which are the most relevant to the expansion. Since here we cared mostly about obtaining the Taylor rank in a consistent way, we refrain from explicitly carrying on such procedure, and we content ourselves with the redundant expression for $C_{uH}$. A graphical depiction of how such algorithm works is shown in Figure~\ref{fig:MFValgorithm}.
	


\section{Hilbert series and $\lambda$ scaling for $U(2)^5$ spurions}\label{section:U(2)5}
To account for the fact that the third-generation Yukawa couplings are large, one can modify the $U(3)^5$ flavor symmetry assumption of MFV and reduce it to a $U(2)^5$ that only involves the first two generations \cite{Barbieri:2011ci, Barbieri:2012uh, Blankenburg:2012nx, Faroughy:2020ina}. In particular, given a fermion species $\psi_f$ with $f=l,\, q,\, e,\, u,\, d$, the first two generations form a doublet under the respective $U(2)_f$, while $\psi_f^3$ transforms as a singlet. Following the notation found in the literature \cite{Faroughy:2020ina} we can denote the five $U(2)$ groups as
\begin{align}
	U(2)^5=U(2)_L\otimes U(2)_Q\otimes U(2)_E\otimes U(2)_U\otimes U(2)_D
\end{align}
where for convenience of notation we indicate with $L,\, Q,\, E,\, U,\, D$ the flavor doublets. Then, to reproduce the observed Yukawa couplings, one needs to break the flavor symmetry via spurions with the transformation properties in Table~\ref{tab:spurionstransform}.

It is worthwhile to employ the tools of the Plethystic program and the Hilbert series to analyze what are the possible independent invariants that can come out of this parametrization, and in particular how many independent parameters it contains. However, since it represents a simple case that can be solved by hand, we address the lepton sector first, and use it as a starting point to expose our methods. 
\subsection{Lepton sector}
In the lepton sector, the relevant spurions are $V_l$ and $\Delta_e$. With an appropriate choice of basis, we can parametrize the lepton Yukawa matrix as
\begin{align}
	Y_e=\begin{pmatrix}
		\Delta_e & 0\\
		0& y_\tau
	\end{pmatrix}\ ,
\end{align}
where $y_\tau$ is a $U(2)^5$ singlet. The spurion $V_l$ does not appear in the Yukawa matrix, but it can be used to allow for lepton flavor violation in the SMEFT expansion \cite{Faroughy:2020ina,Greljo:2022cah}. As such, we cannot determine its $\lambda$ scaling by relating it to Yukawa parameters, and we will just set it to 0 in the following. 
Defining $x_e\equiv \Delta_e^{\vphantom{\dagger}}\Delta_e^\dagger$, we can write 
\begin{align}
	X_e^{\vphantom{\dagger}}=Y_e^{\vphantom{\dagger}}Y_e^\dagger=\begin{pmatrix}
		x_e & 0\\
		0 & y_\tau^2
	\end{pmatrix}\ .
\end{align}
There are only two invariants that can be built with $x_e$, namely $\Tr[x_e]$ and $\Tr[x_e^2]$. On the other hand, with $X_e$ we can build
\begin{align}
	\Tr[X_e]&=\Tr[x_e]+y_\tau^2\ ,\\
	\Tr[X_e^2]&=\Tr[x_e^2]+y_\tau^4\ ,\\
	\Tr[X_e^3]&=\Tr[x_e^3]+y_\tau^6\ .
\end{align}
We can use the following relation for $\Tr[x_e^3]$
\begin{align}
	\Tr[x_e^3]=&\frac{1}{2}\left(3\Tr[x_e]\Tr[x_e^2]-\Tr[x_e]^3\right)\ ,
\end{align}
and the following parametrization
\begin{align}
	\Delta_e=
	\begin{pmatrix}
		\delta_l & 0\\
		0& \delta'_l\\
	\end{pmatrix}\ ,
\end{align}
to get, trivially
\begin{align}
	\delta_l=a_e\lambda^9,\quad \delta_l'=a_\mu\lambda^5, \quad y_\tau=a_\tau\lambda^3\ .
\end{align}

\subsection{Quark sector}
Since $\Delta_u$ is the only object charged under $U(2)_U$, it can only appear in the combination $x_u\equiv \Delta_u\Delta_u^\dagger \in \mathbf{2}_Q\otimes\mathbf{\bar{2}}_Q$. Analogously, we can define $x_d\equiv \Delta_d^{\vphantom{\dagger}}\Delta_d^\dagger \in \mathbf{2}_Q\otimes\mathbf{\bar{2}}_Q$. Then, using the properties of $2\times 2$ matrices, as done e.g. in Section V. (A) of \cite{Jenkins:2009dy}, one could build the following CP-even invariants
\begin{align}
	I_{2,0,0}&=V_qV_q^\dagger  & I_{0,1,0}&=\Tr[x_u] & I_{0,0,1}&=\Tr[x_d]\nonumber\\
	I_{0,2,0}&=\Tr[x_u^2]  & I_{0,0,2}&=\Tr[x_d^2] & I_{2,1,0}&=V_qx_uV_q^\dagger\nonumber\\
	I_{2,0,1}&=V_qx_dV_q^\dagger& I_{0,1,1}&=\Tr[x_ux_d] & I_{2,1,1}^{(+)}&=V_qx_ux_dV_q^\dagger+V_qx_dx_uV_q^\dagger\ ,
\end{align}
and the single CP-odd invariant 
\begin{align}
	I_{2,1,1}^{(-)}&=V_qx_ux_dV_q^\dagger-V_qx_dx_uV_q^\dagger\ .
\end{align}
At the same time, one can build the relevant multi-graded Hilbert series:
\begin{align}
	h(v_q,v_q^\dagger, x_u,x_d)=\int \left[\dd \mu\right]_{U(2)_Q}\prod_{i=\left\{v_q,v_q^\dagger,x_u,x_d\right\}}\text{PE}(\vec{z};v_q,v_q^\dagger, x_u,x_d)\ ,
\end{align}
with obvious association between a spurion and the corresponding building block.
The resulting expression is
\begin{multline}
	h(v_q,v_q^\dagger, x_u,x_d)=\\
	=\frac{1+v_q x_d x_u v_q^{\dagger }}{(1-x_u)(1-x_u^2)(1-x_d)(1-x_d^2) (1-x_d x_u) (1-v_q v_q^{\dagger}) (1-v_q x_d v_q^{\dagger }) (1-v_q x_u v_q^{\dagger })}
	\label{eq:multigradedHSresult}
\end{multline}
whence it is straightforward to count that there are $8$ independent parameters, corresponding to as many algebraically independent invariants. 
These are
\begin{align}
\{I_{2,0,0},\, I_{0,1,0},\, I_{0,0,1},\, I_{0,2,0},\, I_{0,0,2},\, I_{2,1,0},\, I_{2,0,1},\, I_{0,1,1}\}\ .
\end{align}
Indeed, $I_{2,1,1}^{(+)}$ is not independent, as it can be decomposed as
\begin{align}
	I_{2,1,1}^{(+)}=I_{210}I_{001}+I_{201}I_{010}+I_{200}\left(I_{011}-I_{010}I_{001}\right)\ .
\end{align}
Computing the Plethystic Logarithm helps us instead in understanding the role of $I_{2,1,1}^{(-)}$:
\begin{multline}
	PL[h(v_q,v_q^\dagger, x_u,x_d)]=x_u+x_d+x_u^2+x_d^2+v_q v_q^{\dagger }+v_q x_u v_q^{\dagger }+v_q x_d v_q^{\dagger }+v_q x_d x_u v_q^{\dagger}
	+\\
	-v_q^2 x_d^2 x_u^2 (v_q^{\dagger})^2\ .
\end{multline}
The presence of a term with a negative sign $-v_q^2 x_d^2 x_u^2 (v_q^{\dagger})^2$ signals the presence of a syzygy at that order:
\begin{align}
	\left(I_{2,1,1}^{(-)}\right)^2=&+I_{0,0,2}^{\vphantom{2}}I_{0,1,0}^2 I_{2,0,0}^2+I_{0,0,1}^2 I_{0,2,0}^{\vphantom{2}} I_{2,0,0}^2-I_{0,0,1}^2 I_{2,1,0}^2-2 I_{0,0,1}^{\vphantom{2}} I_{0,1,0}^{\vphantom{2}} I_{0,1,1}^{\vphantom{2}} I_{2,0,0}^2\nonumber\\
	&+2 I_{0,0,1}^{\vphantom{2}} I_{0,1,0}^{\vphantom{2}} I_{2,0,1}^{\vphantom{2}} I_{2,1,0}^{\vphantom{2}}+2 I_{0,0,1}^{\vphantom{2}} I_{0,1,1}^{\vphantom{2}} I_{2,0,0}^{\vphantom{2}} I_{2,1,0}^{\vphantom{2}}-2 I_{0,0,1}^{\vphantom{2}} I_{0,2,0}^{\vphantom{2}} I_{2,0,0}^{\vphantom{2}}I_{2,0,1}^{\vphantom{2}}\nonumber\\
	&-2 I_{0,0,2}^{\vphantom{2}} I_{0,1,0}^{\vphantom{2}} I_{2,0,0}^{\vphantom{2}} I_{2,1,0}^{\vphantom{2}}-I_{0,0,2}^{\vphantom{2}} I_{0,2,0}^{\vphantom{2}} I_{2,0,0}^2+2 I_{0,0,2}^{\vphantom{2}} I_{2,1,0}^2-I_{0,1,0}^2 I_{2,0,1}^2\nonumber\\
	&+2 I_{0,1,0}^{\vphantom{2}} I_{0,1,1}^{\vphantom{2}} I_{2,0,0}^{\vphantom{2}} I_{2,0,1}^{\vphantom{2}}+I_{0,1,1}^2 I_{2,0,0}^2-4 I_{0,1,1}^{\vphantom{2}} I_{2,0,1}^{\vphantom{2}} I_{2,1,0}^{\vphantom{2}}+2 I_{0,2,0}^{\vphantom{2}} I_{2,0,1}^2\ .
\end{align}
This means that only the sign carried by $I_{2,1,1}^{(-)}$ is relevant. The 8 independent parameters correspond to 4 masses, 3 angles and 1 phase contained in the SM Yukawa matrices. The remaining two parameters, i.e. the top and bottom quark masses, are captured by two $U(2)^5$ singlets \footnote{\label{foot:LinVsAlgIndependent}As we mentioned in the main text at the end of section~\ref{section:U(2)5maxset}, the rank of the transfer matrix is sensitive to the different linearly independent invariants. These are not the same as those we found using the multi-graded Hilbert series of Eq.~\eqref{eq:multigradedHSresult}, which are the algebraically independent ones, i.e. up to all possible polynomial relations. On the other hand, the linearly independent invariants would be the ones corresponding to the expression given by expanding Eq.~\eqref{eq:multigradedHSresult} at a fixed order in $v_{q}^{\vphantom{\dagger}},v_q^\dagger,x_{u}, x_{d}$.}. 
Using the parametrization \cite{Greljo:2022cah}:
\begin{align}
	Y_u=\begin{pmatrix}
		\Delta_u & V_q\\
		0 & y_t
	\end{pmatrix}
\qquad
Y_d=\begin{pmatrix}
	\Delta_d & 0\\
	0 & y_b
\end{pmatrix}
\end{align}
with $y_{t,b}$ being $U(2)^3$ singlets, we can write 
\begin{align}
	X_u=Y_uY_u^\dagger=\begin{pmatrix}
		x_u + V_q\otimes V_q^\dagger & y_t V_q\\
		y_t V_q^\dagger& y_t^2
	\end{pmatrix} \qquad
X_d=Y_dY_d^\dagger=\begin{pmatrix}
	x_d  &0 \\
	0& y_b^2
\end{pmatrix} \ .
\label{eq:xuxdU25}
\end{align}
This allows us to rewrite the 10 independent invariants one can build with the Yukawas \cite{Jenkins:2009dy} as:
\begin{align}
\Tr[X_u]&=\Tr[x_u]+V_q^\dagger V_q+y_t^2 \label{eq:u3tou2first}\\
\Tr[X_d]&=\Tr[x_d]+y_b^2\\
\Tr[X_u^2]&=\Tr[x_u^2]+2V_q^\dagger x_u V_q+ \left(V_q^\dagger V_q+y_t^2\right)^2\\
\Tr[X_d^2]&=\Tr[x_d^2]+y_b^4\\
\Tr[X_uX_d]&=\Tr[x_ux_d]+V_q^\dagger x_d V_q+y_t^2y_b^2\\
\Tr[X_u^3]&=\Tr[x_u^3]+3V_q^\dagger x_u^2 V_q+3 \left(V_q^\dagger V_q +y_t^2)(V_q^\dagger  x_u V_q\right)+\left(V_q^\dagger V_q +y_t^2\right)^3\\
\Tr[X_d^3]&=\Tr[x_d^3]+y_b^6\\
\Tr[X_uX_d^2]&=\Tr[x_ux_d^2]+V_q^\dagger x_d^2V_q +y_t^2y_b^4\\
\Tr[X_dX_u^2]&=\Tr[x_dx_u^2]+\left(V_q^\dagger x_d V_q\right)\left(V_q^\dagger V_q+ y_t^2\right)+\left(V_q^\dagger V_q +y_t^2 \right)y_t^2 y_b^2\nonumber\\
&+V_q^\dagger \left( x_ux_d+x_dx_u \right)V_q\\
\Tr[X_u^2X_d^2]&=\Tr[x_u^2x_d^2]+V_q^\dagger x_d^2x_uV_q +V_q^\dagger x_ux_d^2V_q +V_q^\dagger x_d^2 V_q \left(V_q^\dagger V_q +y_t^2\right)\nonumber\\
&+ y_t^2y_b^4 \left(V_q^\dagger V_q +y_t^2\right)
\label{eq:u3tou2last}
\end{align}
we see that we need the explicit expression for the additional invariants that appear but we know are not algebraically independent:
\begin{align}
V_q^\dagger \left(x_u^2x_d+x_dx_u^2\right)V_q=&V_q^\dagger x_u V_q \Tr[x_d]\Tr[x_u]+V_q^\dagger x_d V_q \Tr[x_u^2]+\nonumber\\
	&+V_q^\dagger V_q \left(\Tr[x_u]\Tr[x_ux_d]-\Tr[x_d]\Tr[x_u]^2\right)\\
V_q^\dagger \left(x_d^2x_u+x_ux_d^2\right)V_q=&V_q^\dagger x_d V_q \Tr[x_u]\Tr[x_d]+V_q^\dagger x_u V_q \Tr[x_d^2]+\nonumber\\
	&+V_q^\dagger V_q \left((\Tr[x_d]\Tr[x_dx_u]-\Tr[x_u]\Tr[x_d]^2\right)\\	
V_q^\dagger \left(x_ux_d+x_dx_u \right)V_q=&V_q^{\dagger }x_uV_q \Tr[x_d]+V_q^{\dagger }x_dV_q \Tr[x_u]+\nonumber\\
	&+V_q^{\dagger }V_q \left(\Tr[x_ux_d]-\Tr[x_d] \Tr[x_u]\right)\\
V_q^\dagger x_u^2 V_q=&V_q^\dagger x_u V_q \Tr[x_u]+\frac{1}{2}\left(\Tr[x_u^2]-\Tr[x_u]^2\right)\\
	V_q^\dagger x_d^2 V_q=&V_q^\dagger x_d V_q \Tr[x_d]+\frac{1}{2}\left(\Tr[x_d^2]-\Tr[x_d]^2\right)\\
	\Tr[x_u^3]=&\frac{1}{2}\left(3\Tr[x_u]\Tr[x_u^2]-\Tr[x_u]^3\right)\\
	\Tr[x_d^3] =&\frac{1}{2}\left(3\Tr[x_d]\Tr[x_d^2]-\Tr[x_d]^3\right)\\
	\Tr[x_ux_d^2]=&\frac{1}{2}\left(\Tr[x_u]\Tr[x_d^2]+2\Tr[x_d]\Tr[x_ux_d]-\Tr[x_u]\Tr[x_d]^2\right)\\
	\Tr[x_dx_u^2]=&\frac{1}{2}\left(\Tr[x_d]\Tr[x_u^2]+2\Tr[x_u]\Tr[x_dx_u]-\Tr[x_d]\Tr[x_u]^2\right)\\
	\Tr[x_u^2x_d^2]=&\frac{1}{2}\left(\Tr[x_d^2]\Tr[x_u^2]+2\Tr[x_u]\Tr[x_d]\Tr[x_ux_d]-\Tr[x_d]^2\Tr[x_u]^2\right)\ .
\end{align}
Equations~\eqref{eq:u3tou2first}-\eqref{eq:u3tou2last} constitute a system of ten equations in the ten variables represented by $y_t^2$, $y_b^2$ and the eight independent invariants. This system can be solved to all orders in $\lambda$, the leading order solutions being
\begin{align}
	y_t^2&=a_t^2 \label{eq:yt}\\
	y_b^2&=a_b^2 \lambda ^6 \label{eq:yb}\\
	V_q^\dagger V_q&=A^2 a_t^2 \lambda ^4 +\order{\lambda^5}\\
	\Tr[x_d]&=a_s^2 \lambda ^{10}+\order{\lambda^{11}}\\
	\Tr[x_u]&=a_c^2 \lambda ^8+\order{\lambda^9}\\
	\Tr[x_d^2]&=a_s^4 \lambda ^{20}+\order{\lambda^{21}}\\
	\Tr[x_u^2]&=a_c^4 \lambda ^{16}+\order{\lambda^{17}}\\
	V_q^\dagger x_u V_q&= A^2 a_c^2 a_t^2 \lambda ^{12}+\order{\lambda^{13}}\\
	V_q^\dagger x_d V_q&=A^2 a_s^2 a_t^2 \lambda ^{14}+\order{\lambda^{15}}\\
	\Tr[x_u x_d ]&=a_c^2 a_s^2 \lambda ^{18}+\order{\lambda^{19}}\ .
\end{align}
Assuming then that $\Delta_{u,d}$ and $V_q$ are the only spurions breaking the flavor symmetry, we can express each coefficient at dimension-6 as a suitable combination of them. This means, in turn, that all the invariants we build for each operator can be recast as combinations of the algebraically independent ones. Then, using the expression we just obtained, we can obtain a consistent expansion in $\lambda$ for our invariants. 
It is actually more practical to use the $U(2)^3$ rotations to pick a basis for the spurions. In particular, we can choose \cite{Greljo:2022cah}:
\begin{align}
	V_q=\begin{pmatrix}
		0\\
		\epsilon_q
	\end{pmatrix}
	\qquad
	\Delta_u=\begin{pmatrix}
		c_u & -s_u\\
		s_u & c_u\\
	\end{pmatrix}
	\begin{pmatrix}
		\delta_u & 0\\
		0& \delta'_u\\
	\end{pmatrix}\qquad
	\Delta_u=\begin{pmatrix}
		c_d & -s_d e^{i\alpha}\\
		s_d e^{-i\alpha}& c_d\\
	\end{pmatrix}
	\begin{pmatrix}
		\delta_d & 0\\
		0& \delta'_d\\
	\end{pmatrix}\ .
\end{align}
We see that, as expected, after exhausting all $U(2)^3$ rotations we are left with 8 independent objects, i.e. 5 real positive parameters, 2 angles and 1 phase. Inserting this parametrization in the system represented by Eqs.~\eqref{eq:u3tou2first}-\eqref{eq:u3tou2last}, together with the solutions in Eqs~\eqref{eq:yt}-\eqref{eq:yb} we can obtain an expression for these parameters. At leading order in $\lambda$, we found
\begin{align}
	\epsilon_q&=A^2 a_t^2 \lambda ^4\\
	s_u&=\sqrt{\eta ^2+\rho ^2} \lambda\\
	s_d&=\lambda \sqrt{\eta ^2+(\rho -1)^2}\\
	\delta_d&=a_d \lambda^{7}\\
	\delta'_d&=a_s \lambda^{5}\\
	\delta_u&=a_u \lambda ^{8}\\
	\delta'_u	&=a_c\lambda ^4\\
	\cos(\alpha)&=\frac{\eta ^2+(\rho -1) \rho }{\sqrt{\eta ^2+\rho ^2} \sqrt{\eta ^2+\rho ^2-2 \rho +1}}\ .
\end{align}
Notice, in passing, that the fact that we can only determine $\cos(\alpha)$ is linked to the impossibility of determining the sign of $\alpha$ with the invariants in Eqs.~\eqref{eq:u3tou2first}-\eqref{eq:u3tou2last}. To find it, we have to look at the relation obtained by expressing $\Tr[X_u^2X_d^2X_uX_d]-\Tr[X_d^2X_u^2X_dX_u]$ in terms of Eq.~\eqref{eq:xuxdU25}. This equation, too, can be solved to all orders in $\lambda$, the leading order being
\begin{align}
	\sin(\alpha)=-\frac{\eta }{\sqrt{\eta ^2+(\rho -1)^2} \sqrt{\eta ^2+\rho ^2}}\ .
\end{align}
As a check, one can verify that, at this order $\cos^2(\alpha)+\sin^2(\alpha)=1$.

\section{Bilinear operators}
	\label{section:bilinearoperators}
We present here explicitly the maximal set of invariants for each bilinear operator in SMEFT at $\order{{1}/{\Lambda^2}}$.
	\begin{table}[H]
		\centering
		\textbf{Maximal set for all bilinear operators}\\\medskip
		\renewcommand{\arraystretch}{1.2}
		\resizebox{\columnwidth}{!}{%
			\begin{tabular}{l|c|c|c}
				Wilson coefficient& \begin{minipage}{0.07\columnwidth}\centering
					\vspace*{0.1cm}
					\# min set\vspace*{0.1cm}
				\end{minipage}& \begin{minipage}{0.07\columnwidth}\centering
				\vspace*{0.1cm}
				\# max set\vspace*{0.1cm}
			\end{minipage}&Maximal set\\
				\hline
				\begin{minipage}{0.2\textwidth}
					\vspace{0.2cm}
					$C_e\equiv\left\{\begin{matrix}\smash{C_{eH}}\\C_{eW}\\C_{eB}\end{matrix}\right.$
					\vspace{0.2cm}
				\end{minipage}&3&3&
				$\left\{ \ \begin{matrix}
					\colorbox{newlightgrey}{$L_0\(C_eY_e^\dagger\)$}&
					\colorbox{newlightgrey}{$L_1\(C_eY_e^\dagger\)$}&
					\colorbox{newlightgrey}{$L_2\(C_eY_e^\dagger\)$}
				\end{matrix} \ \right\}$\\
				\hline
				\begin{minipage}{0.2\textwidth}
					\vspace{0.2cm}
					$C_u\equiv\left\{\begin{matrix}C_{uH}\\C_{uG}\\C_{uW}\\C_{uB}\end{matrix}\right.$
					\vspace{0.2cm}
				\end{minipage}
				&\multirow{6}{*}{9}&\multirow{6}{*}{18}&			
				$\left\{ \ 
				\begin{array}{cccccc}
					\colorbox{newlightgrey}{$L_{0000}\left(C_uY_u^{\dagger }\right)$} & \colorbox{newlightgrey}{$L_{1000}\left(C_uY_u^{\dagger }\right) $}& L_{2000}\left(C_uY_u^{\dagger }\right) \\\colorbox{newlightgrey}{$ L_{0100}\left(C_uY_u^{\dagger }\right)$} & \colorbox{newlightgrey}{$L_{1100}\left(C_uY_u^{\dagger }\right) $}& \colorbox{newlightgrey}{$L_{0110}\left(C_uY_u^{\dagger }\right) $}\\
					L_{2100}\left(C_uY_u^{\dagger }\right) & L_{0120}\left(C_uY_u^{\dagger }\right) & L_{1120}\left(C_uY_u^{\dagger }\right) \\ L_{0200}\left(C_uY_u^{\dagger }\right) & L_{1200}\left(C_uY_u^{\dagger }\right) & L_{0210}\left(C_uY_u^{\dagger }\right) \\
					\colorbox{newlightgrey}{$L_{2200}\left(C_uY_u^{\dagger }\right) $}& \colorbox{newlightgrey}{$L_{0220}\left(C_uY_u^{\dagger }\right) $}& \colorbox{newlightgrey}{$L_{1220}\left(C_uY_u^{\dagger }\right) $} \\ L_{0112}\left(C_uY_u^{\dagger }\right) & \colorbox{newlightgrey}{$L_{0122}\left(C_uY_u^{\dagger }\right) $}& L_{1122}\left(C_uY_u^{\dagger }\right) \\
				\end{array}
				\ \right\}$\\
				\begin{minipage}{0.2\textwidth}
					\vspace{0.2cm}
					$C_d\equiv\left\{\begin{matrix}C_{dH}\\C_{dG}\\C_{dW}\\C_{dB}\end{matrix}\right.$
					\vspace{0.2cm}
				\end{minipage}
				&&&
				Same with $C_{u} Y_u^\dagger \rightarrow C_{d}^{\mathstrut} Y_d^\dagger$\\
				$C_{Hud}$
				&&&
				Same with $C_{u} Y_u^\dagger \rightarrow Y_u^{\mathstrut}C_{Hud}^{\mathstrut} Y_d^\dagger$\\
				\hline
				$C_{HL}^{(1,3)},C_{He}^{\vphantom{(1,3)}}$
				&0&0&
				$\emptyset$\\
				\hline
				$C_{HQ}^{(1,3)}$
				&\multirow{3}{*}{3}&\multirow{3}{*}{9}&
				$\left\{\ 
				\begin{array}{ccc}
					\colorbox{newlightgrey}{$L_{1100}\left(C_uY_u^{\dagger }\right)$} & L_{2100}\left(C_uY_u^{\dagger }\right) & L_{1200}\left(C_uY_u^{\dagger }\right) \\
					\colorbox{newlightgrey}{$L_{2200}\left(C_uY_u^{\dagger }\right) $}& L_{1120}\left(C_uY_u^{\dagger }\right) & \colorbox{newlightgrey}{$L_{1122}\left(C_uY_u^{\dagger }\right) $}\\
					L_{1221}\left(C_uY_u^{\dagger }\right) & L_{2112}\left(C_uY_u^{\dagger }\right) & L_{0112}\left(C_uY_u^{\dagger }\right) \\
				\end{array}
				\ \right\}$\\
				$C_{Hu}$
				&&&
				Same with $C_{HQ}^{(1,3)} \to Y_u^{\mathstrut} C_{Hu}^{\mathstrut} Y_u^\dagger$\\
				$C_{Hd}$
				&&&
				Same with $C_{HQ}^{(1,3)} \to Y_d^{\mathstrut} C_{Hd}^{\mathstrut} Y_d^\dagger$\\
			\end{tabular}
		}
		\caption{Maximal sets of CP-odd flavor invariants for all SMEFT dimension-six Wilson coefficients associated to operators bilinear in fermion fields. The invariants belonging to the minimal sets of Ref.~\cite{Bonnefoy:2021tbt} are highlighted in gray. We recall that $X_u\equiv Y_uY_u^\dagger$, and similarly for down quarks or electrons. We also recall the definitions in Eqs.~\eqref{bilinearFormulaquark}-\eqref{bilinearFormulaleptons}.}
		\label{tableInvBilinears}
	\end{table}
	
	\section{4-Fermi operators}
	We present here explicitly the maximal set of invariants for each 4-Fermi operator in SMEFT at $\order{{1}/{\Lambda^2}}$.
	\label{section:4fermioperators}
	\begin{table}[h!]
		\hspace*{-1cm}
		\centering
		\textbf{Maximal set for all 4-Fermi operators}\\\medskip
		\small
		\renewcommand{\arraystretch}{1.01}
		\resizebox{0.88\columnwidth}{!}{%
			\begin{tabular}{l|c|c|c}
				Wilson coefficient& \begin{minipage}{0.1\columnwidth}\centering
					\vspace*{0.1cm}
					\# minimal set\vspace*{0.1cm}
				\end{minipage}& \begin{minipage}{0.1\columnwidth}\centering
					\vspace*{0.1cm}
					\# maximal set\vspace*{0.1cm}
				\end{minipage}&Maximal set\\
				\hline
				$C_{Qe}$&\multirow{21}{0.03\textwidth}{\centering 9}&\multirow{21}{0.03\textwidth}{\centering 27}&
				$\left \{ \ 
				\begin{matrix}
					\colorbox{newlightgrey}{$A_{0000}^{1100}\(C_{QQee} \)$}&A_{0000}^{2100}\(C_{QQee} \)&\colorbox{newlightgrey}{$A_1^{1100}\(C_{QQee} \)$}\\
					A_{0000}^{1200}\(C_{QQee} \)&A_1^{2100}\(C_{QQee} \)&\colorbox{newlightgrey}{$A_{0000}^{2200}\(C_{QQee} \)$}\\
					A_{0000}^{1120}\(C_{QQee} \)&\colorbox{newlightgrey}{$A_2^{1100}\(C_{QQee} \)$}&A_1^{1200}\(C_{QQee} \)\\
					A_2^{2100}\(C_{QQee} \)&\colorbox{newlightgrey}{$A_1^{2200}\(C_{QQee} \)$}&\colorbox{newlightgrey}{$A_{0000}^{1122}\(C_{QQee} \)$}\\
					A_{0000}^{1221}\(C_{QQee} \)&A_{0000}^{2112}\(C_{QQee} \)&A_1^{1120}\(C_{QQee} \)\\
					A_{0000}^{0112}\(C_{QQee} \)&A_2^{1200}\(C_{QQee} \)&\colorbox{newlightgrey}{$A_2^{2200}\(C_{QQee} \)$}\\
					\colorbox{newlightgrey}{$A_1^{1122}\(C_{QQee} \)$}&A_1^{1221}\(C_{QQee} \)&A_1^{2112}\(C_{QQee} \)\\
					A_2^{1120}\(C_{QQee} \)&A_1^{0112}\(C_{QQee} \)&\colorbox{newlightgrey}{$A_2^{1122}\(C_{QQee} \)$}\\
					A_2^{1221}\(C_{QQee} \)&A_2^{2112}\(C_{QQee} \)&A_2^{0112}\(C_{QQee} \)
				\end{matrix}
				\ \right \}$\\
				&&&\\
				\multirow{2}{0.1\textwidth}{$C_{ed}$}
				&&&
				\multirow{2}{0.5\textwidth}{Same with $C_{QQee} \to C_{\smash{ee\tilde d\tilde d}}$ (exchanging upper with lower indices and with $Y_e^{\vphantom{\dagger}}\leftrightarrow Y_e^\dagger$)}\\
				&&&\\
				&&&\\
				\multirow{2}{0.1\textwidth}{$C_{eu}$}
				&&&
				\multirow{2}{0.5\textwidth}{Same with $C_{QQee} \to C_{ee\tilde u\tilde u}$ (exchanging upper with lower indices and with $Y_e^{\vphantom{\dagger}}\leftrightarrow Y_e^\dagger$)}\\
				&&&\\
				&&&\\
				$C_{LQ}^{(1,3)}$&&&$\left \{ \ 
				\begin{matrix}
					\colorbox{newlightgrey}{$A_{1100}^{0000}\(C_{LLQQ} \)$}&A_{2100}^{0000}\(C_{LLQQ} \)&A_{1200}^{0000}\(C_{LLQQ} \)\\
						\colorbox{newlightgrey}{$A_{1100}^1\(C_{LLQQ} \)$}&\colorbox{newlightgrey}{$A_{2200}^{0000}\(C_{LLQQ} \)$}&A_{2100}^1\(C_{LLQQ} \)\\
						A_{1120}^{0000}\(C_{LLQQ} \)&A_{1200}^1\(C_{LLQQ} \)&\colorbox{newlightgrey}{$A_{1100}^2\(C_{LLQQ} \)$}\\
						\colorbox{newlightgrey}{$A_{2200}^1\(C_{LLQQ} \)$}&A_{2100}^2\(C_{LLQQ} \)&\colorbox{newlightgrey}{$A_{1122}^{0000}\(C_{LLQQ} \)$}\\
						A_{1221}^{0000}\(C_{LLQQ} \)&A_{2112}^{0000}\(C_{LLQQ} \)&A_{1120}^1\(C_{LLQQ} \)\\
						A_{0112}^{0000}\(C_{LLQQ} \)&A_{1200}^2\(C_{LLQQ} \)&\colorbox{newlightgrey}{$A_{2200}^2\(C_{LLQQ} \)$}\\
						\colorbox{newlightgrey}{$A_{1122}^1\(C_{LLQQ} \)$}&A_{1221}^1\(C_{LLQQ} \)&A_{2112}^1\(C_{LLQQ} \)\\
						A_{1120}^2\(C_{LLQQ} \)&A_{0112}^1\(C_{LLQQ} \)&\colorbox{newlightgrey}{$A_{1122}^2\(C_{LLQQ} \)$}\\
						A_{1221}^2\(C_{LLQQ} \)&A_{2112}^2\(C_{LLQQ} \)&A_{0112}^2\(C_{LLQQ} \)
				\end{matrix}
				\ \right \}$\\
				&&&\\
				$C_{Ld}$
				&&&
				Same with $C_{LQ}^{(1,3)} \to C_{LL\tilde d\tilde d}$\\
				&&&\\
				$C_{Lu}$
				&&&
				Same with $C_{LQ}^{(1,3)} \to C_{LL\tilde u\tilde u}$\\\hline
			\end{tabular}%
		}
		\caption{Maximal sets of CP-odd flavor invariants for all SMEFT dimension-six Wilson coefficients associated to 4-Fermi operators. The invariants belonging to the minimal sets of Ref.~\cite{Bonnefoy:2021tbt} are highlighted in gray. We recall that $X_u\equiv Y_uY_u^\dagger$, and similarly for down quarks or electrons. We also recall the definitions of Eqs.~\eqref{eq:L4Fermi}-\eqref{eq:tildeCoeffs}.}
		\label{tableInv4Fermi1}
	\end{table}
	\begin{table}[h!]
		\hspace*{-1cm}
		\centering
		\small
		\renewcommand{\arraystretch}{1.01}
		\resizebox{0.88\columnwidth}{!}{%
			\begin{tabular}{l|c|c|c}
			Wilson coefficient& \begin{minipage}{0.06\columnwidth}\centering
				\vspace*{0.1cm}
				\# min set\vspace*{0.1cm}
			\end{minipage}& \begin{minipage}{0.06\columnwidth}\centering
				\vspace*{0.1cm}
				\# max set\vspace*{0.1cm}
			\end{minipage}&Maximal set\\
				\hline
				$C_{LeQu}$&\multirow{3}{0.03\textwidth}{\centering 27}&\multirow{3}{0.03\textwidth}{\centering 54}&$\left \{ \ 
				\begin{matrix} 
					\colorbox{newlightgrey}{$A_{0000}^{0000}\(C_{L\tilde{e}Q\tilde{u}} \)$}&\colorbox{newlightgrey}{$A_{1000}^{0000}\(C_{L\tilde{e}Q\tilde{u}} \)$}&A_{2000}^{0000}\(C_{L\tilde{e}Q\tilde{u}} \)\\
					\colorbox{newlightgrey}{$A_{0100}^{0000}\(C_{L\tilde{e}Q\tilde{u}} \)$}&\colorbox{newlightgrey}{$A_{0000}^1\(C_{L\tilde{e}Q\tilde{u}} \)$}&\colorbox{newlightgrey}{$A_{1100}^{0000}\(C_{L\tilde{e}Q\tilde{u}} \)$}\\
					\colorbox{newlightgrey}{$A_{0110}^{0000}\(C_{L\tilde{e}Q\tilde{u}} \)$}&\colorbox{newlightgrey}{$A_{1000}^1\(C_{L\tilde{e}Q\tilde{u}} \)$}&A_{2100}^{0000}\(C_{L\tilde{e}Q\tilde{u}} \)\\
					A_{0120}^{0000}\(C_{L\tilde{e}Q\tilde{u}} \)&A_{2000}^1\(C_{L\tilde{e}Q\tilde{u}} \)&A_{1120}^{0000}\(C_{L\tilde{e}Q\tilde{u}} \)\\
					A_{0200}^{0000}\(C_{L\tilde{e}Q\tilde{u}} \)&\colorbox{newlightgrey}{$A_{0100}^1\(C_{L\tilde{e}Q\tilde{u}} \)$}&\colorbox{newlightgrey}{$A_{0000}^2\(C_{L\tilde{e}Q\tilde{u}} \)$}\\
					A_{1200}^{0000}\(C_{L\tilde{e}Q\tilde{u}} \)&A_{0210}^{0000}\(C_{L\tilde{e}Q\tilde{u}} \)&\colorbox{newlightgrey}{$A_{1100}^1\(C_{L\tilde{e}Q\tilde{u}} \)$}\\
					\colorbox{newlightgrey}{$A_{0110}^1\(C_{L\tilde{e}Q\tilde{u}} \)$}&\colorbox{newlightgrey}{$A_{1000}^2\(C_{L\tilde{e}Q\tilde{u}} \)$}&\colorbox{newlightgrey}{$A_{2200}^{0000}\(C_{L\tilde{e}Q\tilde{u}} \)$}\\
					\colorbox{newlightgrey}{$A_{0220}^{0000}\(C_{L\tilde{e}Q\tilde{u}} \)$}&A_{2100}^1\(C_{L\tilde{e}Q\tilde{u}} \)&A_{0120}^1\(C_{L\tilde{e}Q\tilde{u}} \)\\
					A_{2000}^2\(C_{L\tilde{e}Q\tilde{u}} \)&\colorbox{newlightgrey}{$A_{1220}^{0000}\(C_{L\tilde{e}Q\tilde{u}} \)$}&A_{1120}^1\(C_{L\tilde{e}Q\tilde{u}} \)\\
					A_{0200}^1\(C_{L\tilde{e}Q\tilde{u}} \)&\colorbox{newlightgrey}{$A_{0100}^2\(C_{L\tilde{e}Q\tilde{u}} \)$}&A_{0112}^{0000}\(C_{L\tilde{e}Q\tilde{u}} \)\\
					A_{1200}^1\(C_{L\tilde{e}Q\tilde{u}} \)&A_{0210}^1\(C_{L\tilde{e}Q\tilde{u}} \)&\colorbox{newlightgrey}{$A_{1100}^2\(C_{L\tilde{e}Q\tilde{u}} \)$}\\
					\colorbox{newlightgrey}{$A_{0110}^2\(C_{L\tilde{e}Q\tilde{u}} \)$}&\colorbox{newlightgrey}{$A_{0122}^{0000}\(C_{L\tilde{e}Q\tilde{u}} \)$}&\colorbox{newlightgrey}{$A_{2200}^1\(C_{L\tilde{e}Q\tilde{u}} \)$}\\
					\colorbox{newlightgrey}{$A_{0220}^1\(C_{L\tilde{e}Q\tilde{u}} \)$}&A_{2100}^2\(C_{L\tilde{e}Q\tilde{u}} \)&A_{0120}^2\(C_{L\tilde{e}Q\tilde{u}} \)\\
					A_{1122}^{0000}\(C_{L\tilde{e}Q\tilde{u}} \)&\colorbox{newlightgrey}{$A_{1220}^1\(C_{L\tilde{e}Q\tilde{u}} \)$}&A_{1120}^2\(C_{L\tilde{e}Q\tilde{u}} \)\\
					A_{0200}^2\(C_{L\tilde{e}Q\tilde{u}} \)&A_{0112}^1\(C_{L\tilde{e}Q\tilde{u}} \)&A_{1200}^2\(C_{L\tilde{e}Q\tilde{u}} \)\\
					A_{0210}^2\(C_{L\tilde{e}Q\tilde{u}} \)&\colorbox{newlightgrey}{$A_{0122}^1\(C_{L\tilde{e}Q\tilde{u}} \)$}&\colorbox{newlightgrey}{$A_{2200}^2\(C_{L\tilde{e}Q\tilde{u}} \)$}\\
					\colorbox{newlightgrey}{$A_{0220}^2\(C_{L\tilde{e}Q\tilde{u}} \)$}&A_{1122}^1\(C_{L\tilde{e}Q\tilde{u}} \)&\colorbox{newlightgrey}{$A_{1220}^2\(C_{L\tilde{e}Q\tilde{u}} \)$}\\A_{0112}^2\(C_{L\tilde{e}Q\tilde{u}} \)&\colorbox{newlightgrey}{$A_{0122}^2\(C_{L\tilde{e}Q\tilde{u}} \)$}&A_{1122}^2\(C_{L\tilde{e}Q\tilde{u}} \)
				\end{matrix}
				\ \right \}$\\
				&&&\\
				$C_{LedQ}$
				&&&
				Same with $C_{L\tilde eQ\tilde u} \to C_{L\tilde e\tilde dQ}$ and $A^a_{bcde}\to A^a_{edcb}$\\\hline
			\end{tabular}%
		}
		\caption{Continuation of Table \ref{tableInv4Fermi1}}
		\label{tableInv4Fermi2}
	\end{table}
	\begin{table}[h!]
		\hspace*{-1cm}
		\centering
		\small
		\renewcommand{\arraystretch}{1.01}
		\resizebox{0.85\columnwidth}{!}{%
			\begin{tabular}{l|c|c|c}
				Wilson coefficient& \begin{minipage}{0.06\columnwidth}\centering
					\vspace*{0.1cm}
					\# min set\vspace*{0.1cm}
				\end{minipage}& \begin{minipage}{0.06\columnwidth}\centering
					\vspace*{0.1cm}
					\# max set\vspace*{0.1cm}
				\end{minipage}&Maximal set\\
				\hline
				$C_{QQ}$&18&45&$\left \{ \ 
				\begin{matrix}
					\colorbox{newlightgrey}{$A_{1100}^{0000}\(C_{QQQQ}\)$}&\colorbox{newlightgrey}{$B_{1100}^{0000}\(C_{QQQQ}\)$}&A_{2100}^{0000}\(C_{QQQQ}\)\\
					B_{2100}^{0000}\(C_{QQQQ}\)&\colorbox{newlightgrey}{$A_{1100}^{1000}\(C_{QQQQ}\)$}&A_{2100}^{1000}\(C_{QQQQ}\)\\
					B_{2100}^{1000}\(C_{QQQQ}\)&A_{2000}^{1100}\(C_{QQQQ}\)&A_{2100}^{2000}\(C_{QQQQ}\)\\
					A_{1200}^{0000}\(C_{QQQQ}\)&B_{1200}^{0000}\(C_{QQQQ}\)&\colorbox{newlightgrey}{$A_{1100}^{0100}\(C_{QQQQ}\)$}\\
					\colorbox{newlightgrey}{$A_{2200}^{0000}\(C_{QQQQ}\)$}&\colorbox{newlightgrey}{$B_{2200}^{0000}\(C_{QQQQ}\)$}&A_{2100}^{0100}\(C_{QQQQ}\)\\
					A_{1200}^{1000}\(C_{QQQQ}\)&\colorbox{newlightgrey}{$A_{1100}^{1100}\(C_{QQQQ}\)$}&\colorbox{newlightgrey}{$A_{2200}^{1000}\(C_{QQQQ}\)$}\\
					B_{2200}^{1000}\(C_{QQQQ}\)&A_{2100}^{1100}\(C_{QQQQ}\)&A_{2000}^{1200}\(C_{QQQQ}\)\\A_{1120}^{1100}\(C_{QQQQ}\)&A_{2110}^{1100}\(C_{QQQQ}\)&A_{2200}^{2000}\(C_{QQQQ}\)\\
					A_{2100}^{2100}\(C_{QQQQ}\)&A_{1120}^{2100}\(C_{QQQQ}\)&A_{1120}^{0000}\(C_{QQQQ}\)\\
					A_{1120}^{1000}\(C_{QQQQ}\)&A_{1120}^{2000}\(C_{QQQQ}\)&A_{1200}^{0100}\(C_{QQQQ}\)\\
					B_{1200}^{0100}\(C_{QQQQ}\)&A_{1100}^{0200}\(C_{QQQQ}\)&\colorbox{newlightgrey}{$A_{2200}^{0100}\(C_{QQQQ}\)$}\\
					B_{2200}^{0100}\(C_{QQQQ}\)&A_{2100}^{0200}\(C_{QQQQ}\)&A_{1200}^{1100}\(C_{QQQQ}\)\\
					\colorbox{newlightgrey}{$A_{1122}^{0000}\(C_{QQQQ}\)$}&\colorbox{newlightgrey}{$B_{1122}^{0000}\(C_{QQQQ}\)$}&\colorbox{newlightgrey}{$A_{2200}^{1100}\(C_{QQQQ}\)$}\\
					\colorbox{newlightgrey}{$A_{2100}^{1200}\(C_{QQQQ}\)$}&\colorbox{newlightgrey}{$A_{1122}^{1000}\(C_{QQQQ}\)$}&\colorbox{newlightgrey}{$A_{1122}^{0100}\(C_{QQQQ}\)$}\\
					\colorbox{newlightgrey}{$A_{1122}^{1100}\(C_{QQQQ}\)$}&\colorbox{newlightgrey}{$A_{2200}^{2200}\(C_{QQQQ}\)$}&\colorbox{newlightgrey}{$A_{1122}^{2200}\(C_{QQQQ}\)$}
				\end{matrix}
				\  \right \}$\\\hline
				$C_{uu}$& 18&45&
				$\left \{
				\begin{matrix}
					\colorbox{newlightgrey}{$A_{1100}^{0000}\(C_{uu\tilde{u}\tilde{u}}\)$}&\colorbox{newlightgrey}{$B_{1100}^{0000}\(C_{uu\tilde{u}\tilde{u}}\)$}&A_{2100}^{0000}\(C_{uu\tilde{u}\tilde{u}}\)\\
					B_{2100}^{0000}\(C_{uu\tilde{u}\tilde{u}}\)&\colorbox{newlightgrey}{$A_{1100}^{1000}\(C_{uu\tilde{u}\tilde{u}}\)$}&B_{1100}^{1000}\(C_{uu\tilde{u}\tilde{u}}\)\\
					A_{2100}^{1000}\(C_{uu\tilde{u}\tilde{u}}\)&A_{2000}^{1100}\(C_{uu\tilde{u}\tilde{u}}\)&A_{2100}^{2000}\(C_{uu\tilde{u}\tilde{u}}\)\\
					A_{1200}^{0000}\(C_{uu\tilde{u}\tilde{u}}\)&B_{1200}^{0000}\(C_{uu\tilde{u}\tilde{u}}\)&\colorbox{newlightgrey}{$A_{1100}^{0100}\(C_{uu\tilde{u}\tilde{u}}\)$}\\
					\colorbox{newlightgrey}{$B_{1100}^{0100}\(C_{uu\tilde{u}\tilde{u}}\)$}&\colorbox{newlightgrey}{$A_{2200}^{0000}\(C_{uu\tilde{u}\tilde{u}}\)$}&A_{2100}^{0100}\(C_{uu\tilde{u}\tilde{u}}\)\\
					B_{2100}^{0100}\(C_{uu\tilde{u}\tilde{u}}\)&\colorbox{newlightgrey}{$B_{1200}^{1000}\(C_{uu\tilde{u}\tilde{u}}\)$}&\colorbox{newlightgrey}{$A_{1100}^{1100}\(C_{uu\tilde{u}\tilde{u}}\)$}\\
					A_{2200}^{1000}\(C_{uu\tilde{u}\tilde{u}}\)&A_{2100}^{1100}\(C_{uu\tilde{u}\tilde{u}}\)&A_{2000}^{1200}\(C_{uu\tilde{u}\tilde{u}}\)\\
					A_{1120}^{1100}\(C_{uu\tilde{u}\tilde{u}}\)&A_{2110}^{1100}\(C_{uu\tilde{u}\tilde{u}}\)&A_{2200}^{2000}\(C_{uu\tilde{u}\tilde{u}}\)\\
					A_{2100}^{2100}\(C_{uu\tilde{u}\tilde{u}}\)&A_{1120}^{2100}\(C_{uu\tilde{u}\tilde{u}}\)&A_{1120}^{0000}\(C_{uu\tilde{u}\tilde{u}}\)\\
					A_{1120}^{1000}\(C_{uu\tilde{u}\tilde{u}}\)&A_{1120}^{2000}\(C_{uu\tilde{u}\tilde{u}}\)&A_{1200}^{0100}\(C_{uu\tilde{u}\tilde{u}}\)\\
					B_{1200}^{0100}\(C_{uu\tilde{u}\tilde{u}}\)&\colorbox{newlightgrey}{$A_{1100}^{0200}\(C_{uu\tilde{u}\tilde{u}}\)$}&\colorbox{newlightgrey}{$A_{2200}^{0100}\(C_{uu\tilde{u}\tilde{u}}\)$}\\
					B_{2200}^{0100}\(C_{uu\tilde{u}\tilde{u}}\)&\colorbox{newlightgrey}{$B_{2100}^{0200}\(C_{uu\tilde{u}\tilde{u}}\)$}&A_{1200}^{1100}\(C_{uu\tilde{u}\tilde{u}}\)\\
					\colorbox{newlightgrey}{$A_{1122}^{0000}\(C_{uu\tilde{u}\tilde{u}}\)$}&A_{1221}^{0000}\(C_{uu\tilde{u}\tilde{u}}\)&A_{2112}^{0000}\(C_{uu\tilde{u}\tilde{u}}\)\\
					\colorbox{newlightgrey}{$A_{2200}^{1100}\(C_{uu\tilde{u}\tilde{u}}\)$}&\colorbox{newlightgrey}{$A_{1122}^{1000}\(C_{uu\tilde{u}\tilde{u}}\)$}&\colorbox{newlightgrey}{$A_{1122}^{0100}\(C_{uu\tilde{u}\tilde{u}}\)$}\\
					\colorbox{newlightgrey}{$A_{0122}^{1100}\(C_{uu\tilde{u}\tilde{u}}\)$}&\colorbox{newlightgrey}{$A_{2200}^{1200}\(C_{uu\tilde{u}\tilde{u}}\)$}&\colorbox{newlightgrey}{$A_{1122}^{1200}\(C_{uu\tilde{u}\tilde{u}}\)$}
				\end{matrix}
				\ \right \}$\\\hline
				$C_{dd}$&18& 45&
				$\left \{
				\begin{matrix}
					\colorbox{newlightgrey}{$A_{1100}^{0000}\(C_{\smash{dd\tilde{d}\tilde{d}}}\) $}&\colorbox{newlightgrey}{$B_{1100}^{0000}\(C_{\smash{dd\tilde{d}\tilde{d}}}\) $}&A_{2100}^{0000}\(C_{\smash{dd\tilde{d}\tilde{d}}}\) \\
					B_{2100}^{0000}\(C_{\smash{dd\tilde{d}\tilde{d}}}\) &A_{1200}^{0000}\(C_{\smash{dd\tilde{d}\tilde{d}}}\) &B_{1200}^{0000}\(C_{\smash{dd\tilde{d}\tilde{d}}}\) \\
					\colorbox{newlightgrey}{$A_{1100}^{1000}\(C_{\smash{dd\tilde{d}\tilde{d}}}\) $}&\colorbox{newlightgrey}{$B_{1100}^{1000}\(C_{\smash{dd\tilde{d}\tilde{d}}}\) $}&\colorbox{newlightgrey}{$A_{2200}^{0000}\(C_{\smash{dd\tilde{d}\tilde{d}}}\) $}\\
					A_{2100}^{1000}\(C_{\smash{dd\tilde{d}\tilde{d}}}\) &B_{2100}^{1000}\(C_{\smash{dd\tilde{d}\tilde{d}}}\) &\colorbox{newlightgrey}{$A_{2000}^{1100}\(C_{\smash{dd\tilde{d}\tilde{d}}}\) $}\\
					A_{2100}^{2000}\(C_{\smash{dd\tilde{d}\tilde{d}}}\) &A_{1120}^{0000}\(C_{\smash{dd\tilde{d}\tilde{d}}}\) &B_{1120}^{0000}\(C_{\smash{dd\tilde{d}\tilde{d}}}\) \\
					\colorbox{newlightgrey}{$A_{1100}^{0100}\(C_{\smash{dd\tilde{d}\tilde{d}}}\) $}&B_{1100}^{0100}\(C_{\smash{dd\tilde{d}\tilde{d}}}\) &A_{2100}^{0100}\(C_{\smash{dd\tilde{d}\tilde{d}}}\) \\
					\colorbox{newlightgrey}{$B_{2100}^{0100}\(C_{\smash{dd\tilde{d}\tilde{d}}}\) $}&A_{1200}^{1000}\(C_{\smash{dd\tilde{d}\tilde{d}}}\) &\colorbox{newlightgrey}{$A_{1100}^{1100}\(C_{\smash{dd\tilde{d}\tilde{d}}}\) $}\\
					\colorbox{newlightgrey}{$A_{2200}^{1000}\(C_{\smash{dd\tilde{d}\tilde{d}}}\) $}&B_{2200}^{1000}\(C_{\smash{dd\tilde{d}\tilde{d}}}\) &A_{2100}^{1100}\(C_{\smash{dd\tilde{d}\tilde{d}}}\) \\
					\colorbox{newlightgrey}{$B_{2000}^{1200}\(C_{\smash{dd\tilde{d}\tilde{d}}}\) $}&\colorbox{newlightgrey}{$A_{1122}^{0000}\(C_{\smash{dd\tilde{d}\tilde{d}}}\) $}&A_{1221}^{0000}\(C_{\smash{dd\tilde{d}\tilde{d}}}\) \\
					A_{2112}^{0000}\(C_{\smash{dd\tilde{d}\tilde{d}}}\) &A_{1120}^{1100}\(C_{\smash{dd\tilde{d}\tilde{d}}}\) &A_{2110}^{1100}\(C_{\smash{dd\tilde{d}\tilde{d}}}\) \\
					A_{2200}^{2000}\(C_{\smash{dd\tilde{d}\tilde{d}}}\) &A_{2100}^{2100}\(C_{\smash{dd\tilde{d}\tilde{d}}}\) &A_{1120}^{2100}\(C_{\smash{dd\tilde{d}\tilde{d}}}\) \\
					A_{2110}^{2100}\(C_{\smash{dd\tilde{d}\tilde{d}}}\) &A_{1120}^{1000}\(C_{\smash{dd\tilde{d}\tilde{d}}}\) &A_{1120}^{2000}\(C_{\smash{dd\tilde{d}\tilde{d}}}\) \\
					A_{0112}^{0000}\(C_{\smash{dd\tilde{d}\tilde{d}}}\) &A_{1200}^{0100}\(C_{\smash{dd\tilde{d}\tilde{d}}}\) &A_{1100}^{0200}\(C_{\smash{dd\tilde{d}\tilde{d}}}\) \\
					\colorbox{newlightgrey}{$A_{2200}^{1100}\(C_{\smash{dd\tilde{d}\tilde{d}}}\) $}&\colorbox{newlightgrey}{$A_{1122}^{1000}\(C_{\smash{dd\tilde{d}\tilde{d}}}\) $}&\colorbox{newlightgrey}{$A_{1220}^{1100}\(C_{\smash{dd\tilde{d}\tilde{d}}}\) $}\\
					\colorbox{newlightgrey}{$A_{2110}^{1200}\(C_{\smash{dd\tilde{d}\tilde{d}}}\) $}&\colorbox{newlightgrey}{$A_{0122}^{2100}\(C_{\smash{dd\tilde{d}\tilde{d}}}\) $}&\colorbox{newlightgrey}{$A_{1220}^{2200}\(C_{\smash{dd\tilde{d}\tilde{d}}}\) $}
				\end{matrix}
				\ \right \}$\\\hline
			\end{tabular}%
		}
		\caption{Continuation of Table \ref{tableInv4Fermi1}}
		\label{tableInv4Fermi3}
	\end{table}
	\begin{table}[h!]
		\hspace*{-1cm}
		\centering
		\small
		\renewcommand{\arraystretch}{1.01}
		\resizebox{0.88\columnwidth}{!}{%
			\begin{tabular}{l|c|c|c}
				Wilson coefficient& \begin{minipage}{0.06\columnwidth}\centering
					\vspace*{0.1cm}
					\# min set\vspace*{0.1cm}
				\end{minipage}& \begin{minipage}{0.06\columnwidth}\centering
					\vspace*{0.1cm}
					\# max set\vspace*{0.1cm}
				\end{minipage}&Maximal set\\
				\hline
				$C_{Qu}^{(1,8)}$&36&81&$\left \{ \ 
				\begin{matrix}
					\colorbox{newlightgrey}{$B_{1000}^{0000}\(C_{QQ\tilde{u}\tilde{u}}\)$}&B_{2000}^{0000}\(C_{QQ\tilde{u}\tilde{u}}\)&\colorbox{newlightgrey}{$B_{0100}^{0000}\(C_{QQ\tilde{u}\tilde{u}}\)$}\\
						\colorbox{newlightgrey}{$B_{0110}^{0000}\(C_{QQ\tilde{u}\tilde{u}}\)$}&\colorbox{newlightgrey}{$B_{1000}^{0100}\(C_{QQ\tilde{u}\tilde{u}}\)$}&\colorbox{newlightgrey}{$A_{0000}^{1100}\(C_{QQ\tilde{u}\tilde{u}}\)$}\\
						B_{0120}^{0000}\(C_{QQ\tilde{u}\tilde{u}}\)&B_{2000}^{0100}\(C_{QQ\tilde{u}\tilde{u}}\)&\colorbox{newlightgrey}{$B_{0110}^{1000}\(C_{QQ\tilde{u}\tilde{u}}\)$}\\
						A_{1000}^{1100}\(C_{QQ\tilde{u}\tilde{u}}\)&A_{0000}^{2100}\(C_{QQ\tilde{u}\tilde{u}}\)&B_{0120}^{1000}\(C_{QQ\tilde{u}\tilde{u}}\)\\
						A_{2000}^{1100}\(C_{QQ\tilde{u}\tilde{u}}\)&B_{2000}^{1100}\(C_{QQ\tilde{u}\tilde{u}}\)&\colorbox{newlightgrey}{$A_{1100}^{0000}\(C_{QQ\tilde{u}\tilde{u}}\)$}\\
						\colorbox{newlightgrey}{$B_{1100}^{0000}\(C_{QQ\tilde{u}\tilde{u}}\)$}&A_{2100}^{0000}\(C_{QQ\tilde{u}\tilde{u}}\)&B_{2100}^{0000}\(C_{QQ\tilde{u}\tilde{u}}\)\\
						\colorbox{newlightgrey}{$A_{1100}^{1000}\(C_{QQ\tilde{u}\tilde{u}}\)$}&B_{2110}^{0000}\(C_{QQ\tilde{u}\tilde{u}}\)&A_{2100}^{1000}\(C_{QQ\tilde{u}\tilde{u}}\)\\
						B_{2100}^{1000}\(C_{QQ\tilde{u}\tilde{u}}\)&A_{1100}^{2000}\(C_{QQ\tilde{u}\tilde{u}}\)&A_{2100}^{2000}\(C_{QQ\tilde{u}\tilde{u}}\)\\
						B_{0200}^{0000}\(C_{QQ\tilde{u}\tilde{u}}\)&B_{0210}^{0000}\(C_{QQ\tilde{u}\tilde{u}}\)&\colorbox{newlightgrey}{$B_{1100}^{0100}\(C_{QQ\tilde{u}\tilde{u}}\)$}\\
						B_{1000}^{0200}\(C_{QQ\tilde{u}\tilde{u}}\)&B_{2000}^{1000}\(C_{QQ\tilde{u}\tilde{u}}\)&\colorbox{newlightgrey}{$A_{0100}^{1100}\(C_{QQ\tilde{u}\tilde{u}}\)$}\\
						A_{0000}^{1200}\(C_{QQ\tilde{u}\tilde{u}}\)&\colorbox{newlightgrey}{$B_{0220}^{0000}\(C_{QQ\tilde{u}\tilde{u}}\)$}&B_{2100}^{0100}\(C_{QQ\tilde{u}\tilde{u}}\)\\
						\colorbox{newlightgrey}{$B_{2000}^{0200}\(C_{QQ\tilde{u}\tilde{u}}\)$}&B_{0210}^{1000}\(C_{QQ\tilde{u}\tilde{u}}\)&\colorbox{newlightgrey}{$A_{1100}^{1100}\(C_{QQ\tilde{u}\tilde{u}}\)$}\\
						\colorbox{newlightgrey}{$A_{0110}^{1100}\(C_{QQ\tilde{u}\tilde{u}}\)$}&\colorbox{newlightgrey}{$B_{1100}^{1100}\(C_{QQ\tilde{u}\tilde{u}}\)$}&\colorbox{newlightgrey}{$A_{1000}^{1200}\(C_{QQ\tilde{u}\tilde{u}}\)$}\\
						A_{0100}^{2100}\(C_{QQ\tilde{u}\tilde{u}}\)&\colorbox{newlightgrey}{$A_{0000}^{2200}\(C_{QQ\tilde{u}\tilde{u}}\)$}&\colorbox{newlightgrey}{$B_{2110}^{0100}\(C_{QQ\tilde{u}\tilde{u}}\)$}\\
						B_{1120}^{0100}\(C_{QQ\tilde{u}\tilde{u}}\)&\colorbox{newlightgrey}{$B_{0220}^{1000}\(C_{QQ\tilde{u}\tilde{u}}\)$}&A_{2100}^{1100}\(C_{QQ\tilde{u}\tilde{u}}\)\\
						A_{0120}^{1100}\(C_{QQ\tilde{u}\tilde{u}}\)&B_{2100}^{1100}\(C_{QQ\tilde{u}\tilde{u}}\)&A_{2000}^{1200}\(C_{QQ\tilde{u}\tilde{u}}\)\\
						B_{2000}^{1200}\(C_{QQ\tilde{u}\tilde{u}}\)&A_{1100}^{2100}\(C_{QQ\tilde{u}\tilde{u}}\)&A_{0110}^{2100}\(C_{QQ\tilde{u}\tilde{u}}\)\\
						A_{1120}^{1100}\(C_{QQ\tilde{u}\tilde{u}}\)&B_{2110}^{1100}\(C_{QQ\tilde{u}\tilde{u}}\)&B_{1120}^{1100}\(C_{QQ\tilde{u}\tilde{u}}\)\\
						A_{2100}^{2100}\(C_{QQ\tilde{u}\tilde{u}}\)&B_{2100}^{2100}\(C_{QQ\tilde{u}\tilde{u}}\)&A_{1120}^{2100}\(C_{QQ\tilde{u}\tilde{u}}\)\\
						B_{2110}^{2100}\(C_{QQ\tilde{u}\tilde{u}}\)&A_{0000}^{1120}\(C_{QQ\tilde{u}\tilde{u}}\)&B_{1120}^{1000}\(C_{QQ\tilde{u}\tilde{u}}\)\\
						B_{0211}^{0000}\(C_{QQ\tilde{u}\tilde{u}}\)&B_{1200}^{0100}\(C_{QQ\tilde{u}\tilde{u}}\)&B_{1100}^{0200}\(C_{QQ\tilde{u}\tilde{u}}\)\\
						A_{0200}^{1100}\(C_{QQ\tilde{u}\tilde{u}}\)&\colorbox{newlightgrey}{$B_{0221}^{0000}\(C_{QQ\tilde{u}\tilde{u}}\)$}&\colorbox{newlightgrey}{$B_{2200}^{0100}\(C_{QQ\tilde{u}\tilde{u}}\)$}\\
						\colorbox{newlightgrey}{$B_{2100}^{0200}\(C_{QQ\tilde{u}\tilde{u}}\)$}&B_{0211}^{1000}\(C_{QQ\tilde{u}\tilde{u}}\)&\colorbox{newlightgrey}{$B_{2110}^{0200}\(C_{QQ\tilde{u}\tilde{u}}\)$}\\
						\colorbox{newlightgrey}{$B_{0221}^{1000}\(C_{QQ\tilde{u}\tilde{u}}\)$}&\colorbox{newlightgrey}{$A_{2200}^{1100}\(C_{QQ\tilde{u}\tilde{u}}\)$}&\colorbox{newlightgrey}{$A_{0220}^{1100}\(C_{QQ\tilde{u}\tilde{u}}\)$}\\
						\colorbox{newlightgrey}{$B_{2200}^{1100}\(C_{QQ\tilde{u}\tilde{u}}\)$}&\colorbox{newlightgrey}{$B_{2100}^{1200}\(C_{QQ\tilde{u}\tilde{u}}\)$}&\colorbox{newlightgrey}{$B_{1200}^{2100}\(C_{QQ\tilde{u}\tilde{u}}\)$}\\
						\colorbox{newlightgrey}{$A_{0110}^{2200}\(C_{QQ\tilde{u}\tilde{u}}\)$}&\colorbox{newlightgrey}{$B_{0221}^{0110}\(C_{QQ\tilde{u}\tilde{u}}\)$}&\colorbox{newlightgrey}{$A_{1122}^{1100}\(C_{QQ\tilde{u}\tilde{u}}\)$}\\
						\colorbox{newlightgrey}{$A_{1220}^{1200}\(C_{QQ\tilde{u}\tilde{u}}\)$}&\colorbox{newlightgrey}{$B_{2210}^{1200}\(C_{QQ\tilde{u}\tilde{u}}\)$}&\colorbox{newlightgrey}{$A_{1122}^{2200}\(C_{QQ\tilde{u}\tilde{u}}\)$}
				\end{matrix}
				\ \right \}$\\\hline
			\end{tabular}%
		}
		\caption{Continuation of Table \ref{tableInv4Fermi1}}
		\label{tableInv4Fermi4}
	\end{table}
	\begin{table}[h!]
		\hspace*{-1cm}
		\centering
		\small
		\renewcommand{\arraystretch}{1.01}
		\resizebox{0.88\columnwidth}{!}{%
			\begin{tabular}{l|c|c|c}
				Wilson coefficient& \begin{minipage}{0.06\columnwidth}\centering
					\vspace*{0.1cm}
					\# min set\vspace*{0.1cm}
				\end{minipage}& \begin{minipage}{0.06\columnwidth}\centering
					\vspace*{0.1cm}
					\# max set\vspace*{0.1cm}
				\end{minipage}&Maximal set\\\hline
				$C_{Qd}^{1,8}$&36&81&$\left \{ \ 
				\begin{matrix}
					\colorbox{newlightgrey}{$B_{1000}^{0000}\(C_{\smash{QQ\tilde{d}\tilde{d}}}\) $}&B_{2000}^{0000}\(C_{\smash{QQ\tilde{d}\tilde{d}}}\) &\colorbox{newlightgrey}{$A_{0000}^{1100}\(C_{\smash{QQ\tilde{d}\tilde{d}}}\) $}\\
					A_{0000}^{2100}\(C_{\smash{QQ\tilde{d}\tilde{d}}}\) &\colorbox{newlightgrey}{$B_{0100}^{0000}\(C_{\smash{QQ\tilde{d}\tilde{d}}}\) $}&\colorbox{newlightgrey}{$A_{1100}^{0000}\(C_{\smash{QQ\tilde{d}\tilde{d}}}\) $}\\
					\colorbox{newlightgrey}{$B_{0110}^{0000}\(C_{\smash{QQ\tilde{d}\tilde{d}}}\) $}&\colorbox{newlightgrey}{$B_{1100}^{0000}\(C_{\smash{QQ\tilde{d}\tilde{d}}}\) $}&\colorbox{newlightgrey}{$B_{1000}^{0100}\(C_{\smash{QQ\tilde{d}\tilde{d}}}\) $}\\
					A_{2100}^{0000}\(C_{\smash{QQ\tilde{d}\tilde{d}}}\) &B_{0120}^{0000}\(C_{\smash{QQ\tilde{d}\tilde{d}}}\) &B_{2100}^{0000}\(C_{\smash{QQ\tilde{d}\tilde{d}}}\) \\
					B_{2000}^{0100}\(C_{\smash{QQ\tilde{d}\tilde{d}}}\) &\colorbox{newlightgrey}{$A_{1100}^{1000}\(C_{\smash{QQ\tilde{d}\tilde{d}}}\) $}&\colorbox{newlightgrey}{$B_{0110}^{1000}\(C_{\smash{QQ\tilde{d}\tilde{d}}}\) $}\\
					\colorbox{newlightgrey}{$A_{1000}^{1100}\(C_{\smash{QQ\tilde{d}\tilde{d}}}\) $}&A_{0000}^{1200}\(C_{\smash{QQ\tilde{d}\tilde{d}}}\) &B_{2110}^{0000}\(C_{\smash{QQ\tilde{d}\tilde{d}}}\) \\
					A_{2100}^{1000}\(C_{\smash{QQ\tilde{d}\tilde{d}}}\) &B_{0120}^{1000}\(C_{\smash{QQ\tilde{d}\tilde{d}}}\) &B_{2100}^{1000}\(C_{\smash{QQ\tilde{d}\tilde{d}}}\) \\
					A_{2000}^{1100}\(C_{\smash{QQ\tilde{d}\tilde{d}}}\) &B_{2000}^{1100}\(C_{\smash{QQ\tilde{d}\tilde{d}}}\) &A_{1100}^{2000}\(C_{\smash{QQ\tilde{d}\tilde{d}}}\) \\
					\colorbox{newlightgrey}{$A_{0000}^{2200}\(C_{\smash{QQ\tilde{d}\tilde{d}}}\) $}&A_{2100}^{2000}\(C_{\smash{QQ\tilde{d}\tilde{d}}}\) &B_{0120}^{2000}\(C_{\smash{QQ\tilde{d}\tilde{d}}}\) \\
					A_{2000}^{2100}\(C_{\smash{QQ\tilde{d}\tilde{d}}}\) &B_{2000}^{1000}\(C_{\smash{QQ\tilde{d}\tilde{d}}}\) &A_{0000}^{1120}\(C_{\smash{QQ\tilde{d}\tilde{d}}}\) \\
					B_{0200}^{0000}\(C_{\smash{QQ\tilde{d}\tilde{d}}}\) &A_{1200}^{0000}\(C_{\smash{QQ\tilde{d}\tilde{d}}}\) &B_{0210}^{0000}\(C_{\smash{QQ\tilde{d}\tilde{d}}}\) \\
					B_{1200}^{0000}\(C_{\smash{QQ\tilde{d}\tilde{d}}}\) &\colorbox{newlightgrey}{$A_{1100}^{0100}\(C_{\smash{QQ\tilde{d}\tilde{d}}}\) $}&B_{0110}^{0100}\(C_{\smash{QQ\tilde{d}\tilde{d}}}\) \\
					\colorbox{newlightgrey}{$B_{1100}^{0100}\(C_{\smash{QQ\tilde{d}\tilde{d}}}\) $}&A_{0100}^{1100}\(C_{\smash{QQ\tilde{d}\tilde{d}}}\) &\colorbox{newlightgrey}{$A_{2200}^{0000}\(C_{\smash{QQ\tilde{d}\tilde{d}}}\) $}\\
					\colorbox{newlightgrey}{$B_{0220}^{0000}\(C_{\smash{QQ\tilde{d}\tilde{d}}}\) $}&\colorbox{newlightgrey}{$B_{2200}^{0000}\(C_{\smash{QQ\tilde{d}\tilde{d}}}\) $}&A_{2100}^{0100}\(C_{\smash{QQ\tilde{d}\tilde{d}}}\) \\
					\colorbox{newlightgrey}{$B_{0120}^{0100}\(C_{\smash{QQ\tilde{d}\tilde{d}}}\) $}&B_{2100}^{0100}\(C_{\smash{QQ\tilde{d}\tilde{d}}}\) &B_{0210}^{1000}\(C_{\smash{QQ\tilde{d}\tilde{d}}}\) \\
					\colorbox{newlightgrey}{$B_{1200}^{1000}\(C_{\smash{QQ\tilde{d}\tilde{d}}}\) $}&\colorbox{newlightgrey}{$A_{1100}^{1100}\(C_{\smash{QQ\tilde{d}\tilde{d}}}\) $}&A_{0110}^{1100}\(C_{\smash{QQ\tilde{d}\tilde{d}}}\) \\
					\colorbox{newlightgrey}{$B_{2210}^{0000}\(C_{\smash{QQ\tilde{d}\tilde{d}}}\) $}&B_{2110}^{0100}\(C_{\smash{QQ\tilde{d}\tilde{d}}}\) &A_{2200}^{1000}\(C_{\smash{QQ\tilde{d}\tilde{d}}}\) \\
					\colorbox{newlightgrey}{$B_{0220}^{1000}\(C_{\smash{QQ\tilde{d}\tilde{d}}}\) $}&B_{2200}^{1000}\(C_{\smash{QQ\tilde{d}\tilde{d}}}\) &\colorbox{newlightgrey}{$A_{2100}^{1100}\(C_{\smash{QQ\tilde{d}\tilde{d}}}\) $}\\
					A_{0120}^{1100}\(C_{\smash{QQ\tilde{d}\tilde{d}}}\) &B_{2100}^{1100}\(C_{\smash{QQ\tilde{d}\tilde{d}}}\) &B_{2000}^{1200}\(C_{\smash{QQ\tilde{d}\tilde{d}}}\) \\
					A_{1200}^{2000}\(C_{\smash{QQ\tilde{d}\tilde{d}}}\) &A_{1120}^{1100}\(C_{\smash{QQ\tilde{d}\tilde{d}}}\) &B_{2110}^{1100}\(C_{\smash{QQ\tilde{d}\tilde{d}}}\) \\
					A_{2200}^{2000}\(C_{\smash{QQ\tilde{d}\tilde{d}}}\) &A_{2100}^{2100}\(C_{\smash{QQ\tilde{d}\tilde{d}}}\) &\colorbox{newlightgrey}{$A_{0000}^{1122}\(C_{\smash{QQ\tilde{d}\tilde{d}}}\) $}\\
					A_{0000}^{1221}\(C_{\smash{QQ\tilde{d}\tilde{d}}}\) &A_{0000}^{2112}\(C_{\smash{QQ\tilde{d}\tilde{d}}}\) &A_{1120}^{2100}\(C_{\smash{QQ\tilde{d}\tilde{d}}}\) \\
					A_{1120}^{1000}\(C_{\smash{QQ\tilde{d}\tilde{d}}}\) &A_{1120}^{2000}\(C_{\smash{QQ\tilde{d}\tilde{d}}}\) &\colorbox{newlightgrey}{$B_{0221}^{0000}\(C_{\smash{QQ\tilde{d}\tilde{d}}}\) $}\\
					\colorbox{newlightgrey}{$A_{1122}^{0000}\(C_{\smash{QQ\tilde{d}\tilde{d}}}\) $}&\colorbox{newlightgrey}{$B_{2210}^{0100}\(C_{\smash{QQ\tilde{d}\tilde{d}}}\) $}&\colorbox{newlightgrey}{$B_{0221}^{1000}\(C_{\smash{QQ\tilde{d}\tilde{d}}}\) $}\\
					\colorbox{newlightgrey}{$A_{2200}^{1100}\(C_{\smash{QQ\tilde{d}\tilde{d}}}\) $}&\colorbox{newlightgrey}{$A_{0220}^{1100}\(C_{\smash{QQ\tilde{d}\tilde{d}}}\) $}&\colorbox{newlightgrey}{$B_{2200}^{1100}\(C_{\smash{QQ\tilde{d}\tilde{d}}}\) $}\\
					\colorbox{newlightgrey}{$B_{2100}^{1200}\(C_{\smash{QQ\tilde{d}\tilde{d}}}\) $}&\colorbox{newlightgrey}{$A_{1122}^{1000}\(C_{\smash{QQ\tilde{d}\tilde{d}}}\) $}&\colorbox{newlightgrey}{$B_{2210}^{1100}\(C_{\smash{QQ\tilde{d}\tilde{d}}}\) $}\\
					\colorbox{newlightgrey}{$A_{1122}^{1100}\(C_{\smash{QQ\tilde{d}\tilde{d}}}\) $}&\colorbox{newlightgrey}{$A_{0122}^{2100}\(C_{\smash{QQ\tilde{d}\tilde{d}}}\) $}&\colorbox{newlightgrey}{$B_{2211}^{2100}\(C_{\smash{QQ\tilde{d}\tilde{d}}}\) $}
				\end{matrix}
				\ \right \}$\\\hline
			\end{tabular}%
		}
		\caption{Continuation of Table \ref{tableInv4Fermi1}}
		\label{tableInv4Fermi5}
	\end{table}
	\begin{table}[h!]
		\hspace*{-1cm}
		\centering
		\small
		\renewcommand{\arraystretch}{1.01}
		\resizebox{0.88\columnwidth}{!}{%
			\begin{tabular}{l|c|c|c}
			Wilson coefficient& \begin{minipage}{0.06\columnwidth}\centering
				\vspace*{0.1cm}
				\# min set\vspace*{0.1cm}
			\end{minipage}& \begin{minipage}{0.06\columnwidth}\centering
				\vspace*{0.1cm}
				\# max set\vspace*{0.1cm}
			\end{minipage}&Maximal set\\
			\hline
				$C_{ud}^{(1,8)}$&36&81&$\left \{ \ 
				\begin{matrix}
					\colorbox{newlightgrey}{$B_{1000}^{0000}\(C_{\smash{\tilde{u}\tilde{u}\tilde{d}\tilde{d}}}\)$}&B_{2000}^{0000}\(C_{\smash{\tilde{u}\tilde{u}\tilde{d}\tilde{d}}}\)&\colorbox{newlightgrey}{$A_{0000}^{1100}\(C_{\smash{\tilde{u}\tilde{u}\tilde{d}\tilde{d}}}\)$}\\
					A_{0000}^{2100}\(C_{\smash{\tilde{u}\tilde{u}\tilde{d}\tilde{d}}}\)&\colorbox{newlightgrey}{$B_{0100}^{0000}\(C_{\smash{\tilde{u}\tilde{u}\tilde{d}\tilde{d}}}\)$}&\colorbox{newlightgrey}{$A_{1100}^{0000}\(C_{\smash{\tilde{u}\tilde{u}\tilde{d}\tilde{d}}}\)$}\\
					\colorbox{newlightgrey}{$B_{1100}^{0000}\(C_{\smash{\tilde{u}\tilde{u}\tilde{d}\tilde{d}}}\)$}&\colorbox{newlightgrey}{$B_{1000}^{0100}\(C_{\smash{\tilde{u}\tilde{u}\tilde{d}\tilde{d}}}\)$}&A_{2100}^{0000}\(C_{\smash{\tilde{u}\tilde{u}\tilde{d}\tilde{d}}}\)\\
					B_{2100}^{0000}\(C_{\smash{\tilde{u}\tilde{u}\tilde{d}\tilde{d}}}\)&B_{2000}^{0100}\(C_{\smash{\tilde{u}\tilde{u}\tilde{d}\tilde{d}}}\)&\colorbox{newlightgrey}{$A_{1100}^{1000}\(C_{\smash{\tilde{u}\tilde{u}\tilde{d}\tilde{d}}}\)$}\\
					B_{1100}^{1000}\(C_{\smash{\tilde{u}\tilde{u}\tilde{d}\tilde{d}}}\)&A_{2100}^{1000}\(C_{\smash{\tilde{u}\tilde{u}\tilde{d}\tilde{d}}}\)&B_{2100}^{1000}\(C_{\smash{\tilde{u}\tilde{u}\tilde{d}\tilde{d}}}\)\\
					A_{1100}^{2000}\(C_{\smash{\tilde{u}\tilde{u}\tilde{d}\tilde{d}}}\)&\colorbox{newlightgrey}{$B_{0110}^{0000}\(C_{\smash{\tilde{u}\tilde{u}\tilde{d}\tilde{d}}}\)$}&B_{0120}^{0000}\(C_{\smash{\tilde{u}\tilde{u}\tilde{d}\tilde{d}}}\)\\
					\colorbox{newlightgrey}{$B_{0110}^{1000}\(C_{\smash{\tilde{u}\tilde{u}\tilde{d}\tilde{d}}}\)$}&\colorbox{newlightgrey}{$A_{1000}^{1100}\(C_{\smash{\tilde{u}\tilde{u}\tilde{d}\tilde{d}}}\)$}&A_{0000}^{1200}\(C_{\smash{\tilde{u}\tilde{u}\tilde{d}\tilde{d}}}\)\\
					B_{2110}^{0000}\(C_{\smash{\tilde{u}\tilde{u}\tilde{d}\tilde{d}}}\)&B_{0120}^{1000}\(C_{\smash{\tilde{u}\tilde{u}\tilde{d}\tilde{d}}}\)&A_{2000}^{1100}\(C_{\smash{\tilde{u}\tilde{u}\tilde{d}\tilde{d}}}\)\\
					B_{2000}^{1100}\(C_{\smash{\tilde{u}\tilde{u}\tilde{d}\tilde{d}}}\)&\colorbox{newlightgrey}{$A_{0000}^{2200}\(C_{\smash{\tilde{u}\tilde{u}\tilde{d}\tilde{d}}}\)$}&B_{0120}^{2000}\(C_{\smash{\tilde{u}\tilde{u}\tilde{d}\tilde{d}}}\)\\
					A_{2000}^{2100}\(C_{\smash{\tilde{u}\tilde{u}\tilde{d}\tilde{d}}}\)&B_{0200}^{0000}\(C_{\smash{\tilde{u}\tilde{u}\tilde{d}\tilde{d}}}\)&A_{1200}^{0000}\(C_{\smash{\tilde{u}\tilde{u}\tilde{d}\tilde{d}}}\)\\
					B_{1200}^{0000}\(C_{\smash{\tilde{u}\tilde{u}\tilde{d}\tilde{d}}}\)&\colorbox{newlightgrey}{$A_{1100}^{0100}\(C_{\smash{\tilde{u}\tilde{u}\tilde{d}\tilde{d}}}\)$}&\colorbox{newlightgrey}{$B_{0110}^{0100}\(C_{\smash{\tilde{u}\tilde{u}\tilde{d}\tilde{d}}}\)$}\\
					B_{1000}^{0200}\(C_{\smash{\tilde{u}\tilde{u}\tilde{d}\tilde{d}}}\)&B_{2000}^{1000}\(C_{\smash{\tilde{u}\tilde{u}\tilde{d}\tilde{d}}}\)&\colorbox{newlightgrey}{$A_{2200}^{0000}\(C_{\smash{\tilde{u}\tilde{u}\tilde{d}\tilde{d}}}\)$}\\
					\colorbox{newlightgrey}{$B_{2200}^{0000}\(C_{\smash{\tilde{u}\tilde{u}\tilde{d}\tilde{d}}}\)$}&A_{2100}^{0100}\(C_{\smash{\tilde{u}\tilde{u}\tilde{d}\tilde{d}}}\)&B_{0120}^{0100}\(C_{\smash{\tilde{u}\tilde{u}\tilde{d}\tilde{d}}}\)\\
					\colorbox{newlightgrey}{$B_{2000}^{0200}\(C_{\smash{\tilde{u}\tilde{u}\tilde{d}\tilde{d}}}\)$}&A_{1200}^{1000}\(C_{\smash{\tilde{u}\tilde{u}\tilde{d}\tilde{d}}}\)&\colorbox{newlightgrey}{$B_{1200}^{1000}\(C_{\smash{\tilde{u}\tilde{u}\tilde{d}\tilde{d}}}\)$}\\
					\colorbox{newlightgrey}{$A_{1100}^{1100}\(C_{\smash{\tilde{u}\tilde{u}\tilde{d}\tilde{d}}}\)$}&\colorbox{newlightgrey}{$A_{0110}^{1100}\(C_{\smash{\tilde{u}\tilde{u}\tilde{d}\tilde{d}}}\)$}&B_{1100}^{1100}\(C_{\smash{\tilde{u}\tilde{u}\tilde{d}\tilde{d}}}\)\\
					A_{0000}^{1120}\(C_{\smash{\tilde{u}\tilde{u}\tilde{d}\tilde{d}}}\)&\colorbox{newlightgrey}{$B_{2110}^{0100}\(C_{\smash{\tilde{u}\tilde{u}\tilde{d}\tilde{d}}}\)$}&B_{1120}^{0100}\(C_{\smash{\tilde{u}\tilde{u}\tilde{d}\tilde{d}}}\)\\
					\colorbox{newlightgrey}{$A_{2200}^{1000}\(C_{\smash{\tilde{u}\tilde{u}\tilde{d}\tilde{d}}}\)$}&B_{2200}^{1000}\(C_{\smash{\tilde{u}\tilde{u}\tilde{d}\tilde{d}}}\)&\colorbox{newlightgrey}{$A_{2100}^{1100}\(C_{\smash{\tilde{u}\tilde{u}\tilde{d}\tilde{d}}}\)$}\\
					A_{0120}^{1100}\(C_{\smash{\tilde{u}\tilde{u}\tilde{d}\tilde{d}}}\)&B_{2100}^{1100}\(C_{\smash{\tilde{u}\tilde{u}\tilde{d}\tilde{d}}}\)&A_{1200}^{2000}\(C_{\smash{\tilde{u}\tilde{u}\tilde{d}\tilde{d}}}\)\\
					A_{1100}^{2100}\(C_{\smash{\tilde{u}\tilde{u}\tilde{d}\tilde{d}}}\)&A_{0110}^{2100}\(C_{\smash{\tilde{u}\tilde{u}\tilde{d}\tilde{d}}}\)&A_{2100}^{2100}\(C_{\smash{\tilde{u}\tilde{u}\tilde{d}\tilde{d}}}\)\\
					A_{0120}^{2100}\(C_{\smash{\tilde{u}\tilde{u}\tilde{d}\tilde{d}}}\)&B_{2100}^{2100}\(C_{\smash{\tilde{u}\tilde{u}\tilde{d}\tilde{d}}}\)&B_{2110}^{0110}\(C_{\smash{\tilde{u}\tilde{u}\tilde{d}\tilde{d}}}\)\\
					B_{1120}^{0110}\(C_{\smash{\tilde{u}\tilde{u}\tilde{d}\tilde{d}}}\)&B_{2110}^{0120}\(C_{\smash{\tilde{u}\tilde{u}\tilde{d}\tilde{d}}}\)&A_{2100}^{1120}\(C_{\smash{\tilde{u}\tilde{u}\tilde{d}\tilde{d}}}\)\\
					A_{1120}^{0000}\(C_{\smash{\tilde{u}\tilde{u}\tilde{d}\tilde{d}}}\)&A_{1120}^{1000}\(C_{\smash{\tilde{u}\tilde{u}\tilde{d}\tilde{d}}}\)&\colorbox{newlightgrey}{$A_{0000}^{1122}\(C_{\smash{\tilde{u}\tilde{u}\tilde{d}\tilde{d}}}\)$}\\
					A_{0000}^{1221}\(C_{\smash{\tilde{u}\tilde{u}\tilde{d}\tilde{d}}}\)&A_{0000}^{2112}\(C_{\smash{\tilde{u}\tilde{u}\tilde{d}\tilde{d}}}\)&\colorbox{newlightgrey}{$B_{0221}^{0000}\(C_{\smash{\tilde{u}\tilde{u}\tilde{d}\tilde{d}}}\)$}\\
					\colorbox{newlightgrey}{$A_{2200}^{0100}\(C_{\smash{\tilde{u}\tilde{u}\tilde{d}\tilde{d}}}\)$}&\colorbox{newlightgrey}{$A_{1122}^{0000}\(C_{\smash{\tilde{u}\tilde{u}\tilde{d}\tilde{d}}}\)$}&\colorbox{newlightgrey}{$B_{2110}^{0200}\(C_{\smash{\tilde{u}\tilde{u}\tilde{d}\tilde{d}}}\)$}\\
					\colorbox{newlightgrey}{$B_{0221}^{1000}\(C_{\smash{\tilde{u}\tilde{u}\tilde{d}\tilde{d}}}\)$}&\colorbox{newlightgrey}{$A_{2200}^{1100}\(C_{\smash{\tilde{u}\tilde{u}\tilde{d}\tilde{d}}}\)$}&\colorbox{newlightgrey}{$B_{2200}^{1100}\(C_{\smash{\tilde{u}\tilde{u}\tilde{d}\tilde{d}}}\)$}\\
					\colorbox{newlightgrey}{$B_{2100}^{1200}\(C_{\smash{\tilde{u}\tilde{u}\tilde{d}\tilde{d}}}\)$}&\colorbox{newlightgrey}{$B_{1200}^{2100}\(C_{\smash{\tilde{u}\tilde{u}\tilde{d}\tilde{d}}}\)$}&\colorbox{newlightgrey}{$A_{1122}^{1000}\(C_{\smash{\tilde{u}\tilde{u}\tilde{d}\tilde{d}}}\)$}\\
					\colorbox{newlightgrey}{$A_{1122}^{0100}\(C_{\smash{\tilde{u}\tilde{u}\tilde{d}\tilde{d}}}\)$}&\colorbox{newlightgrey}{$A_{1122}^{1100}\(C_{\smash{\tilde{u}\tilde{u}\tilde{d}\tilde{d}}}\)$}&\colorbox{newlightgrey}{$B_{2211}^{1100}\(C_{\smash{\tilde{u}\tilde{u}\tilde{d}\tilde{d}}}\)$}
					\end{matrix}
				\ \right \}$\\\hline
			\end{tabular}%
		}
		\caption{Continuation of Table \ref{tableInv4Fermi1}}
		\label{tableInv4Fermi6}
	\end{table}
	\begin{table}[h!]
		\hspace*{-1cm}
		\centering
		\small
		\renewcommand{\arraystretch}{1.01}
		\resizebox{0.71\columnwidth}{!}{%
			\begin{tabular}{l|c|c|c}
				Wilson coefficient& \begin{minipage}{0.06\columnwidth}\centering
					\vspace*{0.1cm}
					\# min set\vspace*{0.1cm}
				\end{minipage}& \begin{minipage}{0.06\columnwidth}\centering
					\vspace*{0.1cm}
					\# max set\vspace*{0.1cm}
				\end{minipage}&Maximal set\\
				\hline
				$C_{QuQd}$&81&162&$\left \{ \ 
				\begin{matrix}
					\colorbox{newlightgrey}{$A_{0000}^{0000}\(C_{\smash{Q\tilde{u}Q\tilde{d}}}\)$}&\colorbox{newlightgrey}{$B_{0000}^{0000}\(C_{\smash{Q\tilde{u}Q\tilde{d}}}\)$}&\colorbox{newlightgrey}{$A_{1000}^{0000}\(C_{\smash{Q\tilde{u}Q\tilde{d}}}\)$}\\
					\colorbox{newlightgrey}{$B_{1000}^{0000}\(C_{\smash{Q\tilde{u}Q\tilde{d}}}\)$}&\colorbox{newlightgrey}{$A_{0000}^{1000}\(C_{\smash{Q\tilde{u}Q\tilde{d}}}\)$}&\colorbox{newlightgrey}{$B_{0000}^{1000}\(C_{\smash{Q\tilde{u}Q\tilde{d}}}\)$}\\
					A_{2000}^{0000}\(C_{\smash{Q\tilde{u}Q\tilde{d}}}\)&B_{2000}^{0000}\(C_{\smash{Q\tilde{u}Q\tilde{d}}}\)&\colorbox{newlightgrey}{$A_{1000}^{1000}\(C_{\smash{Q\tilde{u}Q\tilde{d}}}\)$}\\
					B_{1000}^{1000}\(C_{\smash{Q\tilde{u}Q\tilde{d}}}\)&A_{0000}^{2000}\(C_{\smash{Q\tilde{u}Q\tilde{d}}}\)&A_{2000}^{1000}\(C_{\smash{Q\tilde{u}Q\tilde{d}}}\)\\
					B_{2000}^{1000}\(C_{\smash{Q\tilde{u}Q\tilde{d}}}\)&A_{1000}^{2000}\(C_{\smash{Q\tilde{u}Q\tilde{d}}}\)&A_{2000}^{2000}\(C_{\smash{Q\tilde{u}Q\tilde{d}}}\)\\
					\colorbox{newlightgrey}{$A_{0100}^{0000}\(C_{\smash{Q\tilde{u}Q\tilde{d}}}\)$}&\colorbox{newlightgrey}{$B_{0100}^{0000}\(C_{\smash{Q\tilde{u}Q\tilde{d}}}\)$}&\colorbox{newlightgrey}{$A_{0000}^{0100}\(C_{\smash{Q\tilde{u}Q\tilde{d}}}\)$}\\
					\colorbox{newlightgrey}{$B_{0000}^{0100}\(C_{\smash{Q\tilde{u}Q\tilde{d}}}\)$}&\colorbox{newlightgrey}{$A_{1100}^{0000}\(C_{\smash{Q\tilde{u}Q\tilde{d}}}\)$}&\colorbox{newlightgrey}{$A_{0110}^{0000}\(C_{\smash{Q\tilde{u}Q\tilde{d}}}\)$}\\
					\colorbox{newlightgrey}{$B_{1100}^{0000}\(C_{\smash{Q\tilde{u}Q\tilde{d}}}\)$}&\colorbox{newlightgrey}{$B_{0110}^{0000}\(C_{\smash{Q\tilde{u}Q\tilde{d}}}\)$}&\colorbox{newlightgrey}{$A_{1000}^{0100}\(C_{\smash{Q\tilde{u}Q\tilde{d}}}\)$}\\
					\colorbox{newlightgrey}{$B_{1000}^{0100}\(C_{\smash{Q\tilde{u}Q\tilde{d}}}\)$}&\colorbox{newlightgrey}{$A_{0100}^{1000}\(C_{\smash{Q\tilde{u}Q\tilde{d}}}\)$}&\colorbox{newlightgrey}{$B_{0100}^{1000}\(C_{\smash{Q\tilde{u}Q\tilde{d}}}\)$}\\
					\colorbox{newlightgrey}{$A_{0000}^{1100}\(C_{\smash{Q\tilde{u}Q\tilde{d}}}\)$}&\colorbox{newlightgrey}{$B_{0000}^{1100}\(C_{\smash{Q\tilde{u}Q\tilde{d}}}\)$}&\colorbox{newlightgrey}{$A_{0000}^{0110}\(C_{\smash{Q\tilde{u}Q\tilde{d}}}\)$}\\
					A_{2100}^{0000}\(C_{\smash{Q\tilde{u}Q\tilde{d}}}\)&A_{0120}^{0000}\(C_{\smash{Q\tilde{u}Q\tilde{d}}}\)&B_{2100}^{0000}\(C_{\smash{Q\tilde{u}Q\tilde{d}}}\)\\
					B_{0120}^{0000}\(C_{\smash{Q\tilde{u}Q\tilde{d}}}\)&A_{2000}^{0100}\(C_{\smash{Q\tilde{u}Q\tilde{d}}}\)&B_{2000}^{0100}\(C_{\smash{Q\tilde{u}Q\tilde{d}}}\)\\
					\colorbox{newlightgrey}{$A_{1100}^{1000}\(C_{\smash{Q\tilde{u}Q\tilde{d}}}\)$}&\colorbox{newlightgrey}{$A_{0110}^{1000}\(C_{\smash{Q\tilde{u}Q\tilde{d}}}\)$}&B_{1100}^{1000}\(C_{\smash{Q\tilde{u}Q\tilde{d}}}\)\\
					\colorbox{newlightgrey}{$B_{0110}^{1000}\(C_{\smash{Q\tilde{u}Q\tilde{d}}}\)$}&\colorbox{newlightgrey}{$A_{1000}^{1100}\(C_{\smash{Q\tilde{u}Q\tilde{d}}}\)$}&B_{1000}^{1100}\(C_{\smash{Q\tilde{u}Q\tilde{d}}}\)\\
					A_{0100}^{2000}\(C_{\smash{Q\tilde{u}Q\tilde{d}}}\)&A_{0000}^{2100}\(C_{\smash{Q\tilde{u}Q\tilde{d}}}\)&A_{1000}^{0110}\(C_{\smash{Q\tilde{u}Q\tilde{d}}}\)\\
					A_{1120}^{0000}\(C_{\smash{Q\tilde{u}Q\tilde{d}}}\)&B_{1120}^{0000}\(C_{\smash{Q\tilde{u}Q\tilde{d}}}\)&A_{2100}^{1000}\(C_{\smash{Q\tilde{u}Q\tilde{d}}}\)\\
					A_{0120}^{1000}\(C_{\smash{Q\tilde{u}Q\tilde{d}}}\)&B_{2100}^{1000}\(C_{\smash{Q\tilde{u}Q\tilde{d}}}\)&B_{0120}^{1000}\(C_{\smash{Q\tilde{u}Q\tilde{d}}}\)\\
					A_{2000}^{1100}\(C_{\smash{Q\tilde{u}Q\tilde{d}}}\)&B_{2000}^{1100}\(C_{\smash{Q\tilde{u}Q\tilde{d}}}\)&A_{1100}^{2000}\(C_{\smash{Q\tilde{u}Q\tilde{d}}}\)\\
					A_{0110}^{2000}\(C_{\smash{Q\tilde{u}Q\tilde{d}}}\)&A_{1000}^{2100}\(C_{\smash{Q\tilde{u}Q\tilde{d}}}\)&A_{2000}^{0110}\(C_{\smash{Q\tilde{u}Q\tilde{d}}}\)\\
					A_{1120}^{1000}\(C_{\smash{Q\tilde{u}Q\tilde{d}}}\)&B_{1120}^{1000}\(C_{\smash{Q\tilde{u}Q\tilde{d}}}\)&A_{2100}^{2000}\(C_{\smash{Q\tilde{u}Q\tilde{d}}}\)\\
					A_{0120}^{2000}\(C_{\smash{Q\tilde{u}Q\tilde{d}}}\)&A_{2000}^{2100}\(C_{\smash{Q\tilde{u}Q\tilde{d}}}\)&A_{1120}^{2000}\(C_{\smash{Q\tilde{u}Q\tilde{d}}}\)\\
					A_{0200}^{0000}\(C_{\smash{Q\tilde{u}Q\tilde{d}}}\)&B_{0200}^{0000}\(C_{\smash{Q\tilde{u}Q\tilde{d}}}\)&\colorbox{newlightgrey}{$A_{0100}^{0100}\(C_{\smash{Q\tilde{u}Q\tilde{d}}}\)$}\\
					B_{0100}^{0100}\(C_{\smash{Q\tilde{u}Q\tilde{d}}}\)&A_{0000}^{0200}\(C_{\smash{Q\tilde{u}Q\tilde{d}}}\)&A_{1200}^{0000}\(C_{\smash{Q\tilde{u}Q\tilde{d}}}\)\\
					A_{0210}^{0000}\(C_{\smash{Q\tilde{u}Q\tilde{d}}}\)&B_{1200}^{0000}\(C_{\smash{Q\tilde{u}Q\tilde{d}}}\)&B_{0210}^{0000}\(C_{\smash{Q\tilde{u}Q\tilde{d}}}\)\\
					\colorbox{newlightgrey}{$A_{1100}^{0100}\(C_{\smash{Q\tilde{u}Q\tilde{d}}}\)$}&\colorbox{newlightgrey}{$A_{0110}^{0100}\(C_{\smash{Q\tilde{u}Q\tilde{d}}}\)$}&\colorbox{newlightgrey}{$B_{1100}^{0100}\(C_{\smash{Q\tilde{u}Q\tilde{d}}}\)$}\\
					B_{0110}^{0100}\(C_{\smash{Q\tilde{u}Q\tilde{d}}}\)&A_{1000}^{0200}\(C_{\smash{Q\tilde{u}Q\tilde{d}}}\)&A_{0200}^{1000}\(C_{\smash{Q\tilde{u}Q\tilde{d}}}\)\\
					B_{0200}^{1000}\(C_{\smash{Q\tilde{u}Q\tilde{d}}}\)&A_{0100}^{1100}\(C_{\smash{Q\tilde{u}Q\tilde{d}}}\)&B_{0100}^{1100}\(C_{\smash{Q\tilde{u}Q\tilde{d}}}\)\\
					\colorbox{newlightgrey}{$A_{0100}^{0110}\(C_{\smash{Q\tilde{u}Q\tilde{d}}}\)$}&A_{0000}^{0210}\(C_{\smash{Q\tilde{u}Q\tilde{d}}}\)&\colorbox{newlightgrey}{$A_{2200}^{0000}\(C_{\smash{Q\tilde{u}Q\tilde{d}}}\)$}\\
					\colorbox{newlightgrey}{$A_{0220}^{0000}\(C_{\smash{Q\tilde{u}Q\tilde{d}}}\)$}&\colorbox{newlightgrey}{$B_{2200}^{0000}\(C_{\smash{Q\tilde{u}Q\tilde{d}}}\)$}&\colorbox{newlightgrey}{$B_{0220}^{0000}\(C_{\smash{Q\tilde{u}Q\tilde{d}}}\)$}\\
					A_{2100}^{0100}\(C_{\smash{Q\tilde{u}Q\tilde{d}}}\)&A_{0120}^{0100}\(C_{\smash{Q\tilde{u}Q\tilde{d}}}\)&\colorbox{newlightgrey}{$B_{2100}^{0100}\(C_{\smash{Q\tilde{u}Q\tilde{d}}}\)$}\\
					\colorbox{newlightgrey}{$B_{0120}^{0100}\(C_{\smash{Q\tilde{u}Q\tilde{d}}}\)$}&\colorbox{newlightgrey}{$A_{2000}^{0200}\(C_{\smash{Q\tilde{u}Q\tilde{d}}}\)$}&\colorbox{newlightgrey}{$B_{1200}^{1000}\(C_{\smash{Q\tilde{u}Q\tilde{d}}}\)$}\\
					\colorbox{newlightgrey}{$B_{0210}^{1000}\(C_{\smash{Q\tilde{u}Q\tilde{d}}}\)$}&\colorbox{newlightgrey}{$A_{1100}^{1100}\(C_{\smash{Q\tilde{u}Q\tilde{d}}}\)$}&\colorbox{newlightgrey}{$A_{0110}^{1100}\(C_{\smash{Q\tilde{u}Q\tilde{d}}}\)$}\\
					\colorbox{newlightgrey}{$B_{1100}^{1100}\(C_{\smash{Q\tilde{u}Q\tilde{d}}}\)$}&\colorbox{newlightgrey}{$B_{0110}^{1100}\(C_{\smash{Q\tilde{u}Q\tilde{d}}}\)$}&\colorbox{newlightgrey}{$A_{0200}^{2000}\(C_{\smash{Q\tilde{u}Q\tilde{d}}}\)$}\\
					\colorbox{newlightgrey}{$A_{0100}^{2100}\(C_{\smash{Q\tilde{u}Q\tilde{d}}}\)$}&\colorbox{newlightgrey}{$B_{0000}^{2200}\(C_{\smash{Q\tilde{u}Q\tilde{d}}}\)$}&\colorbox{newlightgrey}{$A_{1100}^{0110}\(C_{\smash{Q\tilde{u}Q\tilde{d}}}\)$}\\
					\colorbox{newlightgrey}{$A_{0110}^{0110}\(C_{\smash{Q\tilde{u}Q\tilde{d}}}\)$}&\colorbox{newlightgrey}{$A_{1000}^{0210}\(C_{\smash{Q\tilde{u}Q\tilde{d}}}\)$}&\colorbox{newlightgrey}{$A_{1220}^{0000}\(C_{\smash{Q\tilde{u}Q\tilde{d}}}\)$}\\
					B_{1220}^{0000}\(C_{\smash{Q\tilde{u}Q\tilde{d}}}\)&A_{1120}^{0100}\(C_{\smash{Q\tilde{u}Q\tilde{d}}}\)&B_{1120}^{0100}\(C_{\smash{Q\tilde{u}Q\tilde{d}}}\)\\
					A_{2200}^{1000}\(C_{\smash{Q\tilde{u}Q\tilde{d}}}\)&A_{0220}^{1000}\(C_{\smash{Q\tilde{u}Q\tilde{d}}}\)&B_{2200}^{1000}\(C_{\smash{Q\tilde{u}Q\tilde{d}}}\)\\
					B_{0220}^{1000}\(C_{\smash{Q\tilde{u}Q\tilde{d}}}\)&A_{2100}^{1100}\(C_{\smash{Q\tilde{u}Q\tilde{d}}}\)&B_{2100}^{1100}\(C_{\smash{Q\tilde{u}Q\tilde{d}}}\)\\
					\colorbox{newlightgrey}{$A_{2000}^{1200}\(C_{\smash{Q\tilde{u}Q\tilde{d}}}\)$}&A_{1200}^{2000}\(C_{\smash{Q\tilde{u}Q\tilde{d}}}\)&A_{0210}^{2000}\(C_{\smash{Q\tilde{u}Q\tilde{d}}}\)\\
					A_{1100}^{2100}\(C_{\smash{Q\tilde{u}Q\tilde{d}}}\)&A_{2100}^{0110}\(C_{\smash{Q\tilde{u}Q\tilde{d}}}\)&A_{0120}^{0110}\(C_{\smash{Q\tilde{u}Q\tilde{d}}}\)\\
					A_{2000}^{0210}\(C_{\smash{Q\tilde{u}Q\tilde{d}}}\)&A_{1220}^{1000}\(C_{\smash{Q\tilde{u}Q\tilde{d}}}\)&B_{1220}^{1000}\(C_{\smash{Q\tilde{u}Q\tilde{d}}}\)\\
					A_{1120}^{1100}\(C_{\smash{Q\tilde{u}Q\tilde{d}}}\)&A_{2200}^{2000}\(C_{\smash{Q\tilde{u}Q\tilde{d}}}\)&A_{0220}^{2000}\(C_{\smash{Q\tilde{u}Q\tilde{d}}}\)\\
					A_{2100}^{2100}\(C_{\smash{Q\tilde{u}Q\tilde{d}}}\)&A_{1120}^{0110}\(C_{\smash{Q\tilde{u}Q\tilde{d}}}\)&A_{1220}^{2000}\(C_{\smash{Q\tilde{u}Q\tilde{d}}}\)\\
					A_{0200}^{0100}\(C_{\smash{Q\tilde{u}Q\tilde{d}}}\)&B_{0200}^{0100}\(C_{\smash{Q\tilde{u}Q\tilde{d}}}\)&A_{0100}^{0200}\(C_{\smash{Q\tilde{u}Q\tilde{d}}}\)\\
					A_{0112}^{0000}\(C_{\smash{Q\tilde{u}Q\tilde{d}}}\)&\colorbox{newlightgrey}{$A_{0122}^{0000}\(C_{\smash{Q\tilde{u}Q\tilde{d}}}\)$}&\colorbox{newlightgrey}{$B_{0122}^{0000}\(C_{\smash{Q\tilde{u}Q\tilde{d}}}\)$}\\
					\colorbox{newlightgrey}{$A_{1220}^{0100}\(C_{\smash{Q\tilde{u}Q\tilde{d}}}\)$}&\colorbox{newlightgrey}{$B_{1220}^{0100}\(C_{\smash{Q\tilde{u}Q\tilde{d}}}\)$}&\colorbox{newlightgrey}{$B_{1120}^{0200}\(C_{\smash{Q\tilde{u}Q\tilde{d}}}\)$}\\
					\colorbox{newlightgrey}{$A_{0122}^{1000}\(C_{\smash{Q\tilde{u}Q\tilde{d}}}\)$}&\colorbox{newlightgrey}{$B_{0122}^{1000}\(C_{\smash{Q\tilde{u}Q\tilde{d}}}\)$}&\colorbox{newlightgrey}{$A_{2200}^{1100}\(C_{\smash{Q\tilde{u}Q\tilde{d}}}\)$}\\
					\colorbox{newlightgrey}{$A_{0220}^{1100}\(C_{\smash{Q\tilde{u}Q\tilde{d}}}\)$}&\colorbox{newlightgrey}{$B_{2200}^{1100}\(C_{\smash{Q\tilde{u}Q\tilde{d}}}\)$}&\colorbox{newlightgrey}{$B_{0220}^{1100}\(C_{\smash{Q\tilde{u}Q\tilde{d}}}\)$}\\
					\colorbox{newlightgrey}{$A_{2100}^{1200}\(C_{\smash{Q\tilde{u}Q\tilde{d}}}\)$}&\colorbox{newlightgrey}{$B_{2100}^{1200}\(C_{\smash{Q\tilde{u}Q\tilde{d}}}\)$}&\colorbox{newlightgrey}{$A_{1200}^{2100}\(C_{\smash{Q\tilde{u}Q\tilde{d}}}\)$}\\
					\colorbox{newlightgrey}{$A_{0210}^{2100}\(C_{\smash{Q\tilde{u}Q\tilde{d}}}\)$}&\colorbox{newlightgrey}{$A_{0110}^{2200}\(C_{\smash{Q\tilde{u}Q\tilde{d}}}\)$}&\colorbox{newlightgrey}{$A_{2200}^{0110}\(C_{\smash{Q\tilde{u}Q\tilde{d}}}\)$}\\
					\colorbox{newlightgrey}{$A_{0220}^{0110}\(C_{\smash{Q\tilde{u}Q\tilde{d}}}\)$}&\colorbox{newlightgrey}{$A_{2000}^{0112}\(C_{\smash{Q\tilde{u}Q\tilde{d}}}\)$}&\colorbox{newlightgrey}{$A_{1220}^{1100}\(C_{\smash{Q\tilde{u}Q\tilde{d}}}\)$}\\
					\colorbox{newlightgrey}{$A_{0112}^{2100}\(C_{\smash{Q\tilde{u}Q\tilde{d}}}\)$}&\colorbox{newlightgrey}{$B_{1122}^{1100}\(C_{\smash{Q\tilde{u}Q\tilde{d}}}\)$}&\colorbox{newlightgrey}{$A_{1220}^{1200}\(C_{\smash{Q\tilde{u}Q\tilde{d}}}\)$}\\
					\colorbox{newlightgrey}{$B_{0122}^{2100}\(C_{\smash{Q\tilde{u}Q\tilde{d}}}\)$}&\colorbox{newlightgrey}{$A_{2200}^{2200}\(C_{\smash{Q\tilde{u}Q\tilde{d}}}\)$}&\colorbox{newlightgrey}{$A_{1122}^{0110}\(C_{\smash{Q\tilde{u}Q\tilde{d}}}\)$}\\
					\colorbox{newlightgrey}{$A_{2100}^{0122}\(C_{\smash{Q\tilde{u}Q\tilde{d}}}\)$}&\colorbox{newlightgrey}{$A_{0220}^{0220}\(C_{\smash{Q\tilde{u}Q\tilde{d}}}\)$}&\colorbox{newlightgrey}{$A_{1122}^{2200}\(C_{\smash{Q\tilde{u}Q\tilde{d}}}\)$}
				\end{matrix}
				\ \right \}$\\\hline
			\end{tabular}%
		}
		\caption{Continuation of Table \ref{tableInv4Fermi1}}
		\label{tableInv4Fermi7}
	\end{table}
	\clearpage
	\section{Some explicit relations for the maximal set of $\mathcal{O}_{HQ}$}\label{section:coeffvalues}
	
	We present here examples of some explicit expressions announced in the main text, focussing on the maximal set of the bilinear hermitian operator $\mathcal{O}_{HQ}$, formed by $\{L_{1100}\left(C_{HQ}\right), $ $L_{2200}\left(C_{HQ}\right),L_{1122}\left(C_{HQ}\right)\}$ (upon trivial redefinitions, the following can be generalized to any bilinear hermitian operator). First, we exhibit the combinations of invariants of \eqref{eq:CHQmaxsetintro} whose leading terms appear in \eqref{eq:CHQmaxsetexpanded1}. Each line of the latter expression corresponds to the following invariants,
	\beq
	\bead
	L_1\ \ & \ ,\\
	\tilde L_2\equiv& \ L_2-\Tr\(X_u\) L_1 \ ,\\
	\tilde L_3\equiv& \  L_3-\Tr\(X_d\) L_1 \ ,\\
	\tilde L_4\equiv& \  L_4+ \frac{1}{2}\[\Tr\(X_u\)\Tr\(X_u\) - \Tr\(X_u^2\)\] L_1 +  \Tr\(X_u\)\tilde L_2 \ ,\\
	\tilde L_5\equiv& \  L_5 - \Tr\(X_u\)\Tr\(X_d\) L_1 - \Tr\(X_d\) \tilde L_2 -   \Tr\(X_u\)\tilde L_3 \ ,\\
	\tilde L_6\equiv& \  L_6 - \frac{1}{2}\[\Tr\(X_d\)\Tr\(X_d\) - \Tr\(X_d^2\)\]L_1 - \Tr\(X_d\) \tilde L_3 -\frac{\Tr\(X_d\)\Tr\(X_d\) - \Tr\(X_d^2\)}{\Tr\(X_u\)\Tr\(X_u\) - \Tr\(X_u^2\)} \tilde L_4 \\
	&- 2 \frac{\Tr\(X_u\)\Tr\(X_d\) - \Tr\(X_uX_d\)}{\Tr\(X_u\)\Tr\(X_u\) - \Tr\(X_u^2\)} \tilde L_5 \ ,\\
	\tilde L_7\equiv& \  L_7- \frac{1}{4}\Big[\Tr\(X_u^2\)\Tr\(X_d^2\) +\Tr\(X_u^2\)\Tr\(X_d\)^2-\Tr\(X_d^2\)\Tr\(X_u\)^2 \\
	&\qquad\qquad+ 4 \Tr\(X_uX_d\) \Tr\(X_u\)\Tr\(X_d\) - \Tr\(X_u\)^2\Tr\(X_d\)^2\Big]L_1 \\
	&- \frac{1}{2}\[\Tr\(X_u\) \Tr\(X_d\)^2-\Tr\(X_u\)\Tr\(X_d^2\)\]\tilde L_2 \\
	&- \frac{1}{2}\[\Tr\(X_u^2\)\Tr\(X_d\) + 2 \Tr\(X_uX_d\)\Tr\(X_u\) -  \Tr\(X_u\)^2\Tr\(X_d\)\]\tilde L_3 \ ,\\
	\tilde L_8\equiv& \  L_8- \frac{1}{2}\Big[\Tr\(X_uX_d\)^2 -\Tr\(\[X_uX_d\]^2\) +2\Tr\(X_u^2X_d\)\Tr\(X_d\)+2\Tr\(X_uX_d^2\)\Tr\(X_u\) \\
	&\qquad\qquad- 2 \Tr\(X_uX_d\) \Tr\(X_u\)\Tr\(X_d\)\Big]L_1 \\
	&- \frac{1}{2}\[\Tr\(X_u\)\Tr\(X_d^2\)-\Tr\(X_u\) \Tr\(X_d\)^2\]\tilde L_2 \\
	&- \frac{1}{2}\[\Tr\(X_u^2\)\Tr\(X_d\) -  \Tr\(X_u\)^2\Tr\(X_d\)\]\tilde L_3+\Tr\(X_u\)\Tr\(X_d\)\tilde L_5 \ , \\
	\tilde L_9\equiv& \  L_9  - \frac{1}{2}\Big[\Tr\(\[X_uX_d\]^2\) -\Tr\(X_uX_d\)^2 +2\Tr\(X_uX_d\)\Tr\(X_u\)\Tr\(X_d\)\Big]L_1 \\
	&- \Tr\(X_uX_d\)\Tr\(X_d\)\tilde L_2-\Tr\(X_uX_d\)\Tr\(X_u\)\tilde L_3+\Tr\(X_u\)\Tr\(X_uX_d\)\tilde L_5 \ .
	\eead
	\eeq 
	Second, we present the explicit decomposition anticipated in \eqref{eq:maxvsminsketchedexample}. One can check that the additional invariant $L_{2100}\left(C_{HQ}\right)$, belonging to the maximal set, can be expressed as:
	\begin{align}
		\cI^I_{2100}L_{2100}^{\vphantom{I}}&=\cI^I_{1100}L_{1100}^{\vphantom{I}}+\cI^I_{2200}L_{2200}^{\vphantom{I}}+\cI^I_{1122}L_{1122}^{\vphantom{I}}+\nonumber\\
		&+J_4\left(R^{\vphantom{R}}_{0000} \cI^R_{0000}+R^{\vphantom{R}}_{0100} \cI^R_{0100}+R^{\vphantom{R}}_{0112} \cI^R_{0112}+R^{\vphantom{R}}_{01122} \cI^R_{01122}+R^{\vphantom{R}}_{0120} \cI^R_{0120}+\right.\nonumber\\
		&\left.+R^{\vphantom{R}}_{0122} \cI^R_{0122}+R^{\vphantom{R}}_{0200} \cI^R_{0200}+R^{\vphantom{R}}_{1000} \cI^R_{1000}+R^{\vphantom{R}}_{1100} \cI^R_{1100}+R^{\vphantom{R}}_{1120} \cI^R_{1120}+R^{\vphantom{R}}_{1122} \cI^R_{1122}+\right.\nonumber\\
		&\left.R^{\vphantom{R}}_{1200} \cI^R_{1200}+R^{\vphantom{R}}_{1220} \cI^R_{1220}+R^{\vphantom{R}}_{1221} \cI^R_{1221}+R^{\vphantom{R}}_{2000} \cI^R_{2000}+R^{\vphantom{R}}_{2112} \cI^R_{2112}+R^{\vphantom{R}}_{2200} \cI^R_{2200}\right) \ ,
		\label{eq:maxvsminpartcase}
	\end{align}
	where $R_{abcd}$ is defined as $L_{abcd}$ in Eq.~\eqref{bilinearFormulaquark} but with $\Re\to\Im$, and the different coefficients are reported in the next subsections.
	
	\subsection{$\cI^{I}$ coefficients}
	\begin{adjustwidth}{}{2.5cm}
	$\cI^I_{1100}=4(8 I_{0,1}^6 I_{1,0}^7-66 I_{0,2}^3 I_{1,0}^7-6 I_{0,1}^2 I_{0,2}^2 I_{1,0}^7+64 I_{0,3}^2 I_{1,0}^7-18 I_{0,1}^3 I_{0,3} I_{1,0}^7+18 I_{0,1} I_{0,2} I_{0,3} I_{1,0}^7-27 I_{0,1}^5 I_{1,1} I_{1,0}^6+297 I_{0,1} I_{0,2}^2 I_{1,1} I_{1,0}^6-192 I_{0,1}^3 I_{0,2} I_{1,1} I_{1,0}^6+198 I_{0,1}^2 I_{0,3} I_{1,1} I_{1,0}^6-276 I_{0,2} I_{0,3} I_{1,1} I_{1,0}^6+42 I_{0,1}^4 I_{1,2} I_{1,0}^6+252 I_{0,2}^2 I_{1,2} I_{1,0}^6+144 I_{0,1}^2 I_{0,2} I_{1,2} I_{1,0}^6-438 I_{0,1} I_{0,3} I_{1,2} I_{1,0}^6-39 I_{0,1}^4 I_{1,1}^2 I_{1,0}^5-135 I_{0,2}^2 I_{1,1}^2 I_{1,0}^5+450 I_{0,1}^2 I_{0,2} I_{1,1}^2 I_{1,0}^5-96 I_{0,1} I_{0,3} I_{1,1}^2 I_{1,0}^5+198 I_{0,1}^2 I_{1,2}^2 I_{1,0}^5-18 I_{0,2} I_{1,2}^2 I_{1,0}^5+72 I_{0,1}^3 I_{1,1} I_{1,2} I_{1,0}^5-1260 I_{0,1} I_{0,2} I_{1,1} I_{1,2} I_{1,0}^5+828 I_{0,3} I_{1,1} I_{1,2} I_{1,0}^5+9 I_{0,1}^6 I_{2,0} I_{1,0}^5+117 I_{0,2}^3 I_{2,0} I_{1,0}^5+225 I_{0,1}^2 I_{0,2}^2 I_{2,0} I_{1,0}^5-342 I_{0,3}^2 I_{2,0} I_{1,0}^5-141 I_{0,1}^4 I_{0,2} I_{2,0} I_{1,0}^5-36 I_{0,1}^3 I_{0,3} I_{2,0} I_{1,0}^5+168 I_{0,1} I_{0,2} I_{0,3} I_{2,0} I_{1,0}^5+15 I_{0,1}^5 I_{2,1} I_{1,0}^5-603 I_{0,1} I_{0,2}^2 I_{2,1} I_{1,0}^5+198 I_{0,1}^3 I_{0,2} I_{2,1} I_{1,0}^5+228 I_{0,1}^2 I_{0,3} I_{2,1} I_{1,0}^5+162 I_{0,2} I_{0,3} I_{2,1} I_{1,0}^5-144 I_{0,1}^4 I_{2,2} I_{1,0}^5-126 I_{0,2}^2 I_{2,2} I_{1,0}^5+108 I_{0,1}^2 I_{0,2} I_{2,2} I_{1,0}^5+162 I_{0,1} I_{0,3} I_{2,2} I_{1,0}^5+240 I_{0,1}^3 I_{1,1}^3 I_{1,0}^4-240 I_{0,3} I_{1,1}^3 I_{1,0}^4-720 I_{1,2}^3 I_{1,0}^4+1440 I_{0,1} I_{1,1} I_{1,2}^2 I_{1,0}^4-1080 I_{0,1}^2 I_{1,1}^2 I_{1,2} I_{1,0}^4+360 I_{0,2} I_{1,1}^2 I_{1,2} I_{1,0}^4-246 I_{0,1}^5 I_{1,1} I_{2,0} I_{1,0}^4-1062 I_{0,1} I_{0,2}^2 I_{1,1} I_{2,0} I_{1,0}^4+1296 I_{0,1}^3 I_{0,2} I_{1,1} I_{2,0} I_{1,0}^4-456 I_{0,1}^2 I_{0,3} I_{1,1} I_{2,0} I_{1,0}^4+468 I_{0,2} I_{0,3} I_{1,1} I_{2,0} I_{1,0}^4+360 I_{0,1}^4 I_{1,2} I_{2,0} I_{1,0}^4+36 I_{0,2}^2 I_{1,2} I_{2,0} I_{1,0}^4-1728 I_{0,1}^2 I_{0,2} I_{1,2} I_{2,0} I_{1,0}^4+1332 I_{0,1} I_{0,3} I_{1,2} I_{2,0} I_{1,0}^4+531 I_{0,1}^4 I_{1,1} I_{2,1} I_{1,0}^4+603 I_{0,2}^2 I_{1,1} I_{2,1} I_{1,0}^4-1242 I_{0,1}^2 I_{0,2} I_{1,1} I_{2,1} I_{1,0}^4-1584 I_{0,1} I_{0,3} I_{1,1} I_{2,1} I_{1,0}^4-864 I_{0,1}^3 I_{1,2} I_{2,1} I_{1,0}^4+2934 I_{0,1} I_{0,2} I_{1,2} I_{2,1} I_{1,0}^4-378 I_{0,3} I_{1,2} I_{2,1} I_{1,0}^4+108 I_{0,1}^3 I_{1,1} I_{2,2} I_{1,0}^4+990 I_{0,1} I_{0,2} I_{1,1} I_{2,2} I_{1,0}^4-54 I_{0,3} I_{1,1} I_{2,2} I_{1,0}^4-666 I_{0,1}^2 I_{1,2} I_{2,2} I_{1,0}^4-378 I_{0,2} I_{1,2} I_{2,2} I_{1,0}^4-63 I_{0,1}^6 I_{3,0} I_{1,0}^4-39 I_{0,2}^3 I_{3,0} I_{1,0}^4-231 I_{0,1}^2 I_{0,2}^2 I_{3,0} I_{1,0}^4+242 I_{0,3}^2 I_{3,0} I_{1,0}^4+279 I_{0,1}^4 I_{0,2} I_{3,0} I_{1,0}^4-44 I_{0,1}^3 I_{0,3} I_{3,0} I_{1,0}^4-144 I_{0,1} I_{0,2} I_{0,3} I_{3,0} I_{1,0}^4-6 I_{0,1}^6 I_{2,0}^2 I_{1,0}^3+54 I_{0,2}^3 I_{2,0}^2 I_{1,0}^3-432 I_{0,1}^2 I_{0,2}^2 I_{2,0}^2 I_{1,0}^3+264 I_{0,3}^2 I_{2,0}^2 I_{1,0}^3+144 I_{0,1}^4 I_{0,2} I_{2,0}^2 I_{1,0}^3+174 I_{0,1}^3 I_{0,3} I_{2,0}^2 I_{1,0}^3-198 I_{0,1} I_{0,2} I_{0,3} I_{2,0}^2 I_{1,0}^3-504 I_{0,1}^4 I_{2,1}^2 I_{1,0}^3-432 I_{0,2}^2 I_{2,1}^2 I_{1,0}^3+216 I_{0,1}^2 I_{0,2} I_{2,1}^2 I_{1,0}^3+2232 I_{0,1} I_{0,3} I_{2,1}^2 I_{1,0}^3+864 I_{0,1}^2 I_{2,2}^2 I_{1,0}^3-648 I_{0,2} I_{2,2}^2 I_{1,0}^3+558 I_{0,1}^4 I_{1,1}^2 I_{2,0} I_{1,0}^3+270 I_{0,2}^2 I_{1,1}^2 I_{2,0} I_{1,0}^3-1404 I_{0,1}^2 I_{0,2} I_{1,1}^2 I_{2,0} I_{1,0}^3+576 I_{0,1} I_{0,3} I_{1,1}^2 I_{2,0} I_{1,0}^3+540 I_{0,1}^2 I_{1,2}^2 I_{2,0} I_{1,0}^3-540 I_{0,2} I_{1,2}^2 I_{2,0} I_{1,0}^3-1188 I_{0,1}^3 I_{1,1} I_{1,2} I_{2,0} I_{1,0}^3+2484 I_{0,1} I_{0,2} I_{1,1} I_{1,2} I_{2,0} I_{1,0}^3-1296 I_{0,3} I_{1,1} I_{1,2} I_{2,0} I_{1,0}^3-1944 I_{0,1}^3 I_{1,1}^2 I_{2,1} I_{1,0}^3+1944 I_{0,3} I_{1,1}^2 I_{2,1} I_{1,0}^3-3456 I_{0,1} I_{1,2}^2 I_{2,1} I_{1,0}^3+5508 I_{0,1}^2 I_{1,1} I_{1,2} I_{2,1} I_{1,0}^3-2052 I_{0,2} I_{1,1} I_{1,2} I_{2,1} I_{1,0}^3+252 I_{0,1}^5 I_{2,0} I_{2,1} I_{1,0}^3+1728 I_{0,1} I_{0,2}^2 I_{2,0} I_{2,1} I_{1,0}^3-1188 I_{0,1}^3 I_{0,2} I_{2,0} I_{2,1} I_{1,0}^3-576 I_{0,1}^2 I_{0,3} I_{2,0} I_{2,1} I_{1,0}^3-216 I_{0,2} I_{0,3} I_{2,0} I_{2,1} I_{1,0}^3+1296 I_{0,1}^2 I_{1,1}^2 I_{2,2} I_{1,0}^3-216 I_{0,2} I_{1,1}^2 I_{2,2} I_{1,0}^3+1944 I_{1,2}^2 I_{2,2} I_{1,0}^3-3024 I_{0,1} I_{1,1} I_{1,2} I_{2,2} I_{1,0}^3-432 I_{0,2}^2 I_{2,0} I_{2,2} I_{1,0}^3+648 I_{0,1}^2 I_{0,2} I_{2,0} I_{2,2} I_{1,0}^3-216 I_{0,1} I_{0,3} I_{2,0} I_{2,2} I_{1,0}^3+468 I_{0,1}^3 I_{2,1} I_{2,2} I_{1,0}^3-1188 I_{0,1} I_{0,2} I_{2,1} I_{2,2} I_{1,0}^3-1008 I_{0,3} I_{2,1} I_{2,2} I_{1,0}^3+288 I_{0,1}^5 I_{1,1} I_{3,0} I_{1,0}^3+108 I_{0,1} I_{0,2}^2 I_{1,1} I_{3,0} I_{1,0}^3-996 I_{0,1}^3 I_{0,2} I_{1,1} I_{3,0} I_{1,0}^3+756 I_{0,1}^2 I_{0,3} I_{1,1} I_{3,0} I_{1,0}^3-156 I_{0,2} I_{0,3} I_{1,1} I_{3,0} I_{1,0}^3-420 I_{0,1}^4 I_{1,2} I_{3,0} I_{1,0}^3+180 I_{0,2}^2 I_{1,2} I_{3,0} I_{1,0}^3+1368 I_{0,1}^2 I_{0,2} I_{1,2} I_{3,0} I_{1,0}^3-1128 I_{0,1} I_{0,3} I_{1,2} I_{3,0} I_{1,0}^3+189 I_{0,1}^5 I_{1,1} I_{2,0}^2 I_{1,0}^2+1053 I_{0,1} I_{0,2}^2 I_{1,1} I_{2,0}^2 I_{1,0}^2-900 I_{0,1}^3 I_{0,2} I_{1,1} I_{2,0}^2 I_{1,0}^2-162 I_{0,1}^2 I_{0,3} I_{1,1} I_{2,0}^2 I_{1,0}^2-180 I_{0,2} I_{0,3} I_{1,1} I_{2,0}^2 I_{1,0}^2-198 I_{0,1}^4 I_{1,2} I_{2,0}^2 I_{1,0}^2-540 I_{0,2}^2 I_{1,2} I_{2,0}^2 I_{1,0}^2+1080 I_{0,1}^2 I_{0,2} I_{1,2} I_{2,0}^2 I_{1,0}^2-342 I_{0,1} I_{0,3} I_{1,2} I_{2,0}^2 I_{1,0}^2+3384 I_{0,1}^3 I_{1,1} I_{2,1}^2 I_{1,0}^2-3384 I_{0,3} I_{1,1} I_{2,1}^2 I_{1,0}^2-4104 I_{0,1}^2 I_{1,2} I_{2,1}^2 I_{1,0}^2+2052 I_{0,2} I_{1,2} I_{2,1}^2 I_{1,0}^2+324 I_{0,1} I_{1,1} I_{2,2}^2 I_{1,0}^2+1512 I_{1,2} I_{2,2}^2 I_{1,0}^2-1746 I_{0,1}^4 I_{1,1} I_{2,0} I_{2,1} I_{1,0}^2-1026 I_{0,2}^2 I_{1,1} I_{2,0} I_{2,1} I_{1,0}^2+4104 I_{0,1}^2 I_{0,2} I_{1,1} I_{2,0} I_{2,1} I_{1,0}^2+720 I_{0,1} I_{0,3} I_{1,1} I_{2,0} I_{2,1} I_{1,0}^2+1800 I_{0,1}^3 I_{1,2} I_{2,0} I_{2,1} I_{1,0}^2-5184 I_{0,1} I_{0,2} I_{1,2} I_{2,0} I_{2,1} I_{1,0}^2+1332 I_{0,3} I_{1,2} I_{2,0} I_{2,1} I_{1,0}^2+828 I_{0,1}^3 I_{1,1} I_{2,0} I_{2,2} I_{1,0}^2-2268 I_{0,1} I_{0,2} I_{1,1} I_{2,0} I_{2,2} I_{1,0}^2-612 I_{0,3} I_{1,1} I_{2,0} I_{2,2} I_{1,0}^2-216 I_{0,1}^2 I_{1,2} I_{2,0} I_{2,2} I_{1,0}^2+2268 I_{0,2} I_{1,2} I_{2,0} I_{2,2} I_{1,0}^2-5292 I_{0,1}^2 I_{1,1} I_{2,1} I_{2,2} I_{1,0}^2+1080 I_{0,2} I_{1,1} I_{2,1} I_{2,2} I_{1,0}^2+4428 I_{0,1} I_{1,2} I_{2,1} I_{2,2} I_{1,0}^2-96 I_{0,1}^4 I_{1,1}^2 I_{3,0} I_{1,0}^2+360 I_{0,1}^2 I_{0,2} I_{1,1}^2 I_{3,0} I_{1,0}^2-984 I_{0,1} I_{0,3} I_{1,1}^2 I_{3,0} I_{1,0}^2+72 I_{0,1}^2 I_{1,2}^2 I_{3,0} I_{1,0}^2-792 I_{0,2} I_{1,2}^2 I_{3,0} I_{1,0}^2-72 I_{0,1}^3 I_{1,1} I_{1,2} I_{3,0} I_{1,0}^2+720 I_{0,1} I_{0,2} I_{1,1} I_{1,2} I_{3,0} I_{1,0}^2+792 I_{0,3} I_{1,1} I_{1,2} I_{3,0} I_{1,0}^2+36 I_{0,1}^6 I_{2,0} I_{3,0} I_{1,0}^2+18 I_{0,2}^3 I_{2,0} I_{3,0} I_{1,0}^2+612 I_{0,1}^2 I_{0,2}^2 I_{2,0} I_{3,0} I_{1,0}^2-108 I_{0,3}^2 I_{2,0} I_{3,0} I_{1,0}^2-318 I_{0,1}^4 I_{0,2} I_{2,0} I_{3,0} I_{1,0}^2+36 I_{0,1}^3 I_{0,3} I_{2,0} I_{3,0} I_{1,0}^2-276 I_{0,1} I_{0,2} I_{0,3} I_{2,0} I_{3,0} I_{1,0}^2-114 I_{0,1}^5 I_{2,1} I_{3,0} I_{1,0}^2-666 I_{0,1} I_{0,2}^2 I_{2,1} I_{3,0} I_{1,0}^2+1008 I_{0,1}^3 I_{0,2} I_{2,1} I_{3,0} I_{1,0}^2-912 I_{0,1}^2 I_{0,3} I_{2,1} I_{3,0} I_{1,0}^2+684 I_{0,2} I_{0,3} I_{2,1} I_{3,0} I_{1,0}^2+324 I_{0,1}^4 I_{2,2} I_{3,0} I_{1,0}^2+612 I_{0,2}^2 I_{2,2} I_{3,0} I_{1,0}^2-1620 I_{0,1}^2 I_{0,2} I_{2,2} I_{3,0} I_{1,0}^2+684 I_{0,1} I_{0,3} I_{2,2} I_{3,0} I_{1,0}^2+15 I_{0,1}^6 I_{2,0}^3 I_{1,0}-135 I_{0,2}^3 I_{2,0}^3 I_{1,0}+135 I_{0,1}^2 I_{0,2}^2 I_{2,0}^3 I_{1,0}-30 I_{0,3}^2 I_{2,0}^3 I_{1,0}-45 I_{0,1}^4 I_{0,2} I_{2,0}^3 I_{1,0}-120 I_{0,1}^3 I_{0,3} I_{2,0}^3 I_{1,0}+180 I_{0,1} I_{0,2} I_{0,3} I_{2,0}^3 I_{1,0}-2016 I_{0,1}^3 I_{2,1}^3 I_{1,0}+2016 I_{0,3} I_{2,1}^3 I_{1,0}-1728 I_{2,2}^3 I_{1,0}-135 I_{0,1}^4 I_{1,1}^2 I_{2,0}^2 I_{1,0}-135 I_{0,2}^2 I_{1,1}^2 I_{2,0}^2 I_{1,0}+270 I_{0,1}^2 I_{0,2} I_{1,1}^2 I_{2,0}^2 I_{1,0}-270 I_{0,1}^2 I_{1,2}^2 I_{2,0}^2 I_{1,0}+270 I_{0,2} I_{1,2}^2 I_{2,0}^2 I_{1,0}+360 I_{0,1}^3 I_{1,1} I_{1,2} I_{2,0}^2 I_{1,0}-540 I_{0,1} I_{0,2} I_{1,1} I_{1,2} I_{2,0}^2 I_{1,0}+180 I_{0,3} I_{1,1} I_{1,2} I_{2,0}^2 I_{1,0}-540 I_{1,2}^2 I_{2,1}^2 I_{1,0}+1764 I_{0,1}^4 I_{2,0} I_{2,1}^2 I_{1,0}+756 I_{0,2}^2 I_{2,0} I_{2,1}^2 I_{1,0}-2484 I_{0,1}^2 I_{0,2} I_{2,0} I_{2,1}^2 I_{1,0}-2088 I_{0,1} I_{0,3} I_{2,0} I_{2,1}^2 I_{1,0}-540 I_{1,1}^2 I_{2,2}^2 I_{1,0}-324 I_{0,1}^2 I_{2,0} I_{2,2}^2 I_{1,0}+216 I_{0,2} I_{2,0} I_{2,2}^2 I_{1,0}-2376 I_{0,1} I_{2,1} I_{2,2}^2 I_{1,0}-26 I_{0,1}^6 I_{3,0}^2 I_{1,0}-30 I_{0,2}^3 I_{3,0}^2 I_{1,0}-222 I_{0,1}^2 I_{0,2}^2 I_{3,0}^2 I_{1,0}-36 I_{0,3}^2 I_{3,0}^2 I_{1,0}+162 I_{0,1}^4 I_{0,2} I_{3,0}^2 I_{1,0}-100 I_{0,1}^3 I_{0,3} I_{3,0}^2 I_{1,0}+252 I_{0,1} I_{0,2} I_{0,3} I_{3,0}^2 I_{1,0}-423 I_{0,1}^5 I_{2,0}^2 I_{2,1} I_{1,0}-1269 I_{0,1} I_{0,2}^2 I_{2,0}^2 I_{2,1} I_{1,0}+1170 I_{0,1}^3 I_{0,2} I_{2,0}^2 I_{2,1} I_{1,0}+720 I_{0,1}^2 I_{0,3} I_{2,0}^2 I_{2,1} I_{1,0}-198 I_{0,2} I_{0,3} I_{2,0}^2 I_{2,1} I_{1,0}+360 I_{0,1}^3 I_{1,1}^2 I_{2,0} I_{2,1} I_{1,0}-360 I_{0,3} I_{1,1}^2 I_{2,0} I_{2,1} I_{1,0}+1080 I_{0,1} I_{1,2}^2 I_{2,0} I_{2,1} I_{1,0}-1080 I_{0,1}^2 I_{1,1} I_{1,2} I_{2,0} I_{2,1} I_{1,0}+36 I_{0,1}^4 I_{2,0}^2 I_{2,2} I_{1,0}+594 I_{0,2}^2 I_{2,0}^2 I_{2,2} I_{1,0}-108 I_{0,1}^2 I_{0,2} I_{2,0}^2 I_{2,2} I_{1,0}-522 I_{0,1} I_{0,3} I_{2,0}^2 I_{2,2} I_{1,0}+5184 I_{0,1}^2 I_{2,1}^2 I_{2,2} I_{1,0}-1080 I_{0,2} I_{2,1}^2 I_{2,2} I_{1,0}-540 I_{0,1}^2 I_{1,1}^2 I_{2,0} I_{2,2} I_{1,0}+540 I_{0,2} I_{1,1}^2 I_{2,0} I_{2,2} I_{1,0}-1080 I_{1,2}^2 I_{2,0} I_{2,2} I_{1,0}+1080 I_{0,1} I_{1,1} I_{1,2} I_{2,0} I_{2,2} I_{1,0}+1080 I_{1,1} I_{1,2} I_{2,1} I_{2,2} I_{1,0}-2160 I_{0,1}^3 I_{2,0} I_{2,1} I_{2,2} I_{1,0}+3240 I_{0,1} I_{0,2} I_{2,0} I_{2,1} I_{2,2} I_{1,0}+1080 I_{0,3} I_{2,0} I_{2,1} I_{2,2} I_{1,0}-240 I_{0,1}^3 I_{1,1}^3 I_{3,0} I_{1,0}+240 I_{0,3} I_{1,1}^3 I_{3,0} I_{1,0}+720 I_{1,2}^3 I_{3,0} I_{1,0}-1440 I_{0,1} I_{1,1} I_{1,2}^2 I_{3,0} I_{1,0}+1080 I_{0,1}^2 I_{1,1}^2 I_{1,2} I_{3,0} I_{1,0}-360 I_{0,2} I_{1,1}^2 I_{1,2} I_{3,0} I_{1,0}+48 I_{0,1}^5 I_{1,1} I_{2,0} I_{3,0} I_{1,0}-504 I_{0,1} I_{0,2}^2 I_{1,1} I_{2,0} I_{3,0} I_{1,0}+144 I_{0,1}^3 I_{0,2} I_{1,1} I_{2,0} I_{3,0} I_{1,0}+60 I_{0,1}^2 I_{0,3} I_{1,1} I_{2,0} I_{3,0} I_{1,0}+252 I_{0,2} I_{0,3} I_{1,1} I_{2,0} I_{3,0} I_{1,0}-36 I_{0,2}^2 I_{1,2} I_{2,0} I_{3,0} I_{1,0}-324 I_{0,1}^2 I_{0,2} I_{1,2} I_{2,0} I_{3,0} I_{1,0}+360 I_{0,1} I_{0,3} I_{1,2} I_{2,0} I_{3,0} I_{1,0}-360 I_{0,1}^4 I_{1,1} I_{2,1} I_{3,0} I_{1,0}+180 I_{0,2}^2 I_{1,1} I_{2,1} I_{3,0} I_{1,0}-1404 I_{0,1}^2 I_{0,2} I_{1,1} I_{2,1} I_{3,0} I_{1,0}+2304 I_{0,1} I_{0,3} I_{1,1} I_{2,1} I_{3,0} I_{1,0}+756 I_{0,1}^3 I_{1,2} I_{2,1} I_{3,0} I_{1,0}+684 I_{0,1} I_{0,2} I_{1,2} I_{2,1} I_{3,0} I_{1,0}-2160 I_{0,3} I_{1,2} I_{2,1} I_{3,0} I_{1,0}-216 I_{0,1}^3 I_{1,1} I_{2,2} I_{3,0} I_{1,0}+1656 I_{0,1} I_{0,2} I_{1,1} I_{2,2} I_{3,0} I_{1,0}+432 I_{0,3} I_{1,1} I_{2,2} I_{3,0} I_{1,0}+288 I_{0,1}^2 I_{1,2} I_{2,2} I_{3,0} I_{1,0}-2160 I_{0,2} I_{1,2} I_{2,2} I_{3,0} I_{1,0}-648 I_{0,1}^3 I_{1,1} I_{2,0} I_{2,1}^2+648 I_{0,3} I_{1,1} I_{2,0} I_{2,1}^2+972 I_{0,1}^2 I_{1,2} I_{2,0} I_{2,1}^2+972 I_{0,1} I_{1,1} I_{2,0} I_{2,2}^2-1944 I_{1,2} I_{2,0} I_{2,2}^2+1944 I_{1,1} I_{2,1} I_{2,2}^2+72 I_{0,1}^5 I_{1,1} I_{3,0}^2+108 I_{0,1} I_{0,2}^2 I_{1,1} I_{3,0}^2-324 I_{0,1}^3 I_{0,2} I_{1,1} I_{3,0}^2+252 I_{0,1}^2 I_{0,3} I_{1,1} I_{3,0}^2-108 I_{0,2} I_{0,3} I_{1,1} I_{3,0}^2-108 I_{0,1}^4 I_{1,2} I_{3,0}^2+108 I_{0,2}^2 I_{1,2} I_{3,0}^2+432 I_{0,1}^2 I_{0,2} I_{1,2} I_{3,0}^2-432 I_{0,1} I_{0,3} I_{1,2} I_{3,0}^2+243 I_{0,1}^4 I_{1,1} I_{2,0}^2 I_{2,1}+243 I_{0,2}^2 I_{1,1} I_{2,0}^2 I_{2,1}-486 I_{0,1}^2 I_{0,2} I_{1,1} I_{2,0}^2 I_{2,1}-324 I_{0,1}^3 I_{1,2} I_{2,0}^2 I_{2,1}+486 I_{0,1} I_{0,2} I_{1,2} I_{2,0}^2 I_{2,1}-162 I_{0,3} I_{1,2} I_{2,0}^2 I_{2,1}+324 I_{0,1}^3 I_{1,1} I_{2,0}^2 I_{2,2}-486 I_{0,1} I_{0,2} I_{1,1} I_{2,0}^2 I_{2,2}+162 I_{0,3} I_{1,1} I_{2,0}^2 I_{2,2}-486 I_{0,1}^2 I_{1,2} I_{2,0}^2 I_{2,2}+486 I_{0,2} I_{1,2} I_{2,0}^2 I_{2,2}-1944 I_{1,2} I_{2,1}^2 I_{2,2}-972 I_{0,2} I_{1,1} I_{2,0} I_{2,1} I_{2,2}+972 I_{0,1} I_{1,2} I_{2,0} I_{2,1} I_{2,2}+27 I_{0,1}^6 I_{2,0}^2 I_{3,0}+81 I_{0,2}^3 I_{2,0}^2 I_{3,0}-81 I_{0,1}^2 I_{0,2}^2 I_{2,0}^2 I_{3,0}-54 I_{0,3}^2 I_{2,0}^2 I_{3,0}-81 I_{0,1}^4 I_{0,2} I_{2,0}^2 I_{3,0}+108 I_{0,1}^3 I_{0,3} I_{2,0}^2 I_{3,0}-108 I_{0,1}^4 I_{2,1}^2 I_{3,0}-324 I_{0,2}^2 I_{2,1}^2 I_{3,0}+1296 I_{0,1}^2 I_{0,2} I_{2,1}^2 I_{3,0}-864 I_{0,1} I_{0,3} I_{2,1}^2 I_{3,0}+1296 I_{0,2} I_{2,2}^2 I_{3,0}-108 I_{0,1}^4 I_{1,1}^2 I_{2,0} I_{3,0}+324 I_{0,1}^2 I_{0,2} I_{1,1}^2 I_{2,0} I_{3,0}-216 I_{0,1} I_{0,3} I_{1,1}^2 I_{2,0} I_{3,0}+108 I_{0,1}^3 I_{1,1} I_{1,2} I_{2,0} I_{3,0}-$
	\\$324 I_{0,1} I_{0,2} I_{1,1} I_{1,2} I_{2,0} I_{3,0}+216 I_{0,3} I_{1,1} I_{1,2} I_{2,0} I_{3,0}+864 I_{0,1}^3 I_{1,1}^2 I_{2,1} I_{3,0}-864 I_{0,3} I_{1,1}^2 I_{2,1} I_{3,0}+1296 I_{0,1} I_{1,2}^2 I_{2,1} I_{3,0}-2268 I_{0,1}^2 I_{1,1} I_{1,2} I_{2,1} I_{3,0}+972 I_{0,2} I_{1,1} I_{1,2} I_{2,1} I_{3,0}-54 I_{0,1}^5 I_{2,0} I_{2,1} I_{3,0}+810 I_{0,1} I_{0,2}^2 I_{2,0} I_{2,1} I_{3,0}-216 I_{0,1}^3 I_{0,2} I_{2,0} I_{2,1} I_{3,0}-108 I_{0,1}^2 I_{0,3} I_{2,0} I_{2,1} I_{3,0}-432 I_{0,2} I_{0,3} I_{2,0} I_{2,1} I_{3,0}+324 I_{0,1}^2 I_{1,1}^2 I_{2,2} I_{3,0}324 I_{0,2} I_{1,1}^2 I_{2,2} I_{3,0}+1296 I_{1,2}^2 I_{2,2} I_{3,0}-1296 I_{0,1} I_{1,1} I_{1,2} I_{2,2} I_{3,0}+108 I_{0,1}^4 I_{2,0} I_{2,2} I_{3,0}-648 I_{0,2}^2 I_{2,0} I_{2,2} I_{3,0}+540 I_{0,1} I_{0,3} I_{2,0} I_{2,2} I_{3,0}-1944 I_{0,1} I_{0,2} I_{2,1} I_{2,2} I_{3,0}+648 I_{0,3} I_{2,1} I_{2,2} I_{3,0})$
	\vspace*{0.1cm}
	
	\noindent
	$\cI^I_{2200}=4 (54 I_{2,0}^3 I_{0,1}^6-54 I_{1,0}^2 I_{2,0}^2 I_{0,1}^6-18 I_{3,0}^2 I_{0,1}^6-54 I_{1,0}^4 I_{2,0} I_{0,1}^6+180 I_{1,0}^3 I_{3,0} I_{0,1}^6-108 I_{1,0} I_{2,0} I_{3,0} I_{0,1}^6+540 I_{1,0}^3 I_{1,1} I_{2,0} I_{0,1}^5-162 I_{2,0}^2 I_{2,1} I_{0,1}^5-162 I_{1,0}^2 I_{2,0} I_{2,1} I_{0,1}^5-1134 I_{1,0}^2 I_{1,1} I_{3,0} I_{0,1}^5+270 I_{1,1} I_{2,0} I_{3,0} I_{0,1}^5+648 I_{1,0} I_{2,1} I_{3,0} I_{0,1}^5-54 I_{0,2} I_{1,0}^6 I_{0,1}^4-54 I_{1,0}^4 I_{1,1}^2 I_{0,1}^4-162 I_{0,2} I_{1,0}^2 I_{2,0}^2 I_{0,1}^4+162 I_{1,1}^2 I_{2,0}^2 I_{0,1}^4+162 I_{1,0} I_{1,2} I_{2,0}^2 I_{0,1}^4+486 I_{1,0}^2 I_{2,1}^2 I_{0,1}^4-810 I_{2,0} I_{2,1}^2 I_{0,1}^4+54 I_{0,2} I_{3,0}^2 I_{0,1}^4+54 I_{1,0}^5 I_{1,2} I_{0,1}^4+378 I_{0,2} I_{1,0}^4 I_{2,0} I_{0,1}^4-972 I_{1,0}^2 I_{1,1}^2 I_{2,0} I_{0,1}^4-972 I_{1,0}^3 I_{1,2} I_{2,0} I_{0,1}^4-486 I_{1,0}^3 I_{1,1} I_{2,1} I_{0,1}^4+1458 I_{1,0} I_{1,1} I_{2,0} I_{2,1} I_{0,1}^4+324 I_{2,0}^2 I_{2,2} I_{0,1}^4+324 I_{1,0}^2 I_{2,0} I_{2,2} I_{0,1}^4-486 I_{0,2} I_{1,0}^3 I_{3,0} I_{0,1}^4+1512 I_{1,0} I_{1,1}^2 I_{3,0} I_{0,1}^4+1404 I_{1,0}^2 I_{1,2} I_{3,0} I_{0,1}^4+270 I_{0,2} I_{1,0} I_{2,0} I_{3,0} I_{0,1}^4-324 I_{1,2} I_{2,0} I_{3,0} I_{0,1}^4-1296 I_{1,1} I_{2,1} I_{3,0} I_{0,1}^4-972 I_{1,0} I_{2,2} I_{3,0} I_{0,1}^4+90 I_{0,3} I_{1,0}^6 I_{0,1}^3-108 I_{0,3} I_{2,0}^3 I_{0,1}^3+2592 I_{2,1}^3 I_{0,1}^3-54 I_{0,3} I_{1,0}^2 I_{2,0}^2 I_{0,1}^3+810 I_{0,2} I_{1,0} I_{1,1} I_{2,0}^2 I_{0,1}^3-648 I_{1,1} I_{1,2} I_{2,0}^2 I_{0,1}^3-3888 I_{1,0} I_{1,1} I_{2,1}^2 I_{0,1}^3-144 I_{0,3} I_{3,0}^2 I_{0,1}^3+486 I_{0,2} I_{1,0}^5 I_{1,1} I_{0,1}^3-324 I_{1,0}^4 I_{1,1} I_{1,2} I_{0,1}^3-108 I_{0,3} I_{1,0}^4 I_{2,0} I_{0,1}^3-2268 I_{0,2} I_{1,0}^3 I_{1,1} I_{2,0} I_{0,1}^3+2916 I_{1,0}^2 I_{1,1} I_{1,2} I_{2,0} I_{0,1}^3-648 I_{0,2} I_{1,0}^4 I_{2,1} I_{0,1}^3+1944 I_{1,0}^2 I_{1,1}^2 I_{2,1} I_{0,1}^3-648 I_{0,2} I_{2,0}^2 I_{2,1} I_{0,1}^3+1620 I_{1,0}^3 I_{1,2} I_{2,1} I_{0,1}^3+1944 I_{0,2} I_{1,0}^2 I_{2,0} I_{2,1} I_{0,1}^3-648 I_{1,1}^2 I_{2,0} I_{2,1} I_{0,1}^3-2916 I_{1,0} I_{1,2} I_{2,0} I_{2,1} I_{0,1}^3+648 I_{1,0}^3 I_{1,1} I_{2,2} I_{0,1}^3-1620 I_{1,0} I_{1,1} I_{2,0} I_{2,2} I_{0,1}^3-1620 I_{1,0}^2 I_{2,1} I_{2,2} I_{0,1}^3+1944 I_{2,0} I_{2,1} I_{2,2} I_{0,1}^3-108 I_{0,3} I_{1,0}^3 I_{3,0} I_{0,1}^3+1944 I_{0,2} I_{1,0}^2 I_{1,1} I_{3,0} I_{0,1}^3-2592 I_{1,0} I_{1,1} I_{1,2} I_{3,0} I_{0,1}^3+432 I_{0,3} I_{1,0} I_{2,0} I_{3,0} I_{0,1}^3-1620 I_{0,2} I_{1,0} I_{2,1} I_{3,0} I_{0,1}^3+1296 I_{1,2} I_{2,1} I_{3,0} I_{0,1}^3+1296 I_{1,1} I_{2,2} I_{3,0} I_{0,1}^3-810 I_{0,2} I_{1,0}^4 I_{1,1}^2 I_{0,1}^2+162 I_{1,0}^4 I_{1,2}^2 I_{0,1}^2+486 I_{0,2}^2 I_{1,0}^2 I_{2,0}^2 I_{0,1}^2-486 I_{0,2} I_{1,1}^2 I_{2,0}^2 I_{0,1}^2+486 I_{1,2}^2 I_{2,0}^2 I_{0,1}^2-486 I_{0,3} I_{1,0} I_{1,1} I_{2,0}^2 I_{0,1}^2-972 I_{0,2} I_{1,0} I_{1,2} I_{2,0}^2 I_{0,1}^2-972 I_{0,2} I_{1,0}^2 I_{2,1}^2 I_{0,1}^2+6804 I_{1,0} I_{1,2} I_{2,1}^2 I_{0,1}^2+1944 I_{0,2} I_{2,0} I_{2,1}^2 I_{0,1}^2+162 I_{0,2}^2 I_{3,0}^2 I_{0,1}^2-810 I_{0,3} I_{1,0}^5 I_{1,1} I_{0,1}^2-324 I_{0,2} I_{1,0}^5 I_{1,2} I_{0,1}^2+1296 I_{1,0}^3 I_{1,1}^2 I_{1,2} I_{0,1}^2-324 I_{0,2}^2 I_{1,0}^4 I_{2,0} I_{0,1}^2+1944 I_{0,2} I_{1,0}^2 I_{1,1}^2 I_{2,0} I_{0,1}^2-1944 I_{1,0}^2 I_{1,2}^2 I_{2,0} I_{0,1}^2+1728 I_{0,3} I_{1,0}^3 I_{1,1} I_{2,0} I_{0,1}^2+2916 I_{0,2} I_{1,0}^3 I_{1,2} I_{2,0} I_{0,1}^2+324 I_{0,3} I_{1,0}^4 I_{2,1} I_{0,1}^2-324 I_{0,3} I_{2,0}^2 I_{2,1} I_{0,1}^2+2592 I_{0,2} I_{1,0}^3 I_{1,1} I_{2,1} I_{0,1}^2-8748 I_{1,0}^2 I_{1,1} I_{1,2} I_{2,1} I_{0,1}^2-324 I_{0,3} I_{1,0}^2 I_{2,0} I_{2,1} I_{0,1}^2-4860 I_{0,2} I_{1,0} I_{1,1} I_{2,0} I_{2,1} I_{0,1}^2+2916 I_{1,1} I_{1,2} I_{2,0} I_{2,1} I_{0,1}^2+324 I_{0,2} I_{1,0}^4 I_{2,2} I_{0,1}^2-2916 I_{1,0}^2 I_{1,1}^2 I_{2,2} I_{0,1}^2-9720 I_{2,1}^2 I_{2,2} I_{0,1}^2-648 I_{1,0}^3 I_{1,2} I_{2,2} I_{0,1}^2-1944 I_{0,2} I_{1,0}^2 I_{2,0} I_{2,2} I_{0,1}^2+972 I_{1,1}^2 I_{2,0} I_{2,2} I_{0,1}^2+1944 I_{1,0} I_{1,2} I_{2,0} I_{2,2} I_{0,1}^2+10692 I_{1,0} I_{1,1} I_{2,1} I_{2,2} I_{0,1}^2+324 I_{0,2}^2 I_{1,0}^3 I_{3,0} I_{0,1}^2-648 I_{0,2} I_{1,0} I_{1,1}^2 I_{3,0} I_{0,1}^2+1296 I_{1,0} I_{1,2}^2 I_{3,0} I_{0,1}^2-972 I_{0,3} I_{1,0}^2 I_{1,1} I_{3,0} I_{0,1}^2-2592 I_{0,2} I_{1,0}^2 I_{1,2} I_{3,0} I_{0,1}^2-1296 I_{1,1}^2 I_{1,2} I_{3,0} I_{0,1}^2-648 I_{0,2}^2 I_{1,0} I_{2,0} I_{3,0} I_{0,1}^2-108 I_{0,3} I_{1,1} I_{2,0} I_{3,0} I_{0,1}^2+972 I_{0,3} I_{1,0} I_{2,1} I_{3,0} I_{0,1}^2+1296 I_{0,2} I_{1,1} I_{2,1} I_{3,0} I_{0,1}^2+2592 I_{0,2} I_{1,0} I_{2,2} I_{3,0} I_{0,1}^2-1296 I_{1,2} I_{2,2} I_{3,0} I_{0,1}^2-54 I_{0,2} I_{0,3} I_{1,0}^6 I_{0,1}+1188 I_{0,3} I_{1,0}^4 I_{1,1}^2 I_{0,1}-2592 I_{1,0}^3 I_{1,1} I_{1,2}^2 I_{0,1}+162 I_{0,2} I_{0,3} I_{1,0}^2 I_{2,0}^2 I_{0,1}+324 I_{0,3} I_{1,1}^2 I_{2,0}^2 I_{0,1}-486 I_{0,2}^2 I_{1,0} I_{1,1} I_{2,0}^2 I_{0,1}+324 I_{0,3} I_{1,0} I_{1,2} I_{2,0}^2 I_{0,1}+972 I_{0,2} I_{1,1} I_{1,2} I_{2,0}^2 I_{0,1}-1944 I_{0,3} I_{1,0}^2 I_{2,1}^2 I_{0,1}+3240 I_{0,3} I_{2,0} I_{2,1}^2 I_{0,1}-3888 I_{1,0} I_{1,1} I_{2,2}^2 I_{0,1}+7776 I_{2,1} I_{2,2}^2 I_{0,1}-162 I_{0,2}^2 I_{1,0}^5 I_{1,1} I_{0,1}+756 I_{0,3} I_{1,0}^5 I_{1,2} I_{0,1}+1296 I_{0,2} I_{1,0}^4 I_{1,1} I_{1,2} I_{0,1}-216 I_{0,2} I_{0,3} I_{1,0}^4 I_{2,0} I_{0,1}-1944 I_{0,3} I_{1,0}^2 I_{1,1}^2 I_{2,0} I_{0,1}+648 I_{0,2}^2 I_{1,0}^3 I_{1,1} I_{2,0} I_{0,1}-1944 I_{0,3} I_{1,0}^3 I_{1,2} I_{2,0} I_{0,1}-2916 I_{0,2} I_{1,0}^2 I_{1,1} I_{1,2} I_{2,0} I_{0,1}+648 I_{0,2}^2 I_{1,0}^4 I_{2,1} I_{0,1}+6804 I_{1,0}^2 I_{1,2}^2 I_{2,1} I_{0,1}+486 I_{0,2}^2 I_{2,0}^2 I_{2,1} I_{0,1}-4212 I_{0,2} I_{1,0}^3 I_{1,2} I_{2,1} I_{0,1}-1458 I_{0,2}^2 I_{1,0}^2 I_{2,0} I_{2,1} I_{0,1}-2916 I_{1,2}^2 I_{2,0} I_{2,1} I_{0,1}+$\\$6804 I_{0,2} I_{1,0} I_{1,2} I_{2,0} I_{2,1} I_{0,1}-648 I_{0,3} I_{1,0}^4 I_{2,2} I_{0,1}-324 I_{0,3} I_{2,0}^2 I_{2,2} I_{0,1}-1944 I_{0,2} I_{1,0}^3 I_{1,1} I_{2,2} I_{0,1}+8748 I_{1,0}^2 I_{1,1} I_{1,2} I_{2,2} I_{0,1}+1620 I_{0,3} I_{1,0}^2 I_{2,0} I_{2,2} I_{0,1}+2916 I_{0,2} I_{1,0} I_{1,1} I_{2,0} I_{2,2} I_{0,1}-2916 I_{1,1} I_{1,2} I_{2,0} I_{2,2} I_{0,1}+2916 I_{0,2} I_{1,0}^2 I_{2,1} I_{2,2} I_{0,1}-13608 I_{1,0} I_{1,2} I_{2,1} I_{2,2} I_{0,1}-3888 I_{0,2} I_{2,0} I_{2,1} I_{2,2} I_{0,1}+540 I_{0,2} I_{0,3} I_{1,0}^3 I_{3,0} I_{0,1}+1728 I_{0,3} I_{1,0} I_{1,1}^2 I_{3,0} I_{0,1}+2592 I_{1,1} I_{1,2}^2 I_{3,0} I_{0,1}+162 I_{0,2}^2 I_{1,0}^2 I_{1,1} I_{3,0} I_{0,1}+1188 I_{0,3} I_{1,0}^2 I_{1,2} I_{3,0} I_{0,1}-1296 I_{0,2} I_{1,0} I_{1,1} I_{1,2} I_{3,0} I_{0,1}-432 I_{0,2} I_{0,3} I_{1,0} I_{2,0} I_{3,0} I_{0,1}-162 I_{0,2}^2 I_{1,1} I_{2,0} I_{3,0} I_{0,1}+324 I_{0,3} I_{1,2} I_{2,0} I_{3,0} I_{0,1}+324 I_{0,2}^2 I_{1,0} I_{2,1} I_{3,0} I_{0,1}-2592 I_{0,3} I_{1,1} I_{2,1} I_{3,0} I_{0,1}-648 I_{0,2} I_{1,2} I_{2,1} I_{3,0} I_{0,1}-1296 I_{0,3} I_{1,0} I_{2,2} I_{3,0} I_{0,1}+108 I_{0,2}^3 I_{1,0}^6-90 I_{0,3}^2 I_{1,0}^6+1296 I_{1,0}^3 I_{1,2}^3+54 I_{0,3}^2 I_{2,0}^3-2592 I_{0,3} I_{2,1}^3+162 I_{0,2} I_{1,0}^4 I_{1,2}^2-378 I_{0,3}^2 I_{1,0}^2 I_{2,0}^2-486 I_{0,2} I_{1,2}^2 I_{2,0}^2+162 I_{0,2} I_{0,3} I_{1,0} I_{1,1} I_{2,0}^2+486 I_{0,2}^2 I_{1,0} I_{1,2} I_{2,0}^2-324 I_{0,3} I_{1,1} I_{1,2} I_{2,0}^2+486 I_{0,2}^2 I_{1,0}^2 I_{2,1}^2+1944 I_{1,2}^2 I_{2,1}^2+3888 I_{0,3} I_{1,0} I_{1,1} I_{2,1}^2-2916 I_{0,2} I_{1,0} I_{1,2} I_{2,1}^2-486 I_{0,2}^2 I_{2,0} I_{2,1}^2+3888 I_{1,0} I_{1,2} I_{2,2}^2-54 I_{0,2}^3 I_{3,0}^2+486 I_{0,2} I_{0,3} I_{1,0}^5 I_{1,1}-486 I_{0,2}^2 I_{1,0}^5 I_{1,2}-1620 I_{0,3} I_{1,0}^4 I_{1,1} I_{1,2}-162 I_{0,2}^3 I_{1,0}^4 I_{2,0}+486 I_{0,3}^2 I_{1,0}^4 I_{2,0}+972 I_{0,2} I_{1,0}^2 I_{1,2}^2 I_{2,0}-648 I_{0,2} I_{0,3} I_{1,0}^3 I_{1,1} I_{2,0}+1944 I_{0,3} I_{1,0}^2 I_{1,1} I_{1,2} I_{2,0}-324 I_{0,2} I_{0,3} I_{1,0}^4 I_{2,1}-1944 I_{0,3} I_{1,0}^2 I_{1,1}^2 I_{2,1}+648 I_{0,2} I_{0,3} I_{2,0}^2 I_{2,1}-486 I_{0,2}^2 I_{1,0}^3 I_{1,1} I_{2,1}+972 I_{0,3} I_{1,0}^3 I_{1,2} I_{2,1}+1944 I_{0,2} I_{1,0}^2 I_{1,1} I_{1,2} I_{2,1}+648 I_{0,3} I_{1,1}^2 I_{2,0} I_{2,1}+486 I_{0,2}^2 I_{1,0} I_{1,1} I_{2,0} I_{2,1}-972 I_{0,3} I_{1,0} I_{1,2} I_{2,0} I_{2,1}+324 I_{0,2}^2 I_{1,0}^4 I_{2,2}-5832 I_{1,0}^2 I_{1,2}^2 I_{2,2}+1944 I_{0,2} I_{2,1}^2 I_{2,2}+972 I_{0,3} I_{1,0}^3 I_{1,1} I_{2,2}+972 I_{0,2} I_{1,0}^3 I_{1,2} I_{2,2}+1944 I_{1,2}^2 I_{2,0} I_{2,2}-324 I_{0,3} I_{1,0} I_{1,1} I_{2,0} I_{2,2}-2916 I_{0,2} I_{1,0} I_{1,2} I_{2,0} I_{2,2}+648 I_{0,3} I_{1,0}^2 I_{2,1} I_{2,2}-972 I_{0,2} I_{1,0} I_{1,1} I_{2,1} I_{2,2}-1944 I_{1,1} I_{1,2} I_{2,1} I_{2,2}-1944 I_{0,3} I_{2,0} I_{2,1} I_{2,2}-54 I_{0,2}^3 I_{1,0}^3 I_{3,0}-396 I_{0,3}^2 I_{1,0}^3 I_{3,0}-1296 I_{1,2}^3 I_{3,0}+1296 I_{0,2} I_{1,0} I_{1,2}^2 I_{3,0}-1296 I_{0,3} I_{1,0} I_{1,1} I_{1,2} I_{3,0}+162 I_{0,2}^3 I_{1,0} I_{2,0} I_{3,0}+324 I_{0,3}^2 I_{1,0} I_{2,0} I_{3,0}-324 I_{0,2} I_{0,3} I_{1,0} I_{2,1} I_{3,0}+1944 I_{0,3} I_{1,2} I_{2,1} I_{3,0}-324 I_{0,2}^2 I_{1,0} I_{2,2} I_{3,0})$
	\vspace*{0.1cm}

	\noindent
	$\cI^I_{2100}=-4(-54 I_{0,1} I_{0,2}^2 I_{1,0}^6+54 I_{0,1}^3 I_{0,2} I_{1,0}^6-54 I_{0,1}^2 I_{0,3} I_{1,0}^6+54 I_{0,2} I_{0,3} I_{1,0}^6-162 I_{0,1}^4 I_{1,1} I_{1,0}^5-324 I_{0,2}^2 I_{1,1} I_{1,0}^5+162 I_{0,1}^2 I_{0,2} I_{1,1} I_{1,0}^5+324 I_{0,1} I_{0,3} I_{1,1} I_{1,0}^5-54 I_{0,1}^3 I_{1,2} I_{1,0}^5+162 I_{0,1} I_{0,2} I_{1,2} I_{1,0}^5-108 I_{0,3} I_{1,2} I_{1,0}^5+918 I_{0,1}^3 I_{1,1}^2 I_{1,0}^4-810 I_{0,1} I_{0,2} I_{1,1}^2 I_{1,0}^4-432 I_{0,3} I_{1,1}^2 I_{1,0}^4-324 I_{0,1} I_{1,2}^2 I_{1,0}^4-810 I_{0,1}^2 I_{1,1} I_{1,2} I_{1,0}^4+1458 I_{0,2} I_{1,1} I_{1,2} I_{1,0}^4+486 I_{0,1} I_{0,2}^2 I_{2,0} I_{1,0}^4-270 I_{0,1}^3 I_{0,2} I_{2,0} I_{1,0}^4+162 I_{0,1}^2 I_{0,3} I_{2,0} I_{1,0}^4-378 I_{0,2} I_{0,3} I_{2,0} I_{1,0}^4+54 I_{0,1}^4 I_{2,1} I_{1,0}^4-162 I_{0,2}^2 I_{2,1} I_{1,0}^4+648 I_{0,1}^2 I_{0,2} I_{2,1} I_{1,0}^4-540 I_{0,1} I_{0,3} I_{2,1} I_{1,0}^4-324 I_{0,1} I_{0,2} I_{2,2} I_{1,0}^4+324 I_{0,3} I_{2,2} I_{1,0}^4-1296 I_{0,1}^2 I_{1,1}^3 I_{1,0}^3-1296 I_{1,1} I_{1,2}^2 I_{1,0}^3+2592 I_{0,1} I_{1,1}^2 I_{1,2} I_{1,0}^3+378 I_{0,1}^4 I_{1,1} I_{2,0} I_{1,0}^3+810 I_{0,2}^2 I_{1,1} I_{2,0} I_{1,0}^3-648 I_{0,1}^2 I_{0,2} I_{1,1} I_{2,0} I_{1,0}^3-540 I_{0,1} I_{0,3} I_{1,1} I_{2,0} I_{1,0}^3+324 I_{0,1}^3 I_{1,2} I_{2,0} I_{1,0}^3-648 I_{0,1} I_{0,2} I_{1,2} I_{2,0} I_{1,0}^3+324 I_{0,3} I_{1,2} I_{2,0} I_{1,0}^3-1296 I_{0,1}^3 I_{1,1} I_{2,1} I_{1,0}^3-972 I_{0,1} I_{0,2} I_{1,1} I_{2,1} I_{1,0}^3+1296 I_{0,3} I_{1,1} I_{2,1} I_{1,0}^3-324 I_{0,1}^2 I_{1,2} I_{2,1} I_{1,0}^3+1296 I_{0,2} I_{1,2} I_{2,1} I_{1,0}^3+1296 I_{0,1}^2 I_{1,1} I_{2,2} I_{1,0}^3-1620 I_{0,2} I_{1,1} I_{2,2} I_{1,0}^3+324 I_{0,1} I_{1,2} I_{2,2} I_{1,0}^3-162 I_{0,1}^5 I_{3,0} I_{1,0}^3-378 I_{0,1} I_{0,2}^2 I_{3,0} I_{1,0}^3+324 I_{0,1}^3 I_{0,2} I_{3,0} I_{1,0}^3+108 I_{0,1}^2 I_{0,3} I_{3,0} I_{1,0}^3+108 I_{0,2} I_{0,3} I_{3,0} I_{1,0}^3+162 I_{0,1}^5 I_{2,0}^2 I_{1,0}^2-972 I_{0,1} I_{0,2}^2 I_{2,0}^2 I_{1,0}^2+162 I_{0,1}^3 I_{0,2} I_{2,0}^2 I_{1,0}^2-162 I_{0,1}^2 I_{0,3} I_{2,0}^2 I_{1,0}^2+810 I_{0,2} I_{0,3} I_{2,0}^2 I_{1,0}^2-324 I_{0,1}^3 I_{2,1}^2 I_{1,0}^2+2916 I_{0,1} I_{0,2} I_{2,1}^2 I_{1,0}^2-648 I_{0,3} I_{2,1}^2 I_{1,0}^2-972 I_{0,1}^3 I_{1,1}^2 I_{2,0} I_{1,0}^2+1944 I_{0,1} I_{0,2} I_{1,1}^2 I_{2,0} I_{1,0}^2+972 I_{0,1} I_{1,2}^2 I_{2,0} I_{1,0}^2-1944 I_{0,2} I_{1,1} I_{1,2} I_{2,0} I_{1,0}^2+2916 I_{0,1}^2 I_{1,1}^2 I_{2,1} I_{1,0}^2+972 I_{0,2} I_{1,1}^2 I_{2,1} I_{1,0}^2-2916 I_{1,2}^2 I_{2,1} I_{1,0}^2-972 I_{0,1} I_{1,1} I_{1,2} I_{2,1} I_{1,0}^2-324 I_{0,1}^4 I_{2,0} I_{2,1} I_{1,0}^2+972 I_{0,2}^2 I_{2,0} I_{2,1} I_{1,0}^2-1944 I_{0,1}^2 I_{0,2} I_{2,0} I_{2,1} I_{1,0}^2+1296 I_{0,1} I_{0,3} I_{2,0} I_{2,1} I_{1,0}^2-2916 I_{0,1} I_{1,1}^2 I_{2,2} I_{1,0}^2+2916 I_{1,1} I_{1,2} I_{2,2} I_{1,0}^2+324 I_{0,1}^3 I_{2,0} I_{2,2} I_{1,0}^2+972 I_{0,1} I_{0,2} I_{2,0} I_{2,2} I_{1,0}^2-1296 I_{0,3} I_{2,0} I_{2,2} I_{1,0}^2+972 I_{0,1}^2 I_{2,1} I_{2,2} I_{1,0}^2-2916 I_{0,2} I_{2,1} I_{2,2} I_{1,0}^2+1134 I_{0,1}^4 I_{1,1} I_{3,0} I_{1,0}^2-162 I_{0,2}^2 I_{1,1} I_{3,0} I_{1,0}^2-648 I_{0,1}^2 I_{0,2} I_{1,1} I_{3,0} I_{1,0}^2-324 I_{0,1} I_{0,3} I_{1,1} I_{3,0} I_{1,0}^2-1404 I_{0,1}^3 I_{1,2} I_{3,0} I_{1,0}^2+1296 I_{0,1} I_{0,2} I_{1,2} I_{3,0} I_{1,0}^2+108 I_{0,3} I_{1,2} I_{3,0} I_{1,0}^2-972 I_{0,1}^4 I_{1,1} I_{2,0}^2 I_{1,0}-486 I_{0,2}^2 I_{1,1} I_{2,0}^2 I_{1,0}+1458 I_{0,1}^2 I_{0,2} I_{1,1} I_{2,0}^2 I_{1,0}+486 I_{0,1}^3 I_{1,2} I_{2,0}^2 I_{1,0}-486 I_{0,1} I_{0,2} I_{1,2} I_{2,0}^2 I_{1,0}-1944 I_{0,2} I_{1,1} I_{2,1}^2 I_{1,0}-3888 I_{0,1} I_{1,2} I_{2,1}^2 I_{1,0}-1944 I_{1,1} I_{2,2}^2 I_{1,0}+2592 I_{0,1}^3 I_{1,1} I_{2,0} I_{2,1} I_{1,0}+972 I_{0,1} I_{0,2} I_{1,1} I_{2,0} I_{2,1} I_{1,0}-648 I_{0,3} I_{1,1} I_{2,0} I_{2,1} I_{1,0}-2916 I_{0,2} I_{1,2} I_{2,0} I_{2,1} I_{1,0}-2916 I_{0,1}^2 I_{1,1} I_{2,0} I_{2,2} I_{1,0}+1944 I_{0,2} I_{1,1} I_{2,0} I_{2,2} I_{1,0}+972 I_{0,1} I_{1,2} I_{2,0} I_{2,2} I_{1,0}-1944 I_{0,1} I_{1,1} I_{2,1} I_{2,2} I_{1,0}+9720 I_{1,2} I_{2,1} I_{2,2} I_{1,0}-2376 I_{0,1}^3 I_{1,1}^2 I_{3,0} I_{1,0}-648 I_{0,1} I_{0,2} I_{1,1}^2 I_{3,0} I_{1,0}+432 I_{0,3} I_{1,1}^2 I_{3,0} I_{1,0}-2592 I_{0,1} I_{1,2}^2 I_{3,0} I_{1,0}+5184 I_{0,1}^2 I_{1,1} I_{1,2} I_{3,0} I_{1,0}+162 I_{0,1}^5 I_{2,0} I_{3,0} I_{1,0}+1458 I_{0,1} I_{0,2}^2 I_{2,0} I_{3,0} I_{1,0}-540 I_{0,1}^3 I_{0,2} I_{2,0} I_{3,0} I_{1,0}-324 I_{0,1}^2 I_{0,3} I_{2,0} I_{3,0} I_{1,0}-756 I_{0,2} I_{0,3} I_{2,0} I_{3,0} I_{1,0}-540 I_{0,1}^4 I_{2,1} I_{3,0} I_{1,0}+648 I_{0,2}^2 I_{2,1} I_{3,0} I_{1,0}+324 I_{0,1}^2 I_{0,2} I_{2,1} I_{3,0} I_{1,0}-432 I_{0,1} I_{0,3} I_{2,1} I_{3,0} I_{1,0}+972 I_{0,1}^3 I_{2,2} I_{3,0} I_{1,0}-1620 I_{0,1} I_{0,2} I_{2,2} I_{3,0} I_{1,0}+648 I_{0,3} I_{2,2} I_{3,0} I_{1,0}-162 I_{0,1}^5 I_{2,0}^3+162 I_{0,1}^3 I_{0,2} I_{2,0}^3+162 I_{0,1}^2 I_{0,3} I_{2,0}^3-162 I_{0,2} I_{0,3} I_{2,0}^3-1944 I_{0,1}^2 I_{2,1}^3+1944 I_{0,2} I_{2,1}^3+486 I_{0,1}^3 I_{1,1}^2 I_{2,0}^2-486 I_{0,1} I_{0,2} I_{1,1}^2 I_{2,0}^2-486 I_{0,1}^2 I_{1,1} I_{1,2} I_{2,0}^2+486 I_{0,2} I_{1,1} I_{1,2} I_{2,0}^2-1944 I_{1,1} I_{1,2} I_{2,1}^2-324 I_{0,1}^3 I_{2,0} I_{2,1}^2-4860 I_{0,1} I_{0,2} I_{2,0} I_{2,1}^2-648 I_{0,3} I_{2,0} I_{2,1}^2-1944 I_{0,1} I_{2,0} I_{2,2}^2-7776 I_{2,1} I_{2,2}^2-540 I_{0,1} I_{0,2}^2 I_{3,0}^2+108 I_{0,1}^3 I_{0,2} I_{3,0}^2+108 I_{0,1}^2 I_{0,3} I_{3,0}^2+324 I_{0,2} I_{0,3} I_{3,0}^2+810 I_{0,1}^4 I_{2,0}^2 I_{2,1}-1458 I_{0,2}^2 I_{2,0}^2 I_{2,1}+972 I_{0,1}^2 I_{0,2} I_{2,0}^2 I_{2,1}-324 I_{0,1} I_{0,3} I_{2,0}^2 I_{2,1}-972 I_{0,1}^2 I_{1,1}^2 I_{2,0} I_{2,1}-972 I_{0,2} I_{1,1}^2 I_{2,0} I_{2,1}+972 I_{1,2}^2 I_{2,0} I_{2,1}+972 I_{0,1} I_{1,1} I_{1,2} I_{2,0} I_{2,1}-1296 I_{0,1}^3 I_{2,0}^2 I_{2,2}+972 I_{0,1} I_{0,2} I_{2,0}^2 I_{2,2}+324 I_{0,3} I_{2,0}^2 I_{2,2}+7776 I_{0,1} I_{2,1}^2 I_{2,2}+972 I_{0,1} I_{1,1}^2 I_{2,0} I_{2,2}-972 I_{1,1} I_{1,2} I_{2,0} I_{2,2}+1944 I_{1,1}^2 I_{2,1} I_{2,2}+972 I_{0,1}^2 I_{2,0} I_{2,1} I_{2,2}+6804 I_{0,2} I_{2,0} I_{2,1} I_{2,2}+1296 I_{0,1}^2 I_{1,1}^3 I_{3,0}+1296 I_{1,1} I_{1,2}^2 I_{3,0}-2592 I_{0,1} I_{1,1}^2 I_{1,2} I_{3,0}-378 I_{0,1}^4 I_{1,1} I_{2,0} I_{3,0}+162 I_{0,2}^2 I_{1,1} I_{2,0} I_{3,0}-324 I_{0,1}^2 I_{0,2} I_{1,1} I_{2,0} I_{3,0}+540 I_{0,1} I_{0,3} I_{1,1} I_{2,0} I_{3,0}+648 I_{0,1}^3 I_{1,2} I_{2,0} I_{3,0}-324 I_{0,1} I_{0,2} I_{1,2} I_{2,0} I_{3,0}-324 I_{0,3} I_{1,2} I_{2,0} I_{3,0}+1296 I_{0,1}^3 I_{1,1} I_{2,1} I_{3,0}+1944 I_{0,1} I_{0,2} I_{1,1} I_{2,1} I_{3,0}+648 I_{0,3} I_{1,1} I_{2,1} I_{3,0}-1620 I_{0,1}^2 I_{1,2} I_{2,1} I_{3,0}-2268 I_{0,2} I_{1,2} I_{2,1} I_{3,0}-2268 I_{0,1}^2 I_{1,1} I_{2,2} I_{3,0}-324 I_{0,2} I_{1,1} I_{2,2} I_{3,0}+2592 I_{0,1} I_{1,2} I_{2,2} I_{3,0})$
	\vspace*{0.1cm}

	\noindent
	$\cI^I_{1122}=4 (216 I_{0,1}^4 I_{1,0}^5+324 I_{0,2}^2 I_{1,0}^5-648 I_{0,1}^2 I_{0,2} I_{1,0}^5+108 I_{0,1} I_{0,3} I_{1,0}^5-1296 I_{0,1}^3 I_{1,1} I_{1,0}^4+2268 I_{0,1} I_{0,2} I_{1,1} I_{1,0}^4-324 I_{0,3} I_{1,1} I_{1,0}^4+1620 I_{0,1}^2 I_{1,2} I_{1,0}^4-2268 I_{0,2} I_{1,2} I_{1,0}^4+1944 I_{0,1}^2 I_{1,1}^2 I_{1,0}^3-648 I_{0,2} I_{1,1}^2 I_{1,0}^3+3888 I_{1,2}^2 I_{1,0}^3-5184 I_{0,1} I_{1,1} I_{1,2} I_{1,0}^3-648 I_{0,1}^4 I_{2,0} I_{1,0}^3-648 I_{0,2}^2 I_{2,0} I_{1,0}^3+1944 I_{0,1}^2 I_{0,2} I_{2,0} I_{1,0}^3-648 I_{0,1} I_{0,3} I_{2,0} I_{1,0}^3+1944 I_{0,1}^3 I_{2,1} I_{1,0}^3-3240 I_{0,1} I_{0,2} I_{2,1} I_{1,0}^3-2592 I_{0,1}^2 I_{2,2} I_{1,0}^3+3888 I_{0,2} I_{2,2} I_{1,0}^3+1944 I_{0,1}^3 I_{1,1} I_{2,0} I_{1,0}^2-3888 I_{0,1} I_{0,2} I_{1,1} I_{2,0} I_{1,0}^2+1944 I_{0,3} I_{1,1} I_{2,0} I_{1,0}^2-1944 I_{0,1}^2 I_{1,2} I_{2,0} I_{1,0}^2+1944 I_{0,2} I_{1,2} I_{2,0} I_{1,0}^2-5832 I_{0,1}^2 I_{1,1} I_{2,1} I_{1,0}^2+1944 I_{0,2} I_{1,1} I_{2,1} I_{1,0}^2+7776 I_{0,1} I_{1,2} I_{2,1} I_{1,0}^2+7776 I_{0,1} I_{1,1} I_{2,2} I_{1,0}^2-11664 I_{1,2} I_{2,2} I_{1,0}^2-216 I_{0,1}^4 I_{3,0} I_{1,0}^2-1296 I_{0,2}^2 I_{3,0} I_{1,0}^2+648 I_{0,1}^2 I_{0,2} I_{3,0} I_{1,0}^2+864 I_{0,1} I_{0,3} I_{3,0} I_{1,0}^2+648 I_{0,1}^4 I_{2,0}^2 I_{1,0}+972 I_{0,2}^2 I_{2,0}^2 I_{1,0}-1944 I_{0,1}^2 I_{0,2} I_{2,0}^2 I_{1,0}+324 I_{0,1} I_{0,3} I_{2,0}^2 I_{1,0}+3888 I_{0,1}^2 I_{2,1}^2 I_{1,0}-1944 I_{0,2} I_{2,1}^2 I_{1,0}+7776 I_{2,2}^2 I_{1,0}-3240 I_{0,1}^3 I_{2,0} I_{2,1} I_{1,0}+5832 I_{0,1} I_{0,2} I_{2,0} I_{2,1} I_{1,0}-648 I_{0,3} I_{2,0} I_{2,1} I_{1,0}+3888 I_{0,1}^2 I_{2,0} I_{2,2} I_{1,0}-5832 I_{0,2} I_{2,0} I_{2,2} I_{1,0}-9720 I_{0,1} I_{2,1} I_{2,2} I_{1,0}+1296 I_{0,1}^3 I_{1,1} I_{3,0} I_{1,0}-1296 I_{0,1} I_{0,2} I_{1,1} I_{3,0} I_{1,0}-2592 I_{0,3} I_{1,1} I_{3,0} I_{1,0}-2592 I_{0,1}^2 I_{1,2} I_{3,0} I_{1,0}+5184 I_{0,2} I_{1,2} I_{3,0} I_{1,0}-648 I_{0,1}^3 I_{1,1} I_{2,0}^2+972 I_{0,1} I_{0,2} I_{1,1} I_{2,0}^2-324 I_{0,3} I_{1,1} I_{2,0}^2+972 I_{0,1}^2 I_{1,2} I_{2,0}^2-972 I_{0,2} I_{1,2} I_{2,0}^2+1944 I_{1,2} I_{2,1}^2+1944 I_{0,1}^2 I_{1,1} I_{2,0} I_{2,1}-3888 I_{0,1} I_{1,2} I_{2,0} I_{2,1}-1944 I_{0,1} I_{1,1} I_{2,0} I_{2,2}+3888 I_{1,2} I_{2,0} I_{2,2}-1944 I_{1,1} I_{2,1} I_{2,2}-1944 I_{0,1}^2 I_{1,1}^2 I_{3,0}+648 I_{0,2} I_{1,1}^2 I_{3,0}-3888 I_{1,2}^2 I_{3,0}+5184 I_{0,1} I_{1,1} I_{1,2} I_{3,0}+648 I_{0,2}^2 I_{2,0} I_{3,0}-648 I_{0,1} I_{0,3} I_{2,0} I_{3,0}-648 I_{0,1} I_{0,2} I_{2,1} I_{3,0}+1944 I_{0,3} I_{2,1} I_{3,0}+648 I_{0,1}^2 I_{2,2} I_{3,0}-1944 I_{0,2} I_{2,2} I_{3,0})$
	\subsection{$\cI^{R}$ coefficients}
	$\cI^R_{0000}=264 I_{0,2}^2 I_{1,0}^5-891 I_{1,1}^2 I_{1,2} I_{1,0}^2+432 I_{0,3} I_{1,1} I_{2,0} I_{1,0}^2-1620 I_{0,2} I_{1,2} I_{2,0} I_{1,0}^2+216 I_{0,2} I_{1,1} I_{2,1} I_{1,0}^2+972 I_{1,2} I_{2,2} I_{1,0}^2+2616 I_{0,2}^2 I_{3,0} I_{1,0}^2-1458 I_{0,1}^2 I_{0,2} I_{3,0} I_{1,0}^2-1026 I_{0,1} I_{0,3} I_{3,0} I_{1,0}^2-1404 I_{0,2}^2 I_{2,0}^2 I_{1,0}+513 I_{0,1}^2 I_{0,2} I_{2,0}^2 I_{1,0}+297 I_{0,1} I_{0,3} I_{2,0}^2 I_{1,0}-918 I_{0,2} I_{2,1}^2 I_{1,0}+324 I_{2,2}^2 I_{1,0}-891 I_{0,2} I_{1,1}^2 I_{2,0} I_{1,0}+891 I_{1,1} I_{1,2} I_{2,1} I_{1,0}-108 I_{0,3} I_{2,0} I_{2,1} I_{1,0}+1296 I_{0,2} I_{2,0} I_{2,2} I_{1,0}+5670 I_{0,3} I_{1,1} I_{3,0} I_{1,0}-5472 I_{0,2} I_{1,2} I_{3,0} I_{1,0}+5184 I_{0,1} I_{0,2} I_{1,1} I_{2,0}^2-5400 I_{0,3} I_{1,1} I_{2,0}^2-1728 I_{0,1}^2 I_{1,2} I_{2,0}^2+1944 I_{0,2} I_{1,2} I_{2,0}^2+1998 I_{1,2} I_{2,1}^2-3618 I_{0,1}^2 I_{1,1} I_{2,0} I_{2,1}+999 I_{0,2} I_{1,1} I_{2,0} I_{2,1}+675 I_{0,1} I_{1,2} I_{2,0} I_{2,1}+5832 I_{0,1} I_{1,1} I_{2,0} I_{2,2}-3888 I_{1,2} I_{2,0} I_{2,2}-1998 I_{1,1} I_{2,1} I_{2,2}+3618 I_{0,1}^2 I_{1,1}^2 I_{3,0}+594 I_{0,2} I_{1,1}^2 I_{3,0}+2592 I_{1,2}^2 I_{3,0}-6507 I_{0,1} I_{1,1} I_{1,2} I_{3,0}-1476 I_{0,2}^2 I_{2,0} I_{3,0}+945 I_{0,1}^2 I_{0,2} I_{2,0} I_{3,0}+729 I_{0,1} I_{0,3} I_{2,0} I_{3,0}-5184 I_{0,1} I_{0,2} I_{2,1} I_{3,0}-594 I_{0,3} I_{2,1} I_{3,0}+1728 I_{0,1}^2 I_{2,2} I_{3,0}+3852 I_{0,2} I_{2,2} I_{3,0}$\\
	\vspace*{0.1cm}

	\noindent 
	$\cI^R_{1000}=5820 I_{1,0}^4 I_{0,1}^4-6468 I_{1,0} I_{3,0} I_{0,1}^4-14598 I_{1,0}^3 I_{1,1} I_{0,1}^3-846 I_{1,0}^2 I_{2,1} I_{0,1}^3+1980 I_{1,1} I_{3,0} I_{0,1}^3-8244 I_{0,2} I_{1,0}^4 I_{0,1}^2-5562 I_{2,1}^2 I_{0,1}^2+33426 I_{1,0}^3 I_{1,2} I_{0,1}^2-1971 I_{0,2} I_{1,0}^2 I_{2,0} I_{0,1}^2+1080 I_{1,0} I_{1,2} I_{2,0} I_{0,1}^2+26244 I_{1,0} I_{1,1} I_{2,1} I_{0,1}^2-38178 I_{1,0}^2 I_{2,2} I_{0,1}^2+9828 I_{2,0} I_{2,2} I_{0,1}^2+11646 I_{0,2} I_{1,0} I_{3,0} I_{0,1}^2-1296 I_{1,2} I_{3,0} I_{0,1}^2+2892 I_{0,3} I_{1,0}^4 I_{0,1}-16686 I_{0,2} I_{1,0}^3 I_{1,1} I_{0,1}+891 I_{1,0}^2 I_{1,1} I_{1,2} I_{0,1}-243 I_{0,3} I_{1,0}^2 I_{2,0} I_{0,1}+46224 I_{0,2} I_{1,0}^2 I_{2,1} I_{0,1}-35370 I_{1,0} I_{1,2} I_{2,1} I_{0,1}-2646 I_{0,2} I_{2,0} I_{2,1} I_{0,1}+5913 I_{1,0} I_{1,1} I_{2,2} I_{0,1}+108 I_{2,1} I_{2,2} I_{0,1}-3162 I_{0,3} I_{1,0} I_{3,0} I_{0,1}-2322 I_{0,2} I_{1,1} I_{3,0} I_{0,1}-864 I_{0,2}^2 I_{1,0}^4-1782 I_{0,2} I_{1,0}^2 I_{1,1}^2+1080 I_{0,2} I_{2,1}^2+3780 I_{2,2}^2-2952 I_{0,3} I_{1,0}^3 I_{1,1}+216 I_{0,2} I_{1,0}^3 I_{1,2}+1620 I_{0,2}^2 I_{1,0}^2 I_{2,0}-1080 I_{0,2} I_{1,0} I_{1,2} I_{2,0}-2070 I_{0,3} I_{1,0}^2 I_{2,1}+2079 I_{0,2} I_{1,0} I_{1,1} I_{2,1}-891 I_{1,1} I_{1,2} I_{2,1}-6048 I_{0,3} I_{2,0} I_{2,1}-4140 I_{0,2} I_{1,0}^2 I_{2,2}+1782 I_{1,1}^2 I_{2,2}+1728 I_{1,0} I_{1,2} I_{2,2}-540 I_{0,2} I_{2,0} I_{2,2}-2016 I_{0,2}^2 I_{1,0} I_{3,0}+342 I_{0,3} I_{1,1} I_{3,0}+1296 I_{0,2} I_{1,2} I_{3,0}$\\
	\vspace*{0.1cm}

	\noindent 
	$\cI^R_{0100}=-5184 I_{0,1}^2 I_{1,1} I_{1,0}^4+10800 I_{0,1} I_{1,2} I_{1,0}^4-8424 I_{1,1} I_{1,2} I_{1,0}^3+9666 I_{0,1}^2 I_{2,1} I_{1,0}^3-20628 I_{0,1} I_{2,2} I_{1,0}^3+7344 I_{0,1}^2 I_{1,1} I_{2,0} I_{1,0}^2-15768 I_{0,1} I_{1,2} I_{2,0} I_{1,0}^2+54 I_{1,2} I_{2,1} I_{1,0}^2+24813 I_{1,1} I_{2,2} I_{1,0}^2+630 I_{0,1}^3 I_{3,0} I_{1,0}^2+2916 I_{0,1} I_{0,2} I_{3,0} I_{1,0}^2+396 I_{0,3} I_{3,0} I_{1,0}^2-387 I_{0,1}^3 I_{2,0}^2 I_{1,0}-1026 I_{0,1} I_{0,2} I_{2,0}^2 I_{1,0}+90 I_{0,3} I_{2,0}^2 I_{1,0}-13851 I_{0,1}^2 I_{2,0} I_{2,1} I_{1,0}-432 I_{0,2} I_{2,0} I_{2,1} I_{1,0}+30726 I_{0,1} I_{2,0} I_{2,2} I_{1,0}-16443 I_{2,1} I_{2,2} I_{1,0}-15876 I_{0,1}^2 I_{1,1} I_{3,0} I_{1,0}+18684 I_{0,1} I_{1,2} I_{3,0} I_{1,0}+14256 I_{0,1}^2 I_{1,1} I_{2,0}^2-15984 I_{0,1} I_{1,2} I_{2,0}^2+16929 I_{1,2} I_{2,0} I_{2,1}-16200 I_{1,1} I_{2,0} I_{2,2}-729 I_{1,1} I_{1,2} I_{3,0}-243 I_{0,1}^3 I_{2,0} I_{3,0}-1890 I_{0,1} I_{0,2} I_{2,0} I_{3,0}-486 I_{0,3} I_{2,0} I_{3,0}+3645 I_{0,1}^2 I_{2,1} I_{3,0}+432 I_{0,2} I_{2,1} I_{3,0}-7830 I_{0,1} I_{2,2} I_{3,0}$\\
	\vspace*{0.1cm}

	\noindent  
	$\cI^R_{2000}=-1026 I_{1,0}^3 I_{0,1}^4-360 I_{1,0} I_{2,0} I_{0,1}^4+2034 I_{3,0} I_{0,1}^4-11655 I_{1,0}^2 I_{1,1} I_{0,1}^3+4212 I_{1,1} I_{2,0} I_{0,1}^3+1467 I_{1,0} I_{2,1} I_{0,1}^3+2349 I_{0,2} I_{1,0}^3 I_{0,1}^2+26244 I_{1,0} I_{1,1}^2 I_{0,1}^2-22761 I_{1,0}^2 I_{1,2} I_{0,1}^2+567 I_{0,2} I_{1,0} I_{2,0} I_{0,1}^2+11340 I_{1,2} I_{2,0} I_{0,1}^2-10962 I_{1,1} I_{2,1} I_{0,1}^2+16065 I_{1,0} I_{2,2} I_{0,1}^2-4347 I_{0,2} I_{3,0} I_{0,1}^2+513 I_{0,3} I_{1,0}^3 I_{0,1}+33156 I_{0,2} I_{1,0}^2 I_{1,1} I_{0,1}-24246 I_{1,0} I_{1,1} I_{1,2} I_{0,1}-1017 I_{0,3} I_{1,0} I_{2,0} I_{0,1}-19332 I_{0,2} I_{1,1} I_{2,0} I_{0,1}-28674 I_{0,2} I_{1,0} I_{2,1} I_{0,1}-6831 I_{1,2} I_{2,1} I_{0,1}+14337 I_{1,1} I_{2,2} I_{0,1}+1017 I_{0,3} I_{3,0} I_{0,1}-1440 I_{0,2}^2 I_{1,0}^3+2079 I_{0,2} I_{1,0} I_{1,1}^2-2592 I_{1,0} I_{1,2}^2-4950 I_{0,3} I_{1,0}^2 I_{1,1}+7200 I_{0,2} I_{1,0}^2 I_{1,2}+891 I_{1,1}^2 I_{1,2}+1404 I_{0,2}^2 I_{1,0} I_{2,0}+6858 I_{0,3} I_{1,1} I_{2,0}-2484 I_{0,2} I_{1,2} I_{2,0}+8496 I_{0,3} I_{1,0} I_{2,1}-918 I_{0,2} I_{1,1} I_{2,1}+2052 I_{0,2} I_{1,0} I_{2,2}+3780 I_{1,2} I_{2,2}+1296 I_{0,2}^2 I_{3,0}$\\
	\vspace*{0.1cm}

	\noindent  
	$\cI^R_{0200}=-288 I_{0,1}^2 I_{1,0}^5+24 I_{0,2} I_{1,0}^5+5724 I_{0,1} I_{1,1} I_{1,0}^4-792 I_{1,2} I_{1,0}^4-7830 I_{1,1}^2 I_{1,0}^3+864 I_{0,1}^2 I_{2,0} I_{1,0}^3-864 I_{0,2} I_{2,0} I_{1,0}^3-8910 I_{0,1} I_{2,1} I_{1,0}^3-792 I_{2,2} I_{1,0}^3-9558 I_{0,1} I_{1,1} I_{2,0} I_{1,0}^2+3024 I_{1,2} I_{2,0} I_{1,0}^2+23571 I_{1,1} I_{2,1} I_{1,0}^2-2034 I_{0,1}^2 I_{3,0} I_{1,0}^2-2040 I_{0,2} I_{3,0} I_{1,0}^2+513 I_{0,1}^2 I_{2,0}^2 I_{1,0}+1404 I_{0,2} I_{2,0}^2 I_{1,0}-13743 I_{2,1}^2 I_{1,0}+891 I_{1,1}^2 I_{2,0} I_{1,0}+12285 I_{0,1} I_{2,0} I_{2,1} I_{1,0}+2052 I_{2,0} I_{2,2} I_{1,0}+15390 I_{0,1} I_{1,1} I_{3,0} I_{1,0}-288 I_{1,2} I_{3,0} I_{1,0}-14040 I_{0,1} I_{1,1} I_{2,0}^2-1944 I_{1,2} I_{2,0}^2-1269 I_{1,1} I_{2,0} I_{2,1}-1620 I_{1,1}^2 I_{3,0}+945 I_{0,1}^2 I_{2,0} I_{3,0}+1476 I_{0,2} I_{2,0} I_{3,0}-891 I_{0,1} I_{2,1} I_{3,0}-1260 I_{2,2} I_{3,0}$\\
	\vspace*{0.1cm}

	\noindent   
	$\cI^R_{1100}=-3528 I_{0,1}^3 I_{1,0}^4+16668 I_{0,1} I_{0,2} I_{1,0}^4+17550 I_{0,1}^2 I_{1,1} I_{1,0}^3+2484 I_{0,2} I_{1,1} I_{1,0}^3-67284 I_{0,1} I_{1,2} I_{1,0}^3-1782 I_{1,1} I_{1,2} I_{1,0}^2-7281 I_{0,1}^3 I_{2,0} I_{1,0}^2-22950 I_{0,1} I_{0,2} I_{2,0} I_{1,0}^2-252 I_{0,3} I_{2,0} I_{1,0}^2-21789 I_{0,1}^2 I_{2,1} I_{1,0}^2-3375 I_{0,2} I_{2,1} I_{1,0}^2+108297 I_{0,1} I_{2,2} I_{1,0}^2-891 I_{0,2} I_{1,1} I_{2,0} I_{1,0}-2160 I_{0,1} I_{1,2} I_{2,0} I_{1,0}+2160 I_{1,2} I_{2,1} I_{1,0}-14985 I_{1,1} I_{2,2} I_{1,0}+19404 I_{0,1}^3 I_{3,0} I_{1,0}-8352 I_{0,1} I_{0,2} I_{3,0} I_{1,0}+432 I_{0,3} I_{3,0} I_{1,0}-9504 I_{0,1}^3 I_{2,0}^2+9072 I_{0,1} I_{0,2} I_{2,0}^2+648 I_{0,3} I_{2,0}^2+38448 I_{0,1}^2 I_{2,0} I_{2,1}-999 I_{0,2} I_{2,0} I_{2,1}-57672 I_{0,1} I_{2,0} I_{2,2}-3429 I_{2,1} I_{2,2}-5940 I_{0,1}^2 I_{1,1} I_{3,0}-1296 I_{0,2} I_{1,1} I_{3,0}+5184 I_{0,1} I_{1,2} I_{3,0}$\\
	\vspace*{0.1cm}

	\noindent    
	$\cI^R_{1200}=-15516 I_{0,1}^2 I_{1,0}^4+3168 I_{0,2} I_{1,0}^4+55782 I_{0,1} I_{1,1} I_{1,0}^3-7344 I_{1,2} I_{1,0}^3+3564 I_{1,1}^2 I_{1,0}^2+37071 I_{0,1}^2 I_{2,0} I_{1,0}^2-5400 I_{0,2} I_{2,0} I_{1,0}^2-80163 I_{0,1} I_{2,1} I_{1,0}^2+16200 I_{2,2} I_{1,0}^2+891 I_{0,1} I_{1,1} I_{2,0} I_{1,0}+2160 I_{1,2} I_{2,0} I_{1,0}-24867 I_{1,1} I_{2,1} I_{1,0}-16236 I_{0,1}^2 I_{3,0} I_{1,0}+4752 I_{0,2} I_{3,0} I_{1,0}-216 I_{0,1}^2 I_{2,0}^2-12177 I_{2,1}^2-4941 I_{0,1} I_{2,0} I_{2,1}+2052 I_{0,1} I_{1,1} I_{3,0}$\\
	\vspace*{0.1cm}

	\noindent     
	$\cI^R_{0120}=-7785 I_{1,0}^3 I_{0,1}^3+14796 I_{1,0} I_{2,0} I_{0,1}^3-6102 I_{3,0} I_{0,1}^3+10206 I_{1,0}^2 I_{1,1} I_{0,1}^2-28512 I_{1,1} I_{2,0} I_{0,1}^2-9963 I_{1,0} I_{2,1} I_{0,1}^2-4428 I_{0,2} I_{1,0}^3 I_{0,1}+64341 I_{1,0}^2 I_{1,2} I_{0,1}+4806 I_{0,2} I_{1,0} I_{2,0} I_{0,1}+14472 I_{1,2} I_{2,0} I_{0,1}-35586 I_{1,0} I_{2,2} I_{0,1}+5184 I_{0,2} I_{3,0} I_{0,1}-1368 I_{0,3} I_{1,0}^3-297 I_{0,2} I_{1,0}^2 I_{1,1}-20709 I_{1,0} I_{1,1} I_{1,2}+540 I_{0,3} I_{1,0} I_{2,0}+4374 I_{0,2} I_{1,0} I_{2,1}-4023 I_{1,2} I_{2,1}$\\
	\vspace*{0.1cm}
	
	\noindent      
	$\cI^R_{2200}=10773 I_{0,1}^2 I_{1,0}^3+2016 I_{0,2} I_{1,0}^3-39204 I_{0,1} I_{1,1} I_{1,0}^2-2160 I_{1,2} I_{1,0}^2-23085 I_{1,1}^2 I_{1,0}-20682 I_{0,1}^2 I_{2,0} I_{1,0}-648 I_{0,2} I_{2,0} I_{1,0}+55107 I_{0,1} I_{2,1} I_{1,0}-12744 I_{2,2} I_{1,0}+20412 I_{0,1} I_{1,1} I_{2,0}+3888 I_{1,2} I_{2,0}+6993 I_{1,1} I_{2,1}+4806 I_{0,1}^2 I_{3,0}-3888 I_{0,2} I_{3,0}$\\
	\vspace*{0.1cm}
	
	\noindent      
	$\cI^R_{1120}=27864 I_{1,0}^2 I_{0,1}^3-8424 I_{2,0} I_{0,1}^3-52488 I_{1,0} I_{1,1} I_{0,1}^2+16524 I_{2,1} I_{0,1}^2-38016 I_{0,2} I_{1,0}^2 I_{0,1}+66420 I_{1,0} I_{1,2} I_{0,1}+28296 I_{0,2} I_{2,0} I_{0,1}-28674 I_{2,2} I_{0,1}+3240 I_{0,3} I_{1,0}^2-2376 I_{0,2} I_{1,0} I_{1,1}+1782 I_{1,1} I_{1,2}-2916 I_{0,3} I_{2,0}-162 I_{0,2} I_{2,1}$\\
	\vspace*{0.1cm}
	
	\noindent 
	$\cI^R_{0112}=-10944 I_{0,1} I_{1,0}^4+15660 I_{1,1} I_{1,0}^3+15552 I_{0,1} I_{2,0} I_{1,0}^2+594 I_{2,1} I_{1,0}^2-18108 I_{0,1} I_{3,0} I_{1,0}+17712 I_{0,1} I_{2,0}^2-31860 I_{2,0} I_{2,1}+2052 I_{1,1} I_{3,0}$\\
	\vspace*{0.1cm}
	
	\noindent 
	$\cI^R_{1220}=9720 I_{1,0}^2 I_{0,1}^2-19224 I_{2,0} I_{0,1}^2-13608 I_{1,0} I_{1,1} I_{0,1}+21060 I_{2,1} I_{0,1}-3600 I_{0,2} I_{1,0}^2-3564 I_{1,1}^2+864 I_{1,0} I_{1,2}+1080 I_{0,2} I_{2,0}-7560 I_{2,2}$\\
	\vspace*{0.1cm}
	
	\noindent 
	$\cI^R_{0122}=21060 I_{0,1} I_{1,0}^3-49032 I_{1,1} I_{1,0}^2-31374 I_{0,1} I_{2,0} I_{1,0}+30186 I_{2,1} I_{1,0}+32400 I_{1,1} I_{2,0}+6102 I_{0,1} I_{3,0}$\\
	\vspace*{0.1cm}
	
	\noindent 
	$\cI^R_{1122}=-8991 I_{0,1} I_{1,0}^2+21303 I_{1,1} I_{1,0}+7668 I_{0,1} I_{2,0}+405 I_{2,1}$\\
	\vspace*{0.1cm}
	
	\noindent 
	$\cI^R_{1221}=-11367 I_{0,1} I_{1,0}^2+16389 I_{1,1} I_{1,0}+17496 I_{0,1} I_{2,0}+15201 I_{2,1}$\\
	\vspace*{0.1cm}
	
	\noindent 
	$\cI^R_{01122}=-53649 I_{0,1} I_{1,0}^2+18927 I_{1,1} I_{1,0}+17928 I_{0,1} I_{2,0}-7749 I_{2,1}$\\
	\vspace*{0.1cm}
	
	\noindent 
	$\cI^R_{01122}=-53649 I_{0,1} I_{1,0}^2+18927 I_{1,1} I_{1,0}+17928 I_{0,1} I_{2,0}-7749 I_{2,1}$\\
	\vspace*{0.1cm}
	
	\noindent 
	$\cI^R_{2112}=19575 I_{0,1} I_{1,0}^2+23085 I_{1,1} I_{1,0}-28188 I_{0,1} I_{2,0}+4779 I_{2,1}$\\
	\end{adjustwidth}
	
	\clearpage
	\bibliographystyle{apsrev4-1_title}
	\bibliography{biblio.bib}
\end{document}